\documentclass[3p]{elsarticle}

\pdfoutput=1

\usepackage{graphicx}
\usepackage{amsmath}
\usepackage{amssymb}
\usepackage{bbold}
\usepackage{subfigure}
\usepackage{color}

\DeclareMathOperator{\Expect}{E}

\DeclareMathOperator{\Probability}{P}

\begin{document}

\title{Evolutionary game theory: Temporal and spatial effects\\ beyond
replicator dynamics}

\author[gisc]{Carlos P.\ Roca}
\ead{cproca@math.uc3m.es}
\author[gisc]{Jos\'e A.\ Cuesta}
\ead{cuesta@math.uc3m.es}
\author[gisc,bifi,icmat]{Angel S\'anchez}
\ead{anxo@math.uc3m.es}

\address[gisc]{Grupo Interdisciplinar de Sistemas Complejos (GISC), Departamento de Matem\'aticas,
Universidad Carlos III de Madrid, Avenida de la Universidad 30, 28911 Legan\'es, Madrid, Spain}
\address[bifi]{Instituto de Biocomputaci\'on y F\'{\i}sica de Sistemas Complejos (BIFI), Universidad
de Zaragoza, Corona de Arag\'on 42, 50009 Zaragoza, Spain}
\address[icmat]{Instituto de Ciencias Matem\'aticas CSIC-UAM-UC3M-UCM, 28006 Madrid, Spain}

\begin{abstract}
Evolutionary game dynamics is one of the most fruitful frameworks for studying
evolution in different disciplines, from Biology to Economics. Within this context,
the approach of choice for many researchers is the so-called replicator equation,
that describes mathematically the idea that those individuals performing better
have more offspring and thus their frequency in the population grows. While very many
interesting results have been obtained with this equation in the three decades
elapsed since it was first proposed, it is important to realize the limits of its applicability.
One particularly relevant issue in this respect is that of non-mean-field
effects, that
may arise from temporal fluctuations or from spatial correlations, both neglected in
the replicator equation. This review discusses these temporal and spatial effects
focusing on the non-trivial modifications they induce when compared to the outcome
of replicator dynamics. Alongside this question, the hypothesis of linearity
and its relation to the choice of the rule for strategy update is also
analyzed. The discussion is presented in terms of the emergence of
cooperation, as one of the current key problems in Biology and in other disciplines.
\end{abstract}

\begin{keyword}
evolutionary games, replicator dynamics, mean-field, fluctuations, spatial structure,
network reciprocity, emergence of cooperation

\PACS {02.50.Le} 
 \sep 05.40.-a 	
 \sep 87.23.Ge 	
 \sep 87.23.Kg 	
 \sep {89.65.-s} 
 \sep  {89.75.Fb} 

 \MSC[2008] 91A22 
 \sep 91A43 
 \sep  92D15 
\end{keyword}

\maketitle

\tableofcontents

\section{Introduction}

The importance of evolution can hardly be overstated, in so far as it permeates all
sciences. Indeed, in the 150 years that have passed since the publication of
{\em On the Origin of the Species} \cite{darwin:1859}, the original idea of Darwin
that evolution takes place through descent with modification acted upon by
natural selection has become a key concept in many sciences. Thus, nowadays
one can speak of course of evolutionary biology, but there are also evolutionary
disciplines in economics, psychology, linguistics, or computer science, to name a
few.

Darwin's theory of evolution was based on the idea of natural selection. Natural
selection is the process through which favorable heritable traits become more
common in successive generations of a population of reproducing organisms,
displacing unfavorable traits in the struggle for resources. In order to cast
this process in a mathematically
precise form, J.\ B.\ S.\ Haldane and Sewall Wright introduced, in the so-called
modern evolutionary synthesis of the 1920's, the concept of fitness. They applied
theoretical population ideas to the description of evolution and, in that context, they
defined fitness as the expected number of offspring of an individual that reach
adulthood. In this way they were able to come up with a well-defined measure of
the adaptation of individuals and species to their environment.

The simplest mathematical theory of evolution one can think of arises when one
assumes that the fitness of a species does not depend on the distribution of frequencies
of the different species in the population, i.e., it only depends on factors that are intrinsic
to the species under consideration or on environmental influences. Sewall Wright
formalized this idea in terms of fitness landscapes ca.\ 1930, and in that context
R.\ Fisher proved his celebrated theorem, that states that the mean fitness of a
population is a non-decreasing function of time, which increases proportionally
to variability. Since then, a lot of work has been done on
this kind of models; we refer the reader to
\cite{roughgarden:1979,peliti:1996,drossel:2001,ewens:2004} for
reviews.

The approach in terms of fitness landscapes is, however, too simple and, in general,
it is clear that the fitness of a species will depend on the composition of the population
and will therefore change accordingly as the population evolves. If one wants to
describe evolution at this level, the tool of reference is
evolutionary game theory. Brought
into biology by Maynard Smith \cite{maynard-smith:1982}
as an
``exaptation''\footnote{Borrowing the term introduced by Gould and Vrba in evolutionary
theory, see \cite{gould:1982}.}
of the game theory developed originally for economics \cite{morgenstern:1947},
it has since become a unifying framework
for other disciplines, such as sociology or anthropology \cite{gintis:2000}.
The key feature of this mathematical apparatus is that it allows to deal with evolution on a
frequency-dependent fitness landscape or, in other words, with strategic interactions
between entities, these being individuals, groups, species, etc. Evolutionary game theory
is thus the generic approach to evolutionary dynamics \cite{nowak:2006a} and contains
as a special case constant, or fitness landscape, selection.

In its thirty year history,
a great deal of research in evolutionary game theory
has focused on the properties and applications of the replicator
equation \cite{hofbauer:1998}. The replicator equation was introduced in
1978 by Taylor and Jonker \cite{taylor:1978} and describes the evolution of
the frequencies of population
types taking into account their mutual influence on their fitness. This important
property allows the replicator equation to capture the essence of selection and, among
other key results, it provides a connection between the biological concept of
evolutionarily stable strategies \cite{maynard-smith:1982}
with the economical concept of Nash equilibrium \cite{nash:1950}.

As we will see below, the replicator equation
is derived in a specific framework that involves a number
of assumptions, beginning with that of an infinite, well-mixed population with
no mutations. By well-mixed population it is understood that every individual
either interacts with
every other
one or at least has the same probability to interact with any other individual
in the population.
This hypothesis implies that any individual effectively
interacts with a player which uses the average strategy within the
population (an approach that has been
traditionally
used in physics under the name of mean-field approximation).
Deviations from the well-mixed population scenario
affect strongly and non-trivially the outcome of the evolution, in a way which
is difficult
to apprehend in principle. Such deviations can
arise when one considers, for instance, finite
size populations, alternative learning/reproduction dynamics, or some kind of
structure
(spatial or temporal) in the interactions between individuals.

In this review we will
focus on this last point, and discuss the consequences of relaxing the
hypothesis that
every player interacts or can interact with every other one.  We will address
both
spatial and temporal
limitations in this paper, and refer the reader to
Refs.~\cite{nowak:2006a,hofbauer:1998} for discussions of other perturbations.
For the sake of definiteness, we will consider those effects, that go beyond
replicator dynamics, in the specific context of the emergence of cooperation,
a problem of paramount importance with implications at all levels,
from molecular biology to societies and ecosystems
\cite{pennisi:2005}; many
other applications of evolutionary dynamics
have also been proposed but it would be too lengthy to discuss all of them
here (the interested reader should
see, e.g., \cite{nowak:2006a}). Cooperation, understood as a fitness-decreasing
behavior that increases others' fitness,
is an evolutionary puzzle, and many researchers have considered
alternative approaches to the replicator equation as possible explanations of
its
ubiquity in human (and many animal) societies.
As it turns out, human behavior is unique in nature. Indeed, altruism or
cooperative behavior exists in other species, but it can be
understood in terms of genetic relatedness (kin selection,
introduced by Hamilton \cite{hamilton:1964a,hamilton:1964b}) or of repeated
interactions (as proposed by Trivers \cite{trivers:1971}).
Nevertheless, human cooperation extends to genetically unrelated
individuals and to large groups, characteristics that cannot
be understood within those schemes. Subsequently, a number of
theories based on group and/or cultural evolution have been put
forward in order to explain altruism (see
\cite{hammerstein:2003} for a review). Evolutionary game theory is also being
intensively used for this research, its main virtue being that it allows to
pose the dilemmas involved in cooperation in a simple, mathematically
tractable manner. To date, however, there is not a generally accepted
solution to this puzzle \cite{nowak:2006b}.

Considering temporal and spatial effects means, in the language of
physics, going beyond mean-field
to include fluctuations and correlations. Therefore, a first step is to
understand what are the
basic mean field results. To this end, in Section~\ref{sec:2} we briefly
summarize the main
features of replicator equations and introduce the concepts we will refer to
afterwards.
Subsequently, Section~\ref{sec:3} discusses how fluctuations
can be taken into account in evolutionary game theory, and specifically we will
consider that,
generically, interactions and dynamics (evolution) need not occur at the same
pace. We will
show that the existence of different time scales leads to quite unexpected
results, such as
the survival and predominance of individuals that would be the less fit in the
replicator
description. For games in finite populations with two types of individuals or
strategies,
the problem can be understood in terms of Markov processes and the games can be
classified
according to the influence of the time scales on their equilibrium structure. 
Other situations
can be treated by means of numerical simulations with similarly non-trivial
results.

Section~\ref{sec:4} deals with spatial effects.
The inclusion of population structure in evolutionary game theory has been
the subject of intense research in the last 15 years, and a complete review
would be beyond
our purpose (see e.g.\ \cite{szabo:2007}). The existence of a network describing
the possible interactions
in the population has been identified as one of the factors that may promote
cooperation among
selfish individuals \cite{nowak:2006b}. We will discuss the results available to
date and show how they can
be reconciled by realizing the role played by different networks, different
update rules for the
evolution of strategies and the equilibrium structure of the games. As a result,
we will be
able to provide a clear-cut picture of the extent as to which population
structure promote
cooperation in two strategy games.

Finally, in Section~\ref{sec:5} we discuss the implications of the reviewed
results on a more general
context. Our major conclusion will be the lack of generality of models in
evolutionary game
theory, where details of the dynamics and the interaction modify qualitatively
the results.
A behavior like that is not intuitive to physicists, used to disregard those
details as
unimportant. Therefore, until we are able to discern what is and what is not
relevant, when dealing with problems in other sciences,
modeling properly and accurately specific problems is of utmost importance. We
will also indicate a few directions of research that arise from the
presently available
knowledge and that we believe will be most appealing in the near future.

\section{Basic concepts and results of evolutionary game theory}
\label{sec:2}

In this section, we summarize the main facts about evolutionary game theory
that we are going
to need in the remainder of the paper. The focus is on the stability of
strategies and on the
replicator equation, as an equivalent to the dynamical description of a mean
field approach
which we will be comparing with. This summary is by no means intended to be
comprehensive
and we encourage the reader to consult the review \cite{hofbauer:2003} or, for
full details,
the books \cite{maynard-smith:1982,gintis:2000,hofbauer:1998}.

\subsection{Equilibria and stability}

The simplest type of game has only two players and, as this will be the one we
will be
dealing with, we will not dwell into further complications. Player $i$ is
endowed with a
finite number  $n_i$ of strategies. A game is
defined by listing the strategies available to the players
and the payoffs they yield: When a player, using strategy $s_i$, meets another,
who in
turn uses strategy $s_j$, the former receives a payoff $W_{ij}$ whereas the
latter receives
a payoff $Z_{ij}$. We will restrict ourselves to symmetric games, in which the
roles of both
players are exchangeable (except in the example considered in
Section~\ref{sec:ultimatum}); mathematically, this means that the set of
strategies are the
same for both players and that $W=Z^T$. Matrix $W$ is then called the payoff
matrix of the
normal form of the game. In the original economic formulation
\cite{morgenstern:1947}
payoffs were understood as utilities, but Maynard Smith
\cite{maynard-smith:1982} reinterpreted
them in terms of fitness, i.e.\ in terms of reproductive success of the involved
individuals.

The fundamental step to ``solving'' the game or, in other words, to find what
strategies will
be played, was put forward by John Nash \cite{nash:1950}
by introducing the concept of equilibrium. In $2\times 2$ games, a
pair of
strategies $(s_i,s_j)$ is a Nash equilibrium if no unilateral change of strategy
allows any player to improve her payoff. When we restrict ourselves to symmetric
games, one can say simply, by an abuse of language \cite{hofbauer:2003}, that a
strategy $s_i$ is a
Nash equilibrium if it is a best reply to itself: $W_{ii}\ge W_{ij},\, \forall
s_j$ (a strict Nash equilibrium if the inequality is strict).
This in turn implies that if
both players are playing strategy $s_i$, none of them has any incentive to
deviate unilaterally
by choosing other strategy. As an example, let us consider the famous
\emph{Prisoner's Dilemma game},
which we will be discussing throughout the review. Prisoner's Dilemma was
introduced
by Rapoport and Chammah \cite{rapoport:1965} as a model of the implications of
nuclear
deterrence during the Cold War, and is given by the following payoff matrix (we
use the
traditional convention that the matrix indicates payoffs to the row player)
\begin{equation} \label{eq:anxo1}
\begin{array}{ccc}
 & \mbox{ }\, {\rm C} & \!\!\!\!\!\! {\rm D} \\
\begin{array}{c} {\rm C} \\ {\rm D} \end{array} & \left(\begin{array}{c} 3 \\ 5
\end{array}\right. & \left.\begin{array}{c} 0 \\ 1
\end{array}\right). \end{array} \end{equation}
The strategies are named C and D for cooperating and defecting, respectively.
This game is
referred to as Prisoner's Dilemma because it is usually posed in terms of two
persons
that are arrested accused of a crime. The police separates them and makes the
following
offer to them: If one confesses and incriminates the other, she will receive
a large reduction in the sentence, but if both confess they will only get a
minor
reduction; and if nobody confesses then the police is left only with
circumstancial
evidence, enough to imprison them for a short period. The
amounts of the sentence reductions are given by the payoffs in (\ref{eq:anxo1}).
It is
clear from it that D is a strict Nash equilibrium: To begin
with, it is a dominant strategy, because no matter what the column player
chooses to do, the row
player is always better off by defecting; and when both players defect, none
will improve
her situation by cooperating. In terms of the prisoners, this translates into
the fact that
both will confess if they behave rationally. The dilemma arises when one
realizes that both players would be better off cooperating, i.e.\ not
confessing, but rationality leads them unavoidable to confess.

The above discussion concerns Nash equilibria in pure strategies. However,
players can
also use the so-called mixed strategies, defined by a vector with as many
entries as
available strategies, every entry indicating the probability of using that
strategy. The notation
changes then accordingly: We use vectors ${\bf x}=(x_1\,x_2 \ldots x_n)^T$,
which are
elements of the simplex $S_n$ spanned by the vectors ${\bf e}_i$ of the standard
unit
base (vectors ${\bf e}_i$ are then identified with the $n$ pure strategies).The
definition of a Nash equilibrium in mixed strategies is identical to the
previous one:
The strategy profile ${\bf x}$ is a Nash equilibrium if it is a best reply to
itself in terms
of the expected payoffs, i.e.\ if ${\bf x}^TW{\bf x}\ge {\bf x}^TW{\bf y},\,
\forall {\bf y}\in S_n$.
Once mixed strategies have been introduced, one can prove, following Nash
\cite{nash:1950}
that every normal form game has at least one Nash equilibrium, albeit it need
not necessarily be a
Nash equilibrium in pure strategies. An example we will also be discussing below
is
given by the \emph{Hawk-Dove game} (also called \emph{Snowdrift} or
\emph{Chicken} in the literature  \cite{sugden:1986}),
introduced by Maynard Smith and Price to describe animal conflicts
\cite{maynard-smith:1973}
(strategies are labeled H and D for hawk and dove, respectively)
\begin{equation} \label{eq:anxo2}
\begin{array}{ccc}
 & \mbox{ }\, {\rm D} & \!\!\!\!\!\! {\rm H} \\
\begin{array}{c} {\rm D} \\ {\rm H} \end{array} & \left(\begin{array}{c} 3 \\ 5
\end{array}\right. & \left.\begin{array}{c} 1 \\ 0
\end{array}\right). \end{array} \end{equation}
In this case, neither H nor D are \emph{Nash equilibria}, but there is indeed
one Nash equilibrium
in mixed strategies, that can be shown \cite{maynard-smith:1982} to be given by
playing D with probability 1/3. This makes sense in terms of the meaning of the
game, which is an
anti-coordination game, i.e.\ the best thing to do is the opposite of the
other player.
Indeed, in the Snowdrift interpretation, two people are trapped by a snowdrift
at the two
ends of a road. For every one of them, the best option is not to shovel snow off
to free
the road and let the other person do it; however, if the other person does not
shovel, then
the best option is to shovel oneself. There is, hence, a temptation to defect
that creates
a dilemmatic situation (in which mutual defection leads to the worst possible
outcome).

In the same way as he reinterpreted monetary payoffs in terms of reproductive
success,
Maynard Smith reinterpreted mixed strategies as population frequencies. This
allowed
to leave behind the economic concept of rational individual and move forward to
biological
applications (as well as in other fields). As a consequence, the economic
evolutionary idea
in terms of learning new strategies gives way to a genetic transmission of
behavioral strategies
to offspring. Therefore, Maynard Smith's interpretation of the
above result is that a population consisting of one third of individuals that
always use
the D strategy and two thirds of H-strategists is a stable genetic
polymorphism. At
the core of this concept is his notion of \emph{evolutionarily stable strategy}.
Maynard Smith
defined a strategy as evolutionarily stable if the following two conditions are
satisfied
\begin{eqnarray}
\label{eq:anxo3}
{\bf x}^TW{\bf x} & \geq & {\bf x}^TW{\bf y},\, \forall {\bf y} \in S_n,\\
{\rm if}\,\, {\bf x}\neq {\bf y}\,\, {\rm and}\,\, {\bf x}^TW{\bf y}&=&{\bf
x}^TW{\bf x},\, {\rm then}\,\, {\bf x}^TW{\bf y}>
{\bf y}^TW{\bf y}.
\end{eqnarray}
The rationale behind this definition is again of a population theoretical type:
These are
the conditions that must be fulfilled for a population of ${\bf x}$-strategists
to be non-invadable
by any ${\bf y}$-mutant. Indeed, either ${\bf x}$ performs better against itself
than ${\bf y}$ or,
if they perform equally, ${\bf x}$ performs better against ${\bf y}$ than ${\bf
y}$ itself.
These two conditions guarantee non-invasibility of the population. On the other
hand,
comparing the definitions of evolutionarily stable strategy and Nash equilibrium
one
can immediately see that a strict Nash equilibrium is an evolutionarily stable
strategy
and that an evolutionarily stable strategy is a Nash equilibrium.

\subsection{Replicator dynamics}

After Nash proposed his definition of equilibrium, the main criticism that the
concept has received
relates to how equilibria are reached. In other words, Nash provided a rule to
decide which
are the strategies that rational players should play in a game, but how do
people involved in
actual game-theoretical settings but without knowledge of game theory find the
Nash
equilibrium? Furthermore, in case there is more than one Nash equilibrium, which
one
should be played, i.e., which one is the true ``solution'' of the game? These
questions
started out a great number of works dealing with learning and with refinements
of the
concept that allowed to distinguish among equilibria, particularly within the
field of
economics. This literature is out of the scope of the present review and the
reader is
referred to \cite{vegaredondo:2003} for an in-depth discussion.

One of the answers to the above criticism arises as a bonus from the ideas of
Maynard Smith.
The notion of evolutionarily stable strategy has implicit some kind of dynamics
when we
speak of invasibility by mutants; a population is stable if when a small
proportion of it
mutates it eventually evolves back to the original state. One could therefore
expect
that, starting from some random initial condition, populations would evolve to
an evolutionarily
stable strategy, which, as already stated, is nothing but a Nash equilibrium.
Thus, we would
have solved the question as to how the population ``learns'' to play the Nash
equilibrium and
perhaps the problem of selecting among different Nash equilibria. However, so
far we have
only spoken of an abstract
dynamics; nothing is specified as to what is the evolution of the population or
the strategies
that it contains.

The replicator equation, due to Taylor and Jonker \cite{taylor:1978}, was the
first and most successful proposal of an evolutionary game dynamics. Within the
population dynamics framework,
the state of the population, i.e.\ the distribution of strategy frequencies,
is given by ${\bf x}$ as above. A first key point is that we assume that the
$x_i$ are differentiable functions
of time $t$: This requires in turn assuming that the population is infinitely
large (or that
$x_i$ are expected values for an ensemble of populations). Within this
hypothesis,
we can now postulate a law of
motion for ${\bf x}(t)$. Assuming further that individuals meet randomly,
engaging in a
game with payoff matrix $W$, then $(W {\bf x})_i$ is the expected payoff for an
individual using
strategy $s_i$, and ${\bf x}^T W{	\bf x}$ is the average payoff in the
population state ${\bf x}$.
If we, consistently with our interpretation of payoff as fitness, postulate that
the per capita rate of growth of the subpopulation using strategy $s_i$ is
proportional to its payoff, we arrive at the \emph{replicator equation} (the
name was first proposed in
\cite{schuster:1983})
\begin{eqnarray}
\label{eq:anxo4}
\dot{x}_i=x_i[(W {\bf x})_i-
{\bf x}^TW{\bf x}],
\end{eqnarray}
where the term ${\bf x}^TW{\bf x}$ arises to ensure the constraint $\sum_i x_i
= 1$ ($\dot{x}_i$ denotes the time derivative of $x_i$).
This equation translates into mathematical terms the elementary principle of
natural selection:
Strategies, or individuals using a given strategy, that reproduce more
efficiently spread, displacing
those with smaller fitness. Note also that states with $x_i=1$,
$x_j=0$, $\forall j\neq i$ are solutions of
Eq.\ (\ref{eq:anxo4}) and, in fact, they are absorbing states, playing a
relevant role in the
dynamics of the system in the absence of mutation.

Once an equation has been proposed, one can resort to the tools of dynamical
systems theory
to derive its most important consequences. In this regard, it is interesting to
note that the
replicator equation can be transformed by an appropriate change of variable in a
system of
Lotka-Volterra type \cite{hofbauer:1998}.
For our present purposes, we will focus only on the relation of the
replicator dynamics with the two equilibrium concepts discussed in the preceding
subsection.
The rest points of the replicator equation are those frequency distributions
${\bf x}$ that make the
rhs of Eq.~(\ref{eq:anxo4}) vanish, i.e.\ those that verify either $x_i=0$ or
$(W {\bf x})_i = {\bf x}^TW{\bf x},\, \forall i=1,\dots,n$.
The solutions of this system of equations are all the mixed strategy Nash
equilibria of the game \cite{gintis:2000}. Furthermore, it is not difficult
to show (see e.g.\ \cite{hofbauer:1998}) that strict Nash equilibria are
asymptotically stable, and that stable rest points are Nash equilibria. We thus
see that the replicator equation provides us with an evolutionary mechanism
through which the players, or the population, can arrive at a Nash equilibrium
or, equivalently, to an evolutionarily stable strategy. The different basins of
attraction of the different equilibria further explain which of them is selected
in case there are more than one.

For our present purposes, it is important to stress the hypothesis involved
(explicitly or
implicitly) in the derivation
of the replicator equation:
\begin{enumerate}
\item The population is infinitely large.
\item Individuals meet randomly or play against every other one, such that the
payoff of
strategy $s_i$ is proportional to the payoff averaged over the current
population state ${\bf x}$.
\item There are no mutations, i.e.\ strategies increase or decrease in frequency
only due to reproduction.
\item The variation of the population is linear in the payoff difference.
\end{enumerate}

Assumptions 1 and 2 are, as we stated above, crucial to derive the replicator
equation in order
to replace the fitness of a given strategy by its mean value when the population
is described
in terms of frequencies. Of course, finite populations deviate from the values
of
frequencies corresponding to infinite ones. In a series of recent works,
Traulsen and co-workers have considered this problem
\cite{traulsen:2005,claussen:2005,traulsen:2006}.
They have identified different microscopic stochastic processes that
lead to the standard or the adjusted replicator dynamics, showing that
differences on the individual level can
lead to qualitatively different dynamics in asymmetric conflicts and,
depending on the population size, can
even invert the direction of the evolutionary process. Their analytical
framework, which they
have extended to include an arbitrary number of strategies,  provides
good approximations to simulation results for very small sizes. For a recent
review of these
and related issues, see \cite{claussen:2008}. On the other hand, there has also
been
some work showing that evolutionarily stable strategies in infinite populations
may lose
their stable character when the population is small (a result not totally
unrelated to those
we will discuss in Section~\ref{sec:3}). For examples of this in the context of
Hawk-Dove games, see \cite{fogel:1997,fogel:1998}.

Assumption 3 does not pose any severe problem. In fact, mutations
(or migrations among physically separated groups, whose mathematical description
is
equivalent) can be included, yielding the so-called replicator-mutator equation
\cite{page:2002}. This is in turn equivalent to the Price equation
\cite{price:1970},
in which a term involving the covariance of fitness and strategies appears
explicitly. Mutations have been also included in the framework of finite
size populations \cite{traulsen:2006} mentioned above.
We refer the reader to references \cite{page:2002,frank:1995}
for further analysis of this issue.

Assumption 4 is actually the core of the definition of replicator dynamics. In
Section~\ref{sec:4}
below we will come back to this point, when we discuss the relation of
replicator dynamics
to the rules used for the update of strategies in agent-based models.
Work beyond the hypothesis of
linearity can proceed otherwise in different directions, by considering
generalized
replicator equations of the form
\begin{eqnarray}
\label{eq:anxo5}
\dot{x}_i=x_i[W_i({\bf x})-
{\bf x}^TW({\bf x})].
\end{eqnarray}
The precise choice for the functions $W_i({\bf x})$ depends of course on the
particular situation
one is trying to model.
A number of the results on replicator equation carry on for several such
choices.
This topic is well summarized in \cite{hofbauer:2003} and the interested reader
can proceed
from there to the relevant references.

Assumption 2 is the one to which this review is devoted to and, once again,
there are very many different possibilities in which it may not hold. We will
discuss in depth below the
case in which the time scale of selection is faster than that of interaction,
leading to the
impossibility that a given player can interact with all others. Interactions may
be also
physically limited, either for geographical reasons (individuals interact only
with those
in their surroundings), for social reasons (individuals interact only with those
with whom
they are acquainted) or otherwise. As in previous cases, these variations
prevents one from
using the expected value of fitness of a strategy in the population as a good
approximation
for its growth rate. We will see the consequences this has in the following
sections.

\subsection{The problem of the emergence of cooperation}

One of the most important problems to which evolutionary game theory is being
applied
is the understanding of the emergence of cooperation in human (albeit
non-exclusively)
societies \cite{pennisi:2005}. As we stated in the Introduction, this is an
evolutionary
puzzle that can be accurately expressed within the formalism of game theory. One
of
the games that has been most often used in connection with this problem is the
Prisoner's Dilemma introduced above, Eq.~(\ref{eq:anxo1}). As we have seen,
rational players should unavoidably defect and never cooperate, thus leading
to a very bad outcome for both players. On the other hand, it is evident that if
both players had cooperated they would have been much better off. This is a
prototypical example of a social dilemma \cite{kollock:1998} which is, in fact,
(partially) solved in societies.
Indeed, the very existence of human society, with its highly specialized labor
division, is a proof that cooperation is possible.

In more biological terms, the question can be phrased using again the concept of
fitness.
Why should an individual help other achieve more fitness, implying more
reproductive
success and a chance that the helper is eventually displaced? It is important to
realize
that such a cooperative effort is at the roots of very many biological phenomena,
from
mutualism to the appearance of multicellular organisms
\cite{maynard-smith:1995}.

When one considers this problem in the framework of replicator equation, the conclusion
is immediate and disappointing: Cooperation is simply not possible. As defection is the only
Nash equilibrium of Prisoner's Dilemma, for any initial condition with
a positive fraction of defectors, replicator dynamics will inexorably
take the population to a final state in which they all are defectors. Therefore, one needs to
understand how the replicator equation framework can be supplemented or superseded
for evolutionary game theory to become closer to what is observed in the real world
(note that
there is no hope for classical game theory in this respect as it is based in the perfect
rationality of the players). Relaxing the above discussed assumptions leads, in
some
cases, to possible solutions to this puzzle, and our aim here is to summarize
and
review what has been done along these lines with Assumption 2.

\section{The effect of different time scales}
\label{sec:3}

Evolution is generally supposed to occur at a slow pace: Many generations
may be needed for a noticeable change to arise in a species. This is indeed
how Darwin understood the effect of natural selection, and he always
referred to its cumulative effects over very many years. However, this needs
not be the case and, in fact, selection may occur faster
than the interaction between individuals (or of the individuals with their
environment). Thus, recent
experimental studies have reported observations of fast selection
\cite{hendry:1999,hendry:2000,yoshida:2003}. It is also conceivable
that in man-monitored or laboratory processes one might make
selection be the rapid influence rather than interaction. Another
context where these ideas apply naturally is that of
cultural evolution or social learning, where the time scale of selection is
much closer to the time scale of interaction. Therefore, it is
natural to ask about the consequences of the above assumption and the
effect of relaxing it.

This issue has already been considered from an economic viewpoint in
the context of equilibrium selection (but see an early biological example breaking
the assumption of purely random matching in \cite{fagen:1980}, which considered 
Hawk-Dove games where strategists are more likely to encounter individuals
using their same strategy). This refers
to a situation in which for a game there is more than one equilibrium, like
in the \emph{Stag Hunt game}, given e.g.\ by the following payoff
matrix
\begin{equation} \label{eq:anxo6}
\begin{array}{ccc}
 & \mbox{ }\, {\rm C} & \!\!\!\!\!\! {\rm D} \\
\begin{array}{c} {\rm C} \\ {\rm D} \end{array} & \left(\begin{array}{c} 6 \\ 5
\end{array}\right. & \left.\begin{array}{c} 1\\ 2
\end{array}\right). \end{array} \end{equation}
This game was already posed as a metaphor by Rousseau \cite{skyrms:2003},
which reads as follows: Two people go out hunting for stag, because two
of them are necessary to hunt down such a big animal. However, any one of them
can cheat the other by hunting hare, which one can do alone, leaving the other
one in the impossibility of getting the stag. Therefore, we have a coordination
game, in which the best option is to do as the other: Hunt stag together or both
hunting hare separately.
In game theory, this translates into the fact that
both C and D are Nash equilibria, and in principle one is
not able to determine which one would be selected by the players, i.e.,  which
one is the solution of the game. One rationale to choose was proposed by
Harsanyi and Selten\footnote{Harsanyi and Senten received the Nobel Prize
in Economics for this contribution, along with Nash, in 1994.}
\cite{harsanyi:1988}, who classified C as the Pareto-efficient
equilibrium (hunting stag is more profitable than hunting hare),
because that is the most beneficial for both players, and D as the
risk-dominant equilibrium, because it is the strategy that is better in case the other
player chooses D (one can hunt hare alone). Here the tension arises then from the
risk involved in cooperation, rather than from the temptation to defect of
Snowdrift games
\cite{macy:2002} (note that both tensions are present in Prisoner's Dilemma).

Kandori {\em et al} \cite{kandori:1993} showed that the risk-dominant equilibrium is
selected when using a stochastic evolutionary game dynamics, proposed by Foster and Young
\cite{foster:1990}, that considers that every player interacts with every other one
(implying slow selection). However, fast selection leads to another
result. Indeed, Robson and Vega-Redondo \cite{robson:1996} considered the situation in which every player is matched to another one and therefore they only play one game before selection acts. In that case, they
showed that the outcome changed and that the Pareto-efficient equilibrium is selected.
This result was qualified later by Miekisz \cite{miekisz:2005}, who showed that
the selected equilibrium depended on the population size and the mutation level of the
dynamics. Recently, this issue has also been considered in \cite{traulsen:2009}, which compares the situation where the contribution of the game to the fitness is small (weak selection, see Section 4.6 below) to the one where the game is the main source of the fitness, finding that in the former the results are
equivalent to the well-mixed population, but not in the latter, where the
conclusions of \cite{robson:1996} are recovered. It is also worth noticing in this regards the works by Boylan
\cite{boylan:1992,boylan:1995}, where he studied the types of random matching that can still be
approximated by continuous equations. In any case, even if the above are not
general results and their application is mainly in economics, we have already a
hint that time scales may play a non-trivial role in evolutionary games.

In fact, as we will show below, rapid selection affects evolutionary
dynamics in such a dramatic way that for some games it even changes
the stability of equilibria. We will begin our discussion by briefly summarizing
results on a model for the emergence of altruistic behavior, in which the
dynamics is not replicator-like, but that illustrates nicely the very important
effects of fast selection. We will subsequently proceed to present a
general theory for symmetric $2 \times 2$ games. There, in order to make
explicit the relation
between selection and interaction time scales, we use a discrete-time
dynamics that produces results equivalent to the replicator dynamics
when selection is slow. We will then show that the pace at which selection
acts on the population is crucial for the appearance and stability of
cooperation. Even in non-dilemma games such as the \emph{Harmony game}
\citep{licht:1999}, where cooperation is the only possible rational
outcome, defectors may be selected for if population renewal is very
rapid.

\subsection{Time scales in the Ultimatum game}
\label{sec:ultimatum}

As a first illustration of the importance of time scales
in evolutionary game dynamics, we begin by dealing with this problem in the context of a
specific set of such experiments, related to the \emph{Ultimatum game} \cite{guth:1982,henrich:2004}.
In this game,
under conditions of anonymity, two players are shown a sum
of money. One of the players, the ``proposer'',
is instructed to offer any amount
to the other, the ``responder''. The proposer can make only one
offer, which the responder can accept or reject. If the offer is
accepted, the money is shared accordingly; if rejected, both
players receive nothing. Note that the Ultimatum game is not
symmetric, in so far as proposer and responder have clearly
different roles and are therefore not exchangeable. This will be
our only such an example, and the remainder of the paper will only
deal with symmetric games. Since the game is played only once
(no repeated interactions) and anonymously (no reputation gain;
for more on explanations of altruism relying on reputation
see \cite{nowak:1998}),
a self-interested responder will accept any amount of money
offered. Therefore, self-interested proposers will offer the
minimum possible amount, which will be accepted.

The above prediction, based on the rational character of the players,
contrasts clearly with the results of
actual Ultimatum game experiments with human subjects,
in which average offers do not even approximate the self-interested prediction.
Generally speaking, proposers offer respondents very substantial
amounts (50\% being a typical modal offer) and respondents
frequently reject offers below 30\% \cite{camerer:2003,fehr:2003}. Most of the
experiments have been carried out with university students in
western countries, showing a large degree of individual variability
but a striking uniformity between groups in average behavior.
A large study in 15 small-scale societies \cite{henrich:2004}
found that, in all
cases, respondents or proposers behave in such a reciprocal manner.
Furthermore, the behavioral variability across groups was much
larger than previously observed: While mean offers in the case
of university students are in the range 43\%-48\%, in the
cross-cultural study they ranged from 26\% to 58\%.

How does this fit in our focus topic, namely the emergence of cooperation?
The fact that indirect reciprocity is excluded by the anonymity
condition and that interactions
are one-shot (repeated interaction, the mechanism proposed by Axelrod to
foster cooperation \cite{axelrod:1981,axelrod:1984}, does not apply)
allows one to interpret rejections in terms of the so-called
strong reciprocity \cite{gintis:2000a,fehr:2002}.
This amounts to considering that
these behaviors are truly altruistic, i.e.\ that
they are costly for the individual performing them in so far as
they do not result in direct or indirect benefit. As a consequence,
we return to our evolutionary puzzle: The negative effects of
altruistic acts must decrease the altruist's fitness as compared to
that of the recipients of the benefit, ultimately leading to
the extinction of altruists. Indeed, standard evolutionary game
theory arguments applied to the Ultimatum game lead to the expectation
that, in a well-mixed population, punishers (individuals
who reject low offers) have less chance to survive
than rational
players (individuals who accept any offer) and eventually disappear.
We will now show that this conclusion
depends on the dynamics, and that different dynamics may lead to
the survival of punishers through fluctuations.

Consider a population of $N$ agents playing the Ultimatum game,
with a fixed sum of money $M$ per game.
Random pairs of players are chosen, of which one is the proposer
and another one is the respondent. In its simplest version,
we will assume that
players are capable of other-regarding behavior (empathy); consequently,
in order to optimize their gain,
proposers offer the minimum amount of money
that they would accept. Every agent has her own, fixed
acceptance threshold, $1\leq t_i\leq M$ ($t_i$ are always integer
numbers for simplicity). Agents have only one strategy:
Respondents reject any offer
smaller than their own acceptance threshold, and
accept offers otherwise.
Money
shared as a consequence of accepted offers accumulates to the
capital of each player, and is subsequently
interpreted as fitness as usual.
After $s$ games,
the agent with the overall minimum fitness is
removed (randomly picked if there are several)
and a new agent is introduced by duplicating that
with the maximum fitness, i.e.\ with the same threshold and the
same fitness (again randomly picked if there are
several). Mutation is introduced in the duplication process by
allowing changes of $\pm 1$ in the acceptance threshold of the
newly generated player with probability 1/3 each. Agents
have no memory (interactions are one-shot) and no information
about other agents (no reputation gains are possible). We note that the
dynamics of this model is not equivalent to the replicator
equation, and therefore the results do not apply directly in that context.
In fact, such an extremal dynamics leads to an amplification of the effect
of fluctuations that allows to observe more clearly the influence of time
scales. This is the reason why we believe it will help make our main
point.

\begin{figure}[t]
\centering
\includegraphics[height=.25\textheight]{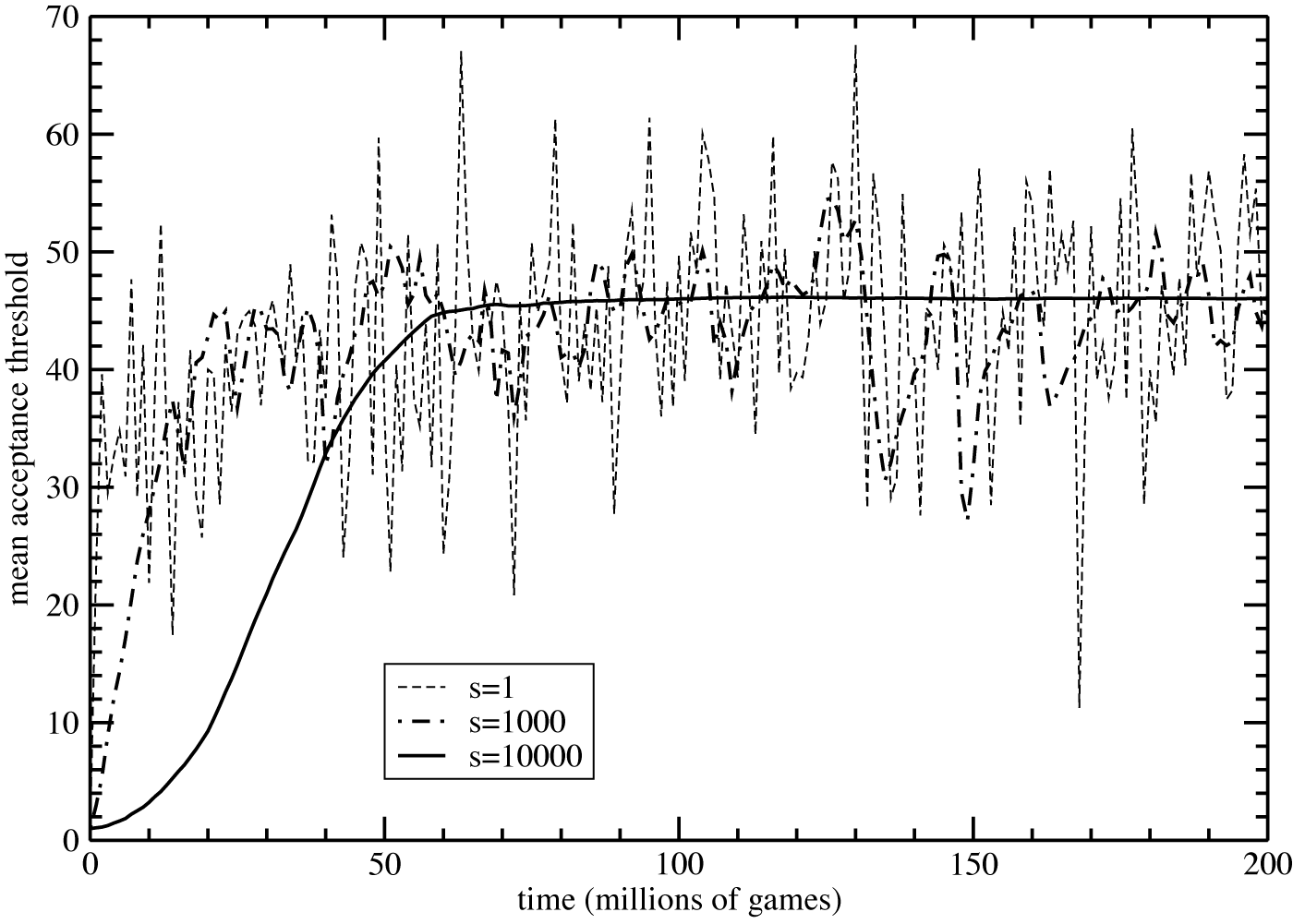}\hspace*{5mm}
\includegraphics[height=.25\textheight]{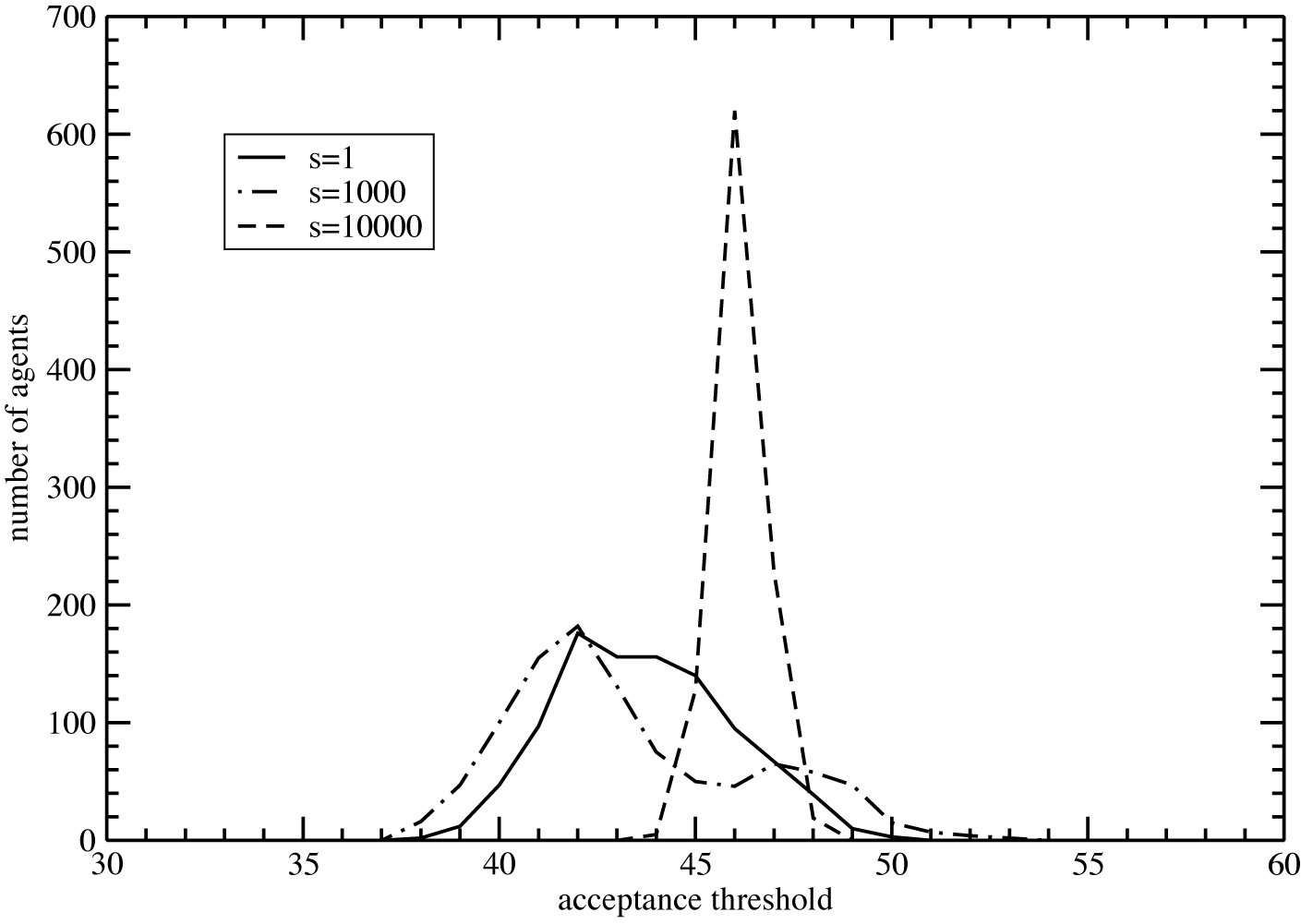}
\caption{Left: mean acceptance threshold as a function of simulation
time. Initial condition is that all agents have $t_i=1$.
Right: acceptance threshold distribution after $10^8$ games (note that this
distribution, for small $s$, is not stationary).
Initial condition is that all agents have uniformly distributed, random
$t_i$.
In both cases, $s$ is as indicated from the plot. \label{figure:anxo1}}
\end{figure}

Fig.~\ref{figure:anxo1} shows the typical outcome of simulations of our model
for
a population of $N=1000$ individuals. An important point to note
is that we are not plotting averages but a single realization for each value of $s$; the
realizations we plot are not specially chosen but rather are representative of the typical
simulation results. We have chosen to plot single realizations instead of averages to
make clear for the reader the large fluctuations arising for small $s$, which are the key
to understand the results and which we discuss below.
As we can see, the mean acceptance threshold rapidly evolves towards
values around 40\%, while the whole
distribution of thresholds converges to a peaked function, with
the range of acceptance thresholds for the agents covering about a
10\% of the available ones.
These are values compatible with the experimental results discussed
above. The
mean acceptance threshold fluctuates
during the length of the simulation, never reaching a stationary value
for the durations we have explored. The width of the peak fluctuates
as well, but in a much smaller scale than the position.
The fluctuations are larger for smaller values of $s$, and when $s$
becomes of the order of $N$ or larger, the evolution of the mean
acceptance threshold is very smooth. As is clear from Fig.~\ref{figure:anxo1}, for very small values
of $s$, the differences in payoff arising from the fact that only  some players play are
amplified by our extreme dynamics, resulting in a very noisy behavior of the
mean threshold. This is a crucial point and will
be discussed in more detail below.
Importantly, the typical evolution
we are describing does not depend on the initial condition. In particular,
a population consisting solely of self-interested agents, i.e.\ all
initial thresholds set to $t_i=1$, evolves in the same fashion.
Indeed, the distributions shown in the left panel of Fig.~\ref{figure:anxo1} (which again correspond to
single realizations) have been obtained with such an initial condition,
and it can be clearly observed that self-interested agents disappear
in the early stages of the evolution.
The number of players and the value $M$ of the capital at stake in every
game are not important either, and increasing $M$ only leads
to a higher resolution of the threshold distribution function, whereas smaller
mutation rates simply change the pace of evolution.

To realize the effect of time scales, it is important to recall
previous studies of the Ultimatum game by Page and
Nowak \cite{page:2000,page:2002a}.
The model introduced in those works has a dynamics completely
different from ours: Following standard evolutionary game theory,
every player plays with every other one in both roles (proponent and
respondent), and afterwards players reproduce with probability
proportional to their payoff (which is fitness in the reproductive
sense). Simulations and adaptive dynamics equations show that the
population ends up composed by players with fair (50\%) thresholds.
Note that this is not what one would expect on a rational basis, but
Page and Nowak traced this result back to empathy, i.e.\ the fact that
the model is constrained to offer what one would accept. In any
event, what we want to stress here is that their findings are also
different from our observations: We only reach an equilibrium for large
$s$. The reason for this difference is that the Page-Nowak model dynamics
describes the $s/N\to\infty$ limit of our model, in which between
death-reproduction events the time average gain obtained by all players is with
high accuracy a constant $O(N)$ times the mean payoff. We thus see that our
model is more general
because it has one free parameter, $s$, that allows selecting different
regimes whereas the Page-Nowak dynamics is only one limiting case.
Those different regimes are what we have described as fluctuation dominated
(when $s/N$ is finite and not too large) and the regime analyzed by
Page and Nowak (when $s/N\to\infty$).
This amounts to saying that by varying $s$ we can
study regimes far from the standard evolutionary game theory
limit. As a result, we find a variability of outcomes for the
acceptance threshold consistent with the observations in real
human societies \cite{henrich:2004,fehr:2003}. Furthermore, if one considers that the acceptance threshold and the offer can be set independently,
the results differ even more \cite{sanchez:2005}: While in the model of Page and
Nowak
both magnitudes evolve to take very low values, close to zero, in
the model presented here the results, when $s$ is small, are very similar to the one-threshold
version, leading again to values compatible with the experimental observations.
This in turn implies that rapid selection may be an alternative to empathy as
an explanation of human behavior in this game.

The main message to be taken from this example is that
 fluctuations due to the finite number of games $s$ are very important.
Among the results summarized above, the
evolution of a population entirely
formed by self-interested players into a diversified population with a
large majority of altruists is the most relevant and surprising one.
One can argue that the underlying reason for this is precisely
the presence of
fluctuations in our model. For the sake of definiteness, let us
consider the case $s=1$ (agent replacement takes place after every game)
although the discussion applies to larger (but finite) values of $s$ as
well. After one or more games, a mutation event will take place
and a ``weak altruistic punisher'' (an agent with $t_i=2$) will appear
in the population,
with a fitness inherited from its ancestor. For this new agent to be
removed at the next iteration, our model rules imply that this agent has to have
the lowest fitness, and also that it does not play as a proposer in
the next game (if playing as a responder the agent will earn nothing
because of her threshold). In any other event this altruistic punisher
will survive at least one cycle, in which an additional one can appear by mutation. It is thus clear that fluctuations indeed help altruists to take
over: As soon as a few altruists are present in the population, it is
easy to see analytically that they will survive and proliferate even
in the limit $s/N\to\infty$.

\subsection{Time scales in symmetric binary games}

The example in the previous subsection suggests that there certainly is an issue of
relative time scales in evolutionary game theory that can have serious implications.
In order to gain insight into this question, it is important to consider a general framework,
and therefore we will now look at the general problem of symmetric $2\times 2$ games.
Asymmetric games can be treated similarly, albeit in a more cumbersome manner, and
their classification involves many more types; we feel, therefore, that the symmetric
case is a much clearer illustration of the effect of time scales. In what follows, we
review and extend previous results of us \cite{roca:2006,roca:2007}, emphasizing
the consequences of the existence of different time scales.

Let us consider a population of $N$ individuals, each of whom plays with a fixed
strategy, that can be either C or D (for ``cooperate'' and ``defect''
respectively, as in Section~\ref{sec:2}). We denote the payoff that an
$X$-strategist gets when confronted to a $Y$-strategist ($X$ and $Y$ are C or
D) by the matrix element $W_{XY}$.
For a certain time individuals interact with other individuals in pairs randomly
chosen from the
population. During these interactions individuals collect payoffs. We shall refer to the
interval between two interaction events as the \emph{interaction time}. Once the interaction
period has finished reproduction occurs, and in steady state selection acts immediately
afterwards restoring
the population size to the maximum allowed by the environment. The time between two of these
reproduction/selection events will be referred to as the \emph{evolution time}.

Reproduction and selection can be implemented in at least two different ways. The first one
is through the Fisher-Wright process \cite{ewens:2004} in which each individual generates a
number of offspring proportional to her payoff. Selection acts by randomly killing
individuals of the new generation until restoring the size of the population back to $N$
individuals. The second option for the evolution is the Moran process \cite{ewens:2004,moran:1962}. It amounts
to randomly choosing an individual for reproduction proportionally to payoffs, whose single offspring replaces another randomly chosen individual, in this case with a probability $1/N$ equal for all. In this manner populations always remains constant. The Fisher-Wright process is
an appropriate model for species which produce a large number of offspring in the next
generation but only a few of them survive, and the next generation replaces the previous one
(like insects or some fishes). The Moran process is a better description for species which
give rise to few offspring and reproduce in continuous time, because individuals neither
reproduce nor die simultaneously, and death occurs at a constant rate. The original process
was generalized to the frequency-dependent fitness context of evolutionary game theory
by Taylor {\em et al.} \cite{taylor:2004}, and used to study the conditions for selection favoring the invasion and/or fixation
of new phenotypes. The results were found to depend on whether the population was infinite or
finite, leading to a classification of the process in three or eight scenarios, respectively.

Both the Fisher-Wright and Moran processes define Markov chains
\cite{karlin:1975,grinstead:1997}
on the population, characterized by the number of its C-strategists $n\in\{0,1,\dots,N\}$,
because in both cases it is assumed that the composition of the next generation is
determined solely by the composition of the current generation. Each process defines a
stochastic matrix $P$ with elements $P_{n,m}=p(m|n)$, the probability that the next generation
has $m$ C-strategists provided the current one has $n$. While for the Fisher-Wright process
all the elements of $P$ may be nonzero, for the Moran process the only nonzero
elements are those
for which $m=n$ or $m=n\pm 1$. Hence Moran is, in the jargon of Markov chains, a
\emph{birth-death process} with two absorbing states, $n=0$ and $n=N$
\cite{karlin:1975,grinstead:1997}. Such a process is mathematically simpler, and for this
reason it will be the one we will choose for our discussion on the effect of time scales.

To introduce explicitly time scales we will implement the Moran process in the
following way, generalizing the proposal by Taylor {\em et al.}
\cite{taylor:2004}. During $s$ time steps pairs of individuals will be chosen to
play, one pair every time step. After
that the above described reproduction/selection process will act according to
the payoffs
collected by players during the $s$ interaction steps. Then, the payoffs of all players are set
to zero and a new cycle starts. Notice that in general players
will play a different number of times ---some not at all--- and this will reflect in the
collected payoffs. If $s$ is too small most players will not have the opportunity to play
and chance will have a more prominent role in driving the evolution of the population.

Quantifying this effect requires that we first compute the probability that, in
a population of $N$ individuals of which $n$ are C-strategists, an
$X$-strategist is chosen
to reproduce after the $s$ interaction steps. Let $n_{XY}$ denote the number of
pairs of $X$- and $Y$-strategists that are chosen to play. The probability of
forming
a given pair, denoted $p_{XY}$, will be
\begin{equation}
p_{\text{CC}}=\frac{n(n-1)}{N(N-1)}, \qquad
p_{\text{CD}}=2\frac{n(N-n)}{N(N-1)}, \qquad
p_{\text{DD}}=\frac{(N-n)(N-n-1)}{N(N-1)}.
\end{equation}
Then the probability of a given set of $n_{XY}$ is dictated by the
multinomial distribution
\begin{equation}
M(\{n_{XY}\};s)=
\begin{cases}
\displaystyle s!\frac{p_{\text{CC}}^{n_{\text{CC}}}}{n_{\text{CC}}!}
\frac{p_{\text{CD}}^{n_{\text{CD}}}}{n_{\text{CD}}!}
\frac{p_{\text{DD}}^{n_{\text{DD}}}}{n_{\text{DD}}!}, &
\text{if $\displaystyle n_{\text{CC}}+n_{\text{CD}}+n_{\text{DD}}=s$,} \\
0, & \text{otherwise.}
\end{cases}
\label{eq:multinomial}
\end{equation}
For a given set of variables $n_{XY}$, the payoffs collected by C- and
D-strategists are
\begin{equation}
W_{\text{C}}=2n_{\text{CC}} W_{\text{CC}}+n_{\text{CD}} W_{\text{CD}}, \qquad
W_{\text{D}}=n_{\text{CD}} W_{\text{DC}}+2n_{\text{DD}} W_{\text{DD}}.
\end{equation}
Then the probabilities of choosing a C- or D-strategist for reproduction are
\begin{equation}
P_{\text{C}}(n)=\Expect_M\left[\frac{W_{\text{C}}}{W_{\text{C}}+W_{\text{D}}}\right], \qquad
P_{\text{D}}(n)=\Expect_M\left[\frac{W_{\text{D}}}{W_{\text{C}}+W_{\text{D}}}\right],
\end{equation}
where the expectations $\Expect_M\left[ \cdot \right]$ are taken over the
probability distribution $M$ (\ref{eq:multinomial}). Notice that we have to
guarantee $W_X \ge 0$ for the above expressions to define
a true probability. This forces us to choose all payoffs $W_{XY}\ge 0$.
In addition, we have studied the
effect of adding a baseline fitness to every player, which is equivalent to a
translation of the payoff matrix $W$, obtaining the same qualitative results
(see below).

Once these probabilities are obtained the Moran process accounts for the transition probabilities
from a state with $n$ C-strategists to another with $n\pm 1$ C-strategists. For $n\to n+1$
a C-strategist must be selected for reproduction (probability $P_{\text{C}}(n)$) and a
D-strategist for being replaced (probability $(N-n)/N$). Thus
\begin{equation}
P_{n,n+1}=p(n+1|n)=\frac{N-n}{N}P_{\text{C}}(n).
\end{equation}
For $n\to n-1$ a D-strategist must be selected for reproduction (probability $P_{\text{D}}(n)$)
and a C-strategist for being replaced (probability $n/N$). Thus
\begin{equation}
P_{n,n-1}=p(n-1|n)=\frac{n}{N}P_{\text{D}}(n).
\end{equation}
Finally, the transition probabilities are completed by
\begin{equation}
P_{n,n}=1-P_{n,n-1}-P_{n,n+1}.
\end{equation}

\subsubsection{Slow selection limit}

Let us assume that $s\to\infty$, i.e.\ the evolution time is much longer than the interaction time.
Then the distribution (\ref{eq:multinomial}) will be peaked at the values
$n_{XY}=
sp_{XY}$, the larger $s$ the sharper the peak. Therefore in this limit
\begin{equation}
P_{\text{C}}(n)\to\frac{\overline{W_{\text{C}}}(n)}{\overline{W_{\text{C}}}
(n)+\overline{W_{\text{D}}}(n)}, \qquad
P_{\text{D}}(n)\to\frac{\overline{W_{\text{D}}}(n)}{\overline{W_{\text{C}}}(n)+\overline{W_{\text{D}}}(n)},
\end{equation}
where
\begin{equation}
\overline{W_{\text{C}}}(n)=\frac{n}{N}\left[\frac{n-1}{N-1}(W_{\text{CC}}-W_{\text{CD}})+W_{\text{CD}}
\right], \qquad
\overline{W_{\text{D}}}(n)=\frac{N-n}{N}\left[\frac{n}{N-1}(W_{\text{DC}}-W_{\text{DD}})+W_{\text{DD}}
\right].
\end{equation}

In general, for a given population size $N$ we have to resort to a numerical
evaluation of the various quantities that characterize a birth-death process,
according to the formulas in Appendix~\ref{app:A}. However, for large $N$ the
transition probabilities can be expressed in terms of
the fraction of C-strategists $x=n/N$ as
\begin{eqnarray}
P_{n,n+1} &=& x(1-x)\frac{w_{\text{C}}(x)}{xw_{\text{C}}(x)+(1-x)w_{\text{D}}(x)}, \\
P_{n,n-1} &=& x(1-x)\frac{w_{\text{D}}(x)}{xw_{\text{C}}(x)+(1-x)w_{\text{D}}(x)},
\end{eqnarray}
where
\begin{equation}
w_{\text{C}}(x)=x(W_{\text{CC}}-W_{\text{CD}})+W_{\text{CD}}, \qquad
w_{\text{D}}(x)=x(W_{\text{DC}}-W_{\text{DD}})+W_{\text{DD}}.
\label{eq:wCwD}
\end{equation}

The terms $w_{\text{C}}$ and $w_{\text{D}}$ are, respectively, the expected payoff of a cooperator and a defector in this case of large $s$ and $N$.
The factor $x(1-x)$ in front of $P_{n,n+1}$ and $P_{n,n-1}$ arises as a consequence of
$n=0$ and $n=N$ being absorbing states of the process. There is another equilibrium $x^*$
where $P_{n,n\pm 1}=P_{n\pm 1,n}$, i.e.\ $w_{\text{C}}(x^*)=w_{\text{D}}(x^*)$,
with $x^*$ given by
\begin{equation}
x^*=\frac{W_{\text{CD}}-W_{\text{DD}}}{W_{\text{DC}}-W_{\text{CC}}+W_{\text{CD}}-W_{\text{DD}}}.
\label{eq:xstar}
\end{equation}
For $x^*$ to be a valid equilibrium $0<x^*<1$ we must have
\begin{equation}
(W_{\text{DC}}-W_{\text{CC}})(W_{\text{CD}}-W_{\text{DD}})>0.
\label{eq:mixedeq}
\end{equation}
This equilibrium is stable\footnote{Here the notion of stability implies that the process will
remain near $x^*$ for an extremely long time, because as long as $N$ is finite, no matter how
large, the process will eventually end up in $x=0$ or $x=1$, the absorbing states.} as long as
the function $w_{\text{C}}(x)-w_{\text{D}}(x)$ is decreasing at $x^*$, for then if $x<x^*$
$P_{n,n+1}>P_{n+1,n}$ and if $x>x^*$ $P_{n,n-1}>P_{n-1,n}$, i.e.\ the process will tend to restore the
equilibrium, whereas if the function is increasing the process will be led out of $x^*$ by
any fluctuation. In terms of (\ref{eq:wCwD}) this implies
\begin{equation}
W_{\text{DC}}-W_{\text{CC}}>W_{\text{DD}}-W_{\text{CD}}.
\label{eq:stability}
\end{equation}
Notice that the two conditions
\begin{equation}
w_{\text{C}}(x^*)=w_{\text{D}}(x^*), \qquad
w'_{\text{C}}(x^*)<w'_{\text{D}}(x^*),
\end{equation}
are precisely the conditions arising from the replicator dynamics for $x^*$ to
be a stable equilibrium \cite{nowak:2006a,hofbauer:1998}, albeit expressed in a
different manner than in Section~\ref{sec:2} ($w'_X$ represents the
derivative of $w_X$ with respect to $x$).
Out of the classic dilemmas, condition (\ref{eq:mixedeq}) holds for Stag Hunt and Snowdrift games,
but condition (\ref{eq:stability}) only holds for the latter. Thus, as we have
already seen, only Snowdrift has a dynamically stable mixed population.

This analysis leads us to conclude that the standard setting of evolutionary games as advanced above, in which the
time scale for reproduction/selection is implicitly (if not explicitly) assumed to be much longer
than the interaction time scale, automatically yields the distribution of equilibria dictated by
the replicator dynamics for that game. We have explicitly shown this to be true for binary games,
but it can be extended to games with an arbitrary number of strategies. In the
next section we will
analyze what happens if this assumption on the time scales does not hold.

\subsubsection{Fast selection limit}

When $s$ is finite, considering all the possible pairings and their payoffs, we
arrive at
\begin{equation}
\begin{split}
P_{\text{C}}(n)=\sum_{j=0}^s \sum_{k=0}^{s-j} &
2^{s-j-k} \frac{s!n^{s-k}(n-1)^j(N-n)^{s-j}(N-n-1)^k}{j!k!(s-j-k)!N^s(N-1)^s} \\
&\times \frac{2jW_{\text{CC}}+(s-j-k)W_{\text{CD}}}{2jW_{\text{CC}}+2kW_{\text{DD}}
+(s-j-k)(W_{\text{CD}}+W_{\text{DC}})},
\end{split}
\end{equation}
and $P_{\text{D}}(n)=1-P_{\text{C}}(n)$. We have not been able to write this
formula in a simpler way, so we have to evaluate it numerically for
every choice of the payoff matrix. However, in order to have
a glimpse at the effect of reducing the number of interactions between
successive
reproduction/selection events, we can examine analytically the extreme case
$s=1$, for which
\begin{eqnarray}
P_{n,n+1} &=& \frac{n(N-n)}{N(N-1)}\left[\frac{2W_{\text{CD}}}{W_{\text{DC}}+W_{\text{CD}}}+
\frac{n}{N}\frac{W_{\text{DC}}-W_{\text{CD}}}{W_{\text{DC}}+W_{\text{CD}}}-\frac{1}{N}\right], \\
P_{n,n-1} &=& \frac{n(N-n)}{N(N-1)}\left[1+\frac{n}{N}\frac{W_{\text{DC}}-W_{\text{CD}}}{W_{\text{DC}}
+W_{\text{CD}}}-\frac{1}{N}\right].
\end{eqnarray}
From these equations we find that
\begin{equation}
\frac{P_{n,n-1}}{P_{n,n+1}}=\frac{Dn+S(N-1)}{D(n+1)+S(N-1)-D(N+1)}, \qquad
D= W_{\text{DC}}-W_{\text{CD}}, \qquad
S= W_{\text{DC}}+W_{\text{CD}},
\end{equation}
and this particular dependence on $n$ allows us to find the following closed-form expression for
$c_n$, the probability that starting with $n$ cooperators the population ends up with all cooperators (see Appendix~\ref{app:B})
\begin{equation}
c_n=\frac{R_n}{R_N}, \qquad
R_n=
\begin{cases}
\displaystyle \prod_{j=1}^n\frac{S(N-1)+Dj}{S(N-1)-D(N+1-j)}-1, &
\text{if $D\ne 0$,} \\
n, & \text{if $D=0$.}
\end{cases}
\label{eq:cn}
\end{equation}

The first thing worth noticing in this expression is that it only depends on the two off-diagonal
elements of the payoff matrix (through their sum, $S$, and difference, $D$). This means that in an
extreme situation in which the evolution time is so short that it only allows a single pair of
players to interact, the outcome of the game only depends on what happens when two players with
different strategies play. The reason is obvious: Only those two players that have been chosen to
play will have a chance to reproduce. If both players have strategy $X$, an
$X$-strategist will be
chosen to reproduce with probability $1$. Only if each player uses a different strategy the choice of the player that reproduces will depend on the payoffs, and in this case they are precisely $W_{\text{CD}}$ and $W_{\text{DC}}$.
Of course, as $s$ increases this effect crosses over to recover the outcome for the case $s\to\infty$.

We can extend our analysis further for the case of large populations. If we denote
$x=n/N$ and $c(x)=c_n$, then we can write, as $N\to\infty$,
\begin{equation}
c(x) \sim \frac{e^{N\phi(x)}-1}{e^{N\phi(1)}-1}, \qquad
\phi(x) = \int_0^x\left[\ln(S+Dt)-\ln(S+D(t-1))\right]\,dt.
\end{equation}
Then
\begin{equation}
\phi'(x)=\ln\left(\frac{S+Dx}{S+D(x-1)}\right),
\end{equation}
which has the same sign as $D$, and hence $\phi(x)$ is increasing for $D>0$ and decreasing for
$D<0$.

Thus if $D>0$, because of the factor $N$ in the argument of the exponentials and the
fact that $\phi(x)>0$ for $x>0$, the exponential will increase sharply with $x$. Then, expanding
around $x=1$,
\begin{equation}
\phi(x)\approx \phi(1)-(1-x)\phi'(1),
\end{equation}
so
\begin{equation}
c(x)\sim \exp\{-N\ln(1+D/S)(1-x)\}.
\label{eq:asympDp}
\end{equation}
The outcome for this case is that absorption will take place at $n=0$ for almost any initial
condition, except if we start very close to the absorbing state $n=N$, namely
for $n \gtrsim N-1/\ln(1+D/S)$.

On the contrary, if $D<0$ then $\phi(x)<0$ for $x>0$ and the exponential will be peaked at $0$.
So expanding around $x=0$,
\begin{equation}
\phi(x)\approx x\phi'(0)
\end{equation}
and
\begin{equation}
c(x)\sim 1-\exp\{-N\ln(1-D/S)x\}.
\label{eq:asympDm}
\end{equation}
The outcome in this case is therefore symmetrical with respect to the case $D>0$, because
now the probability of ending up absorbed into $n=N$ is $1$ for nearly all initial conditions
except for a small range near $n=0$ determined by $n \lesssim 1/\ln(1-D/S)$.
In both cases the
range of exceptional initial conditions increases with decreasing $|D|$, and in
particular when $D=0$ the evolution becomes neutral,\footnote{Notice that if $D=0$ then $W_{\text{DC}}=
W_{\text{CD}}$ and therefore the evolution does not favor any of the two strategies.}
as it is reflected in the fact that in that special case $c_n=n/N$ (cf.\ Eq.~(\ref{eq:cn}))
\cite{ewens:2004}.

In order to illustrate the effect of a finite $s$, even in the case when $s>1$, we will consider
all possible symmetric $2\times 2$ games. These were classified by Rapoport and Guyer
\cite{rapoport:1966} in $12$ non-equivalent classes which, according to their
Nash equilibria
and their dynamical behavior under replicator dynamics, fall into three different categories:
\begin{enumerate}[(i)]
\item Six games have $W_{\text{CC}}>W_{\text{DC}}$ and $W_{\text{CD}}>W_{\text{DD}}$, or
$W_{\text{CC}}<W_{\text{DC}}$ and $W_{\text{CD}}<W_{\text{DD}}$. For them, their unique Nash
equilibrium corresponds to the dominant strategy (C in the first case and D in the second case).
This equilibrium is the global attractor of the replicator dynamics.

\item Three games have $W_{\text{CC}}>W_{\text{DC}}$ and $W_{\text{CD}}<W_{\text{DD}}$.
They have several Nash equilibria, one of them with a mixed strategy, which is an unstable
equilibrium of the replicator dynamics and therefore acts as a separator of the basins of
attractions of two Nash equilibria in pure strategies, which are the attractors.

\item The remaining three games have $W_{\text{CC}}<W_{\text{DC}}$ and $W_{\text{CD}}>W_{\text{DD}}$.
They also have several Nash equilibria, one of them with a mixed strategy, but in this case
this is the global attractor of the replicator dynamics.
\end{enumerate}
Examples of the first category are the Harmony and Prisoner's Dilemma games.
Category (ii) includes the Stag Hunt game, whereas the Snowdrift game belongs
to category (iii).

\begin{figure}[t]
\subfigure[]{\includegraphics[width=55mm,clip=]{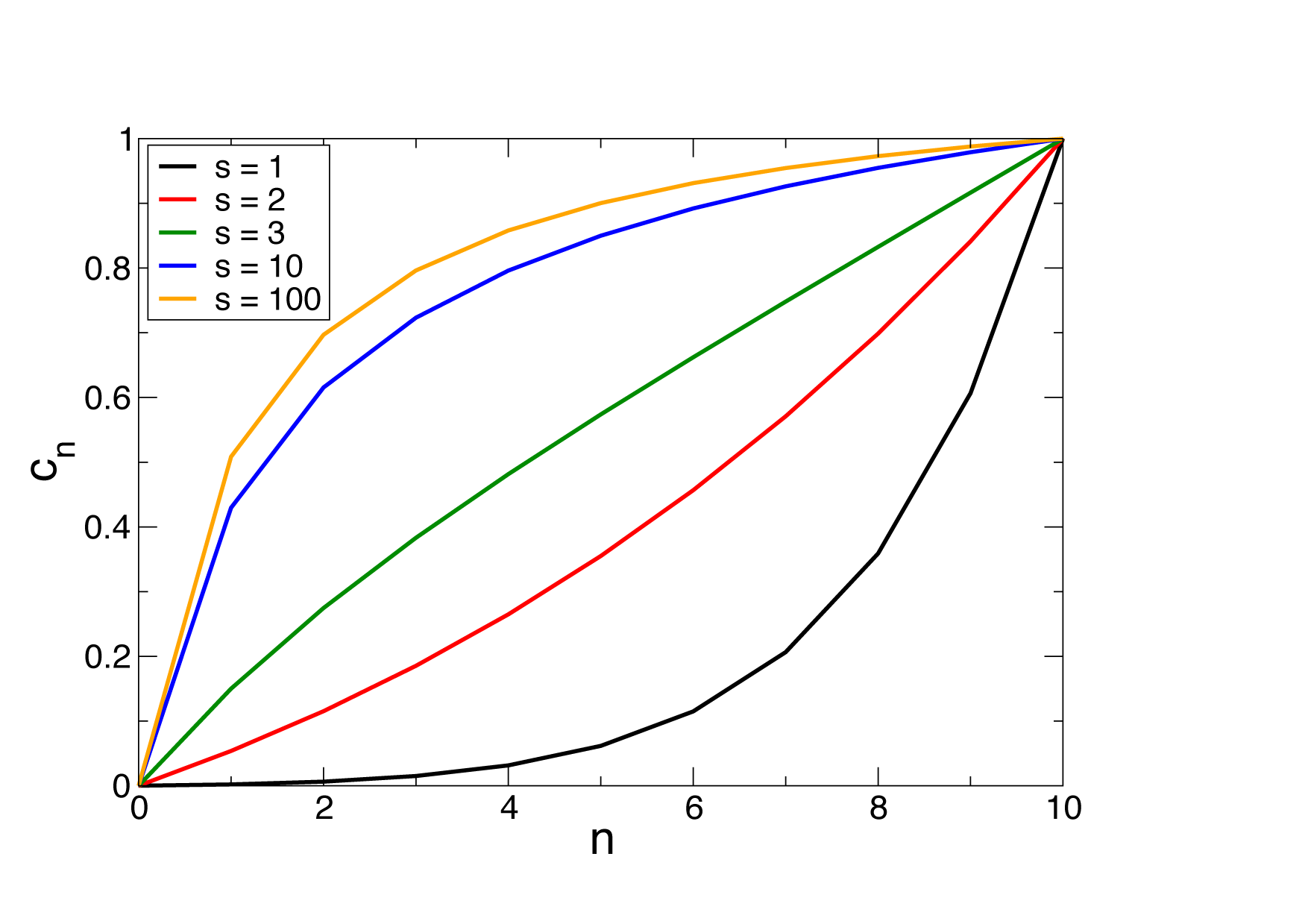}}
\subfigure[]{\includegraphics[width=55mm,clip=]{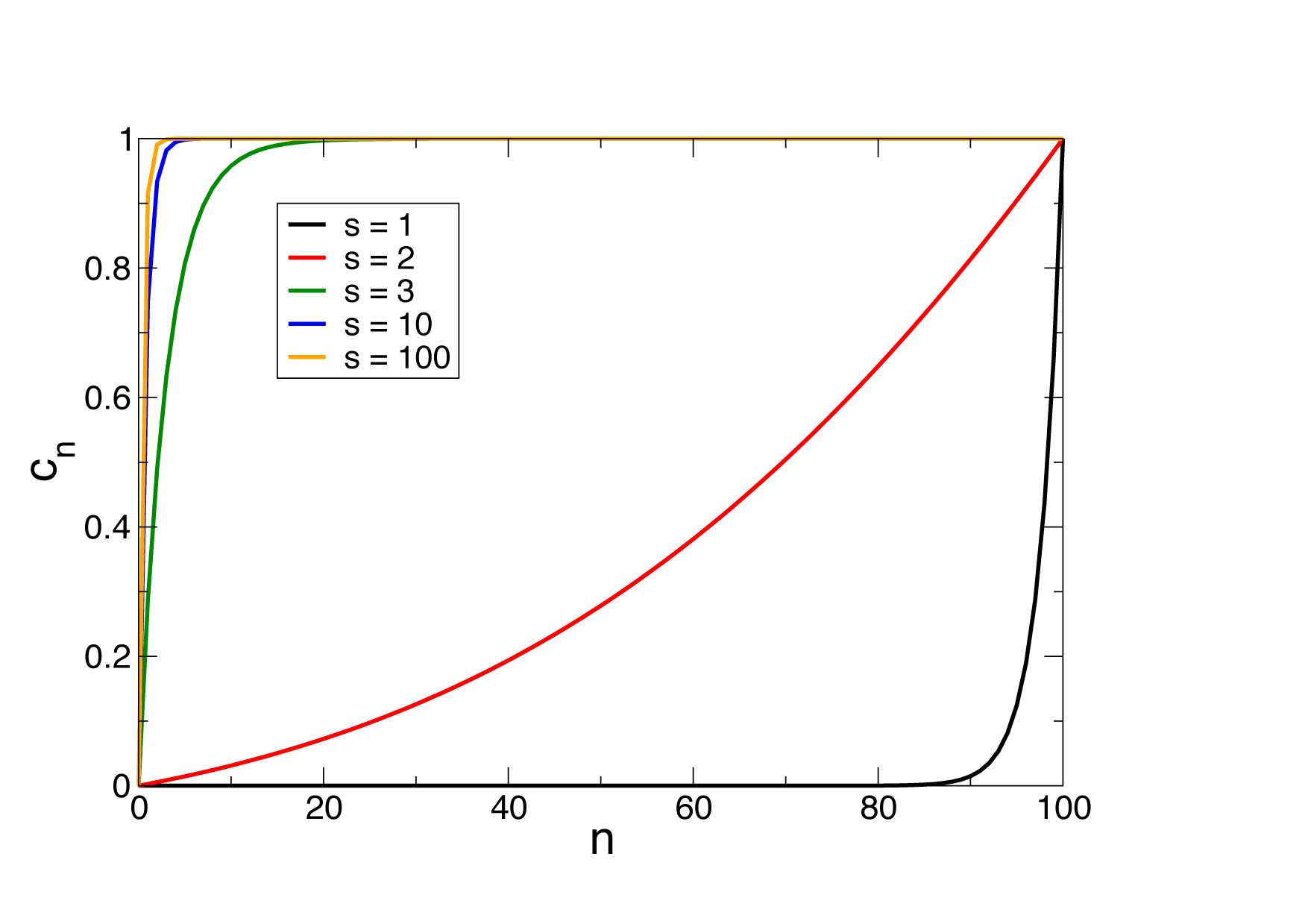}}
\subfigure[]{\includegraphics[width=55mm,clip=]{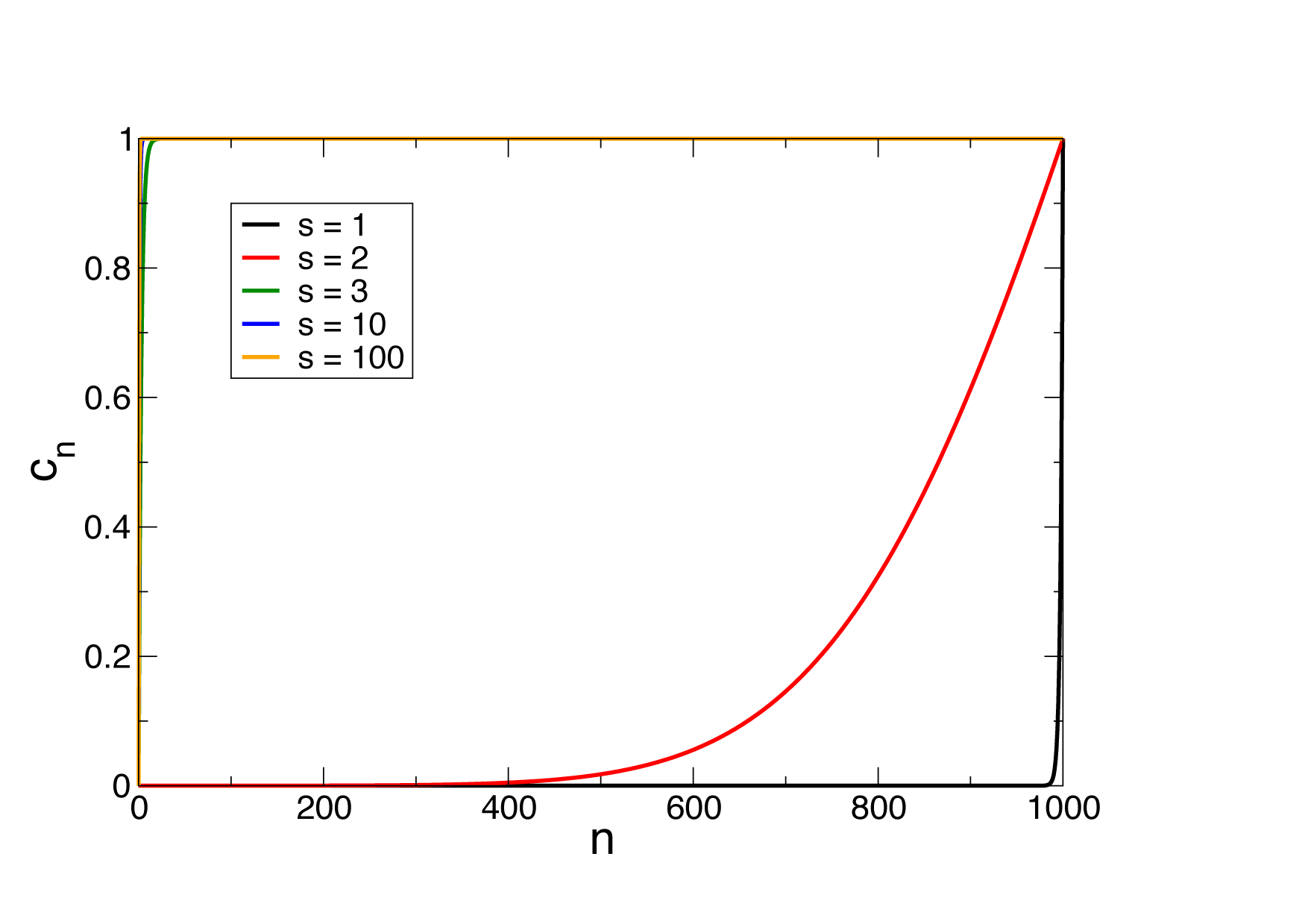}}
\caption[]{Absorption probability $c_n$ to state $n=N$ starting from initial
state $n$, for a Harmony game (payoffs $W_{\text{CC}}=1$,
$W_{\text{CD}}=0.25$, $W_{\text{DC}}=0.75$ and
$W_{\text{DD}}=0.01$), population sizes $N=10$ (a), $N=100$ (b) and $N=1000$
(c), and for values of $s=1$, $2$, $3$, $10$ and $100$. The values for $s=100$
are indistinguishable from the results of replicator dynamics.}
\label{fig:harmony}
\end{figure}

\begin{figure}[t]
\centering{\includegraphics[width=85mm,clip=]{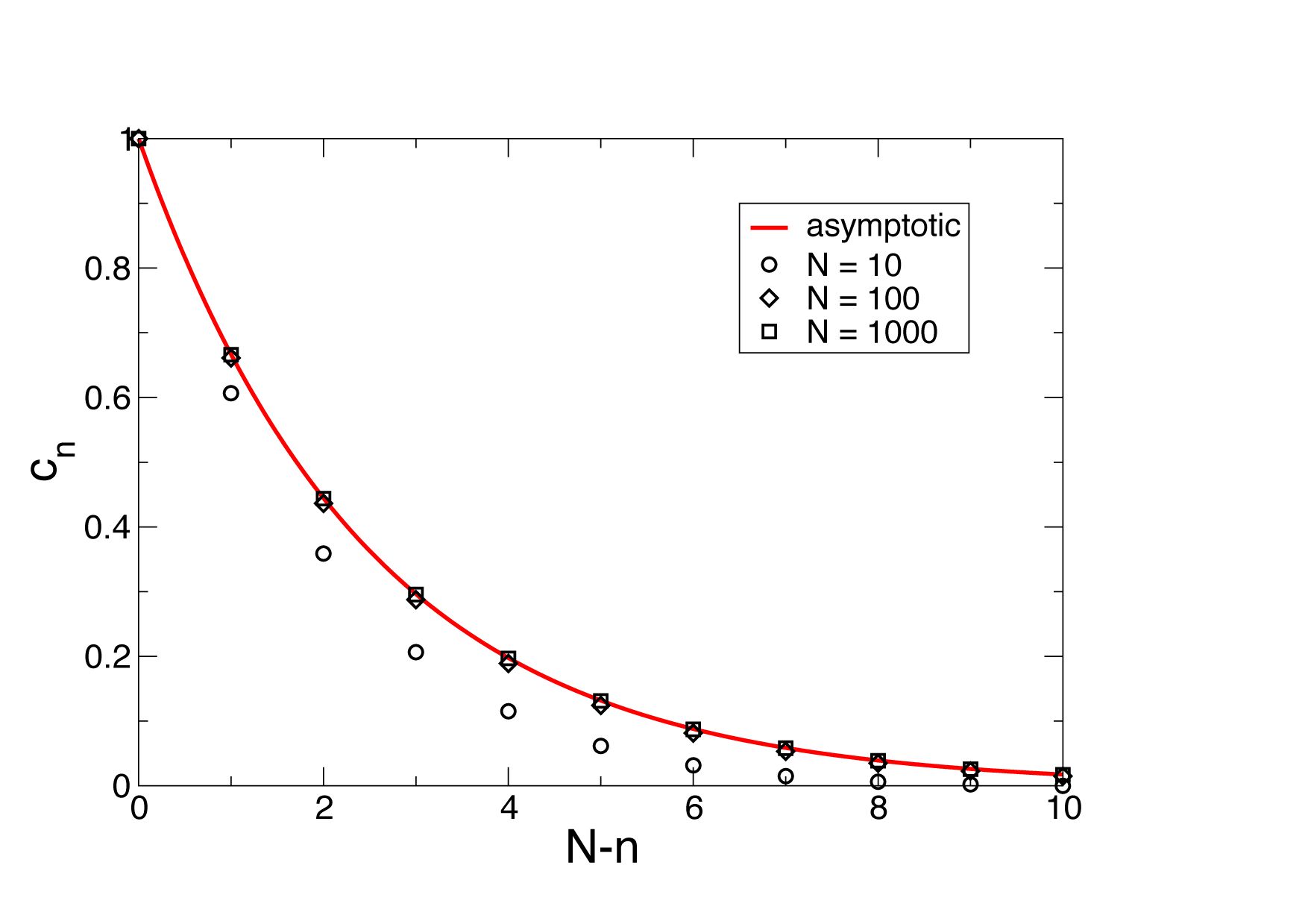}}
\caption[]{Same as in Fig.~\ref{fig:harmony} plotted
against $N-n$, for $s=1$ and $N=10$, $100$ and $1000$. The solid line is the asymptotic prediction
(\ref{eq:asympDp}).}
\label{fig:asympt}
\end{figure}

We will begin by considering one example of category (i): the Harmony game. To that aim we
will choose the parameters $W_{\text{CC}}=1$, $W_{\text{CD}}=0.25$, $W_{\text{DC}}=0.75$
and $W_{\text{DD}}=0.01$. The name of this game refers to the fact that it represents no conflict,
in the sense that all players get the maximum payoff by following strategy C. The values of $c_n$
obtained for different populations $N$ and several values of $s$ are plotted in Fig.~\ref{fig:harmony}.
The curves for large $s$ illustrate the no-conflicting character of this game as the probability
$c_n$ is almost $1$ for every starting initial fraction of C-strategists. The results for small
$s$ also illustrate the effect of fast selection, as the inefficient strategy, D, is
selected for almost any initial fraction of C-strategists. The effect is more pronounced the
larger the population. The crossover between the two regimes takes place at $s=2$ or $3$, but it
depends on the choice of payoffs. A look at Fig.~\ref{fig:harmony} reveals that
the crossing over to the $s\to\infty$ regime as $s$ increases has no connection
whatsoever with $N$, because it occurs nearly at the
same values for any population size $N$. It does depend, however, on the
precise values of the payoffs. As a further check, in Fig.~\ref{fig:asympt} we
plot the results for $s=1$ for different population sizes $N$ and
compare with the asymptotic prediction (\ref{eq:asympDp}), showing its
great accuracy for values of $N=100$ and higher; even for $N=10$ the deviation
from the exact results is not large.

\begin{figure}[t]
\subfigure[]{\includegraphics[width=55mm,clip=]{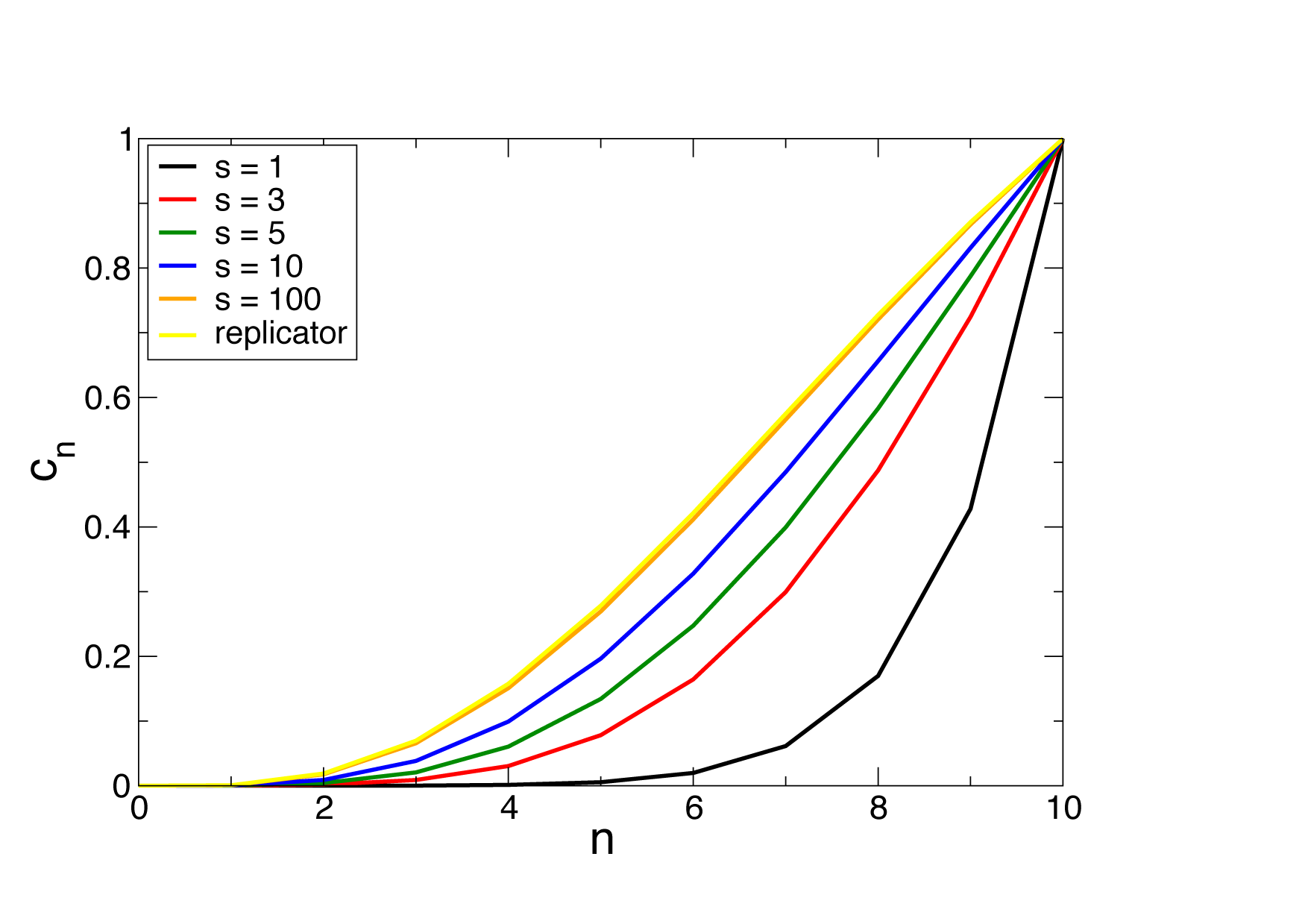}}
\subfigure[]{\includegraphics[width=55mm,clip=]{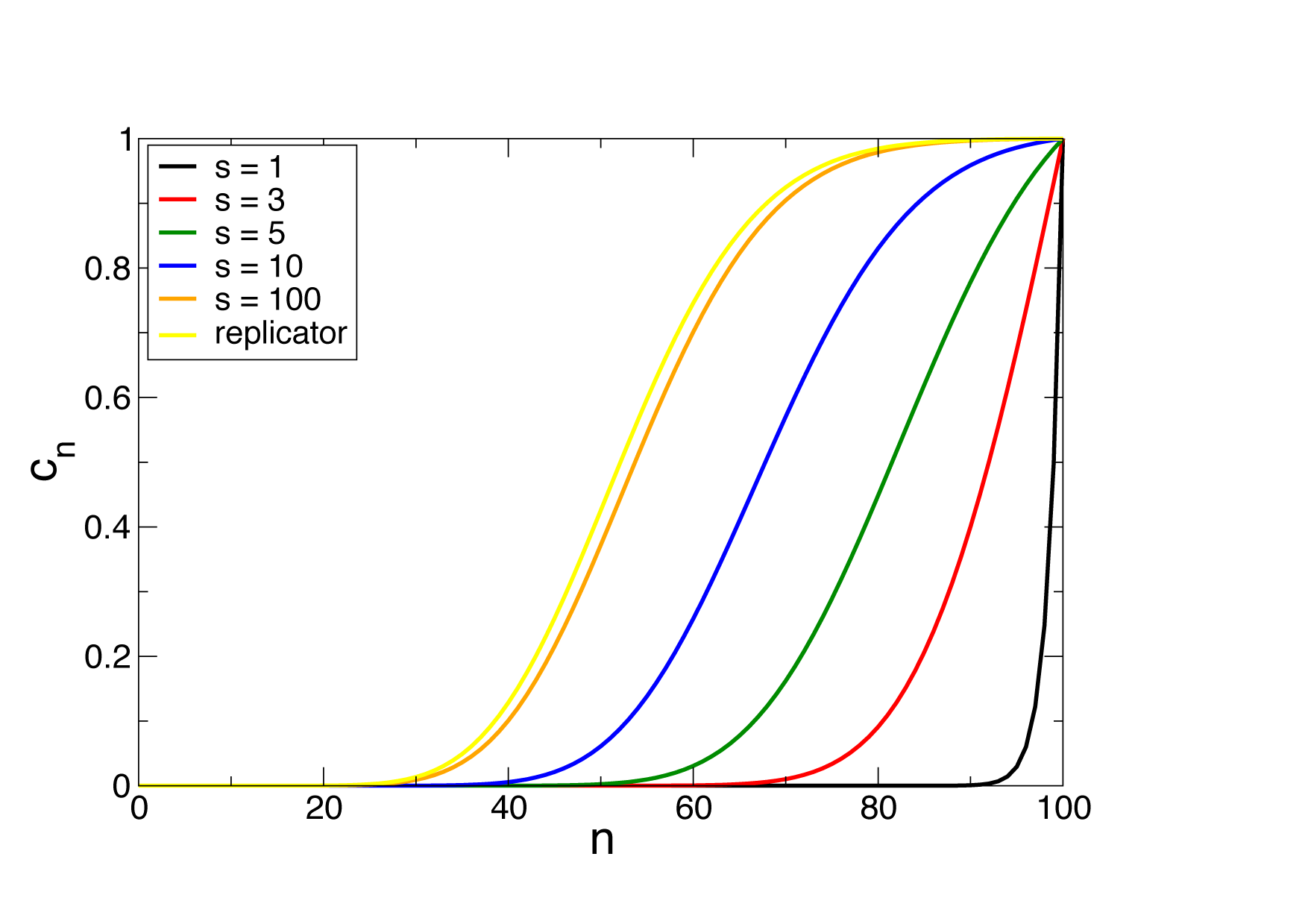}}
\subfigure[]{\includegraphics[width=55mm,clip=]{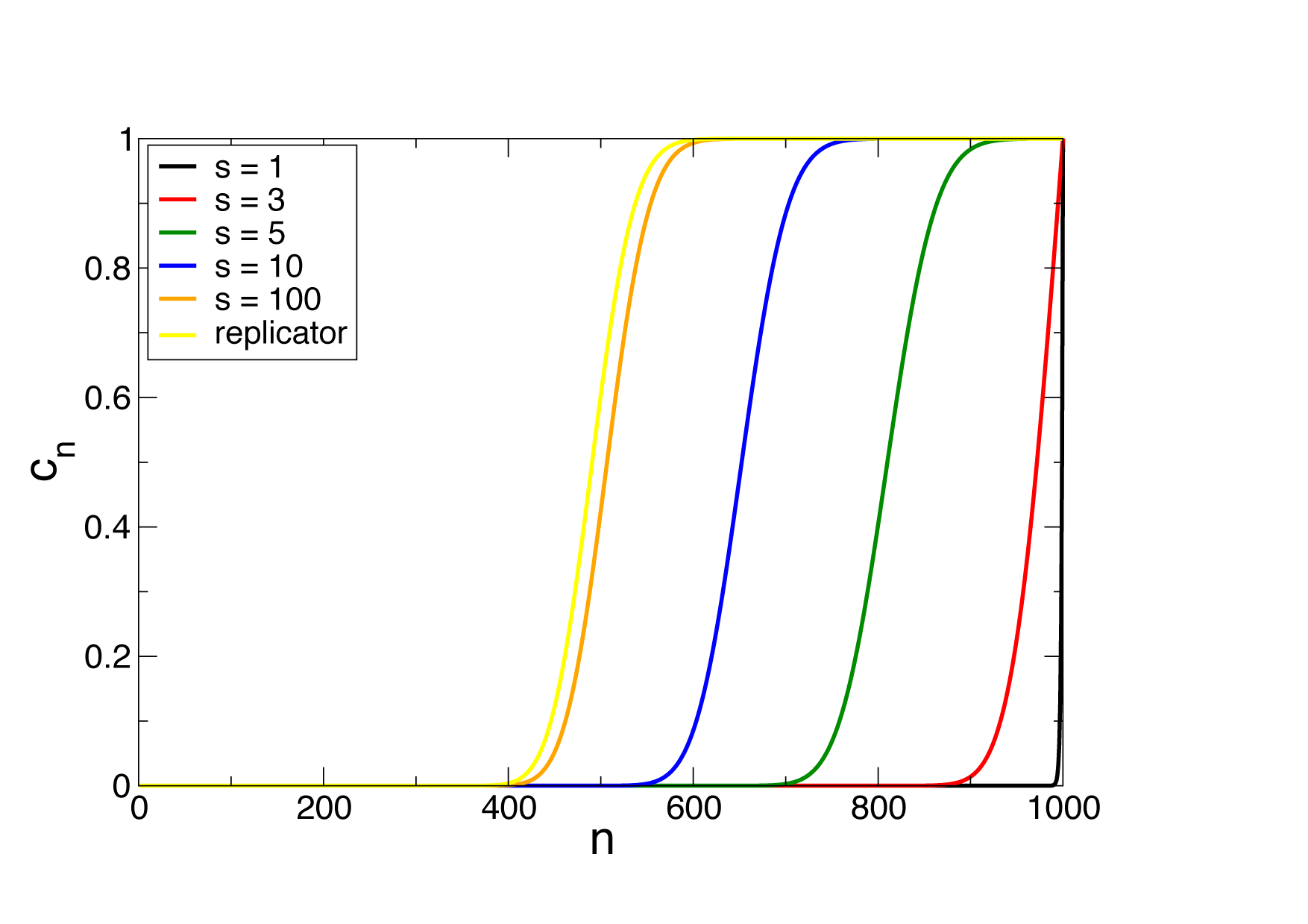}}
\caption[]{Absorption probability $c_n$ to state $n=N$ starting from initial
state $n$, for a Stag Hunt game (payoffs $W_{\text{CC}}=1$,
$W_{\text{CD}}=0.01$, $W_{\text{DC}}=0.8$ and
$W_{\text{DD}}=0.2$), population sizes $N=10$ (a), $N=100$ (b) and $N=1000$
(c), and for values
of $s=1$, $3$, $5$, $10$ and $100$. Results from replicator dynamics are also plotted for
comparison.}
\label{fig:stag-hunt}
\end{figure}

Let us now move to category (ii), well represented by the Stag Hunt game,
discussed
in the preceding subsection.
We will choose for this game the payoffs $W_{\text{CC}}=1$,
$W_{\text{CD}}=0.01$, $W_{\text{DC}}=0.8$ and $W_{\text{DD}}=0.2$. The values of $c_n$ obtained for
different populations $N$ and several values of $s$ are plotted in Fig.~\ref{fig:stag-hunt}. The panel (c) for $s=100$ reveals the behavior of the system according to the
replicator dynamics: Both strategies are attractors, and the crossover fraction
of C-strategists separating the two basins of attraction (given by
Eq.~(\ref{eq:xstar})) is, for this case, $x^*
\approx 0.49$. We can see that the effect of decreasing $s$ amounts to shifting
this crossover
towards $1$, thus increasing the basins of attraction of the risk-dominated strategy. In the extreme
case $s=1$ this strategy is the only attractor. Of course, for small population sizes
(Fig.~\ref{fig:stag-hunt}(a)) all these effects (the existence of the threshold and its shifting
with decreasing $s$) are strongly softened, although still noticeable. An
interesting feature of this game is that the effect of a finite $s$ is more
persistent compared
to what happens to the Harmony game. Whereas in the latter the replicator
dynamics
was practically recovered, for values of $s \geq 10$ we have to go up to $s=100$
to find the same in Stag Hunt.

\begin{figure}[t]
\begin{center}
\subfigure[]{\includegraphics[width=75mm,clip=]{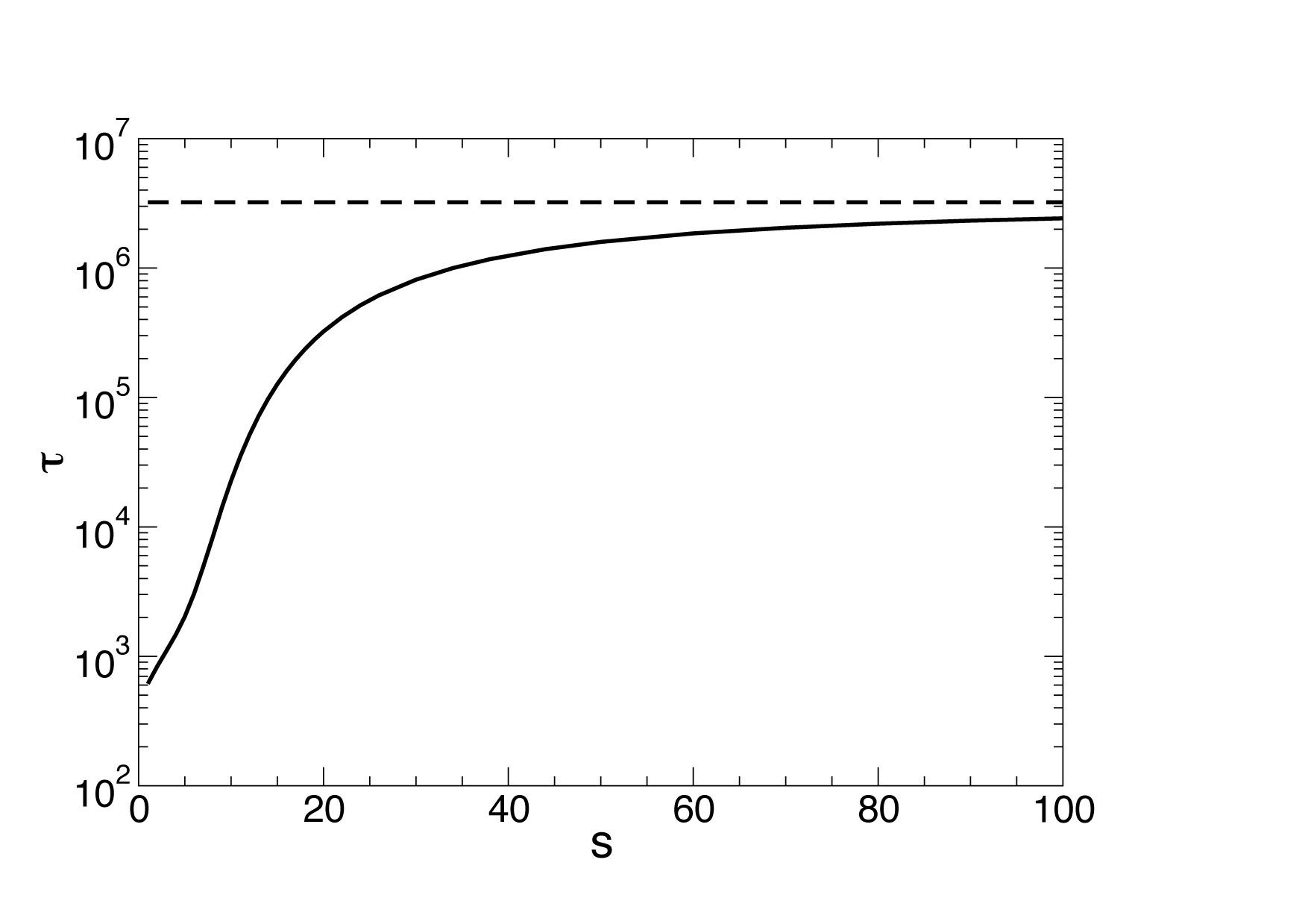}}\qquad
\subfigure[]{\includegraphics[width=75mm,clip=]{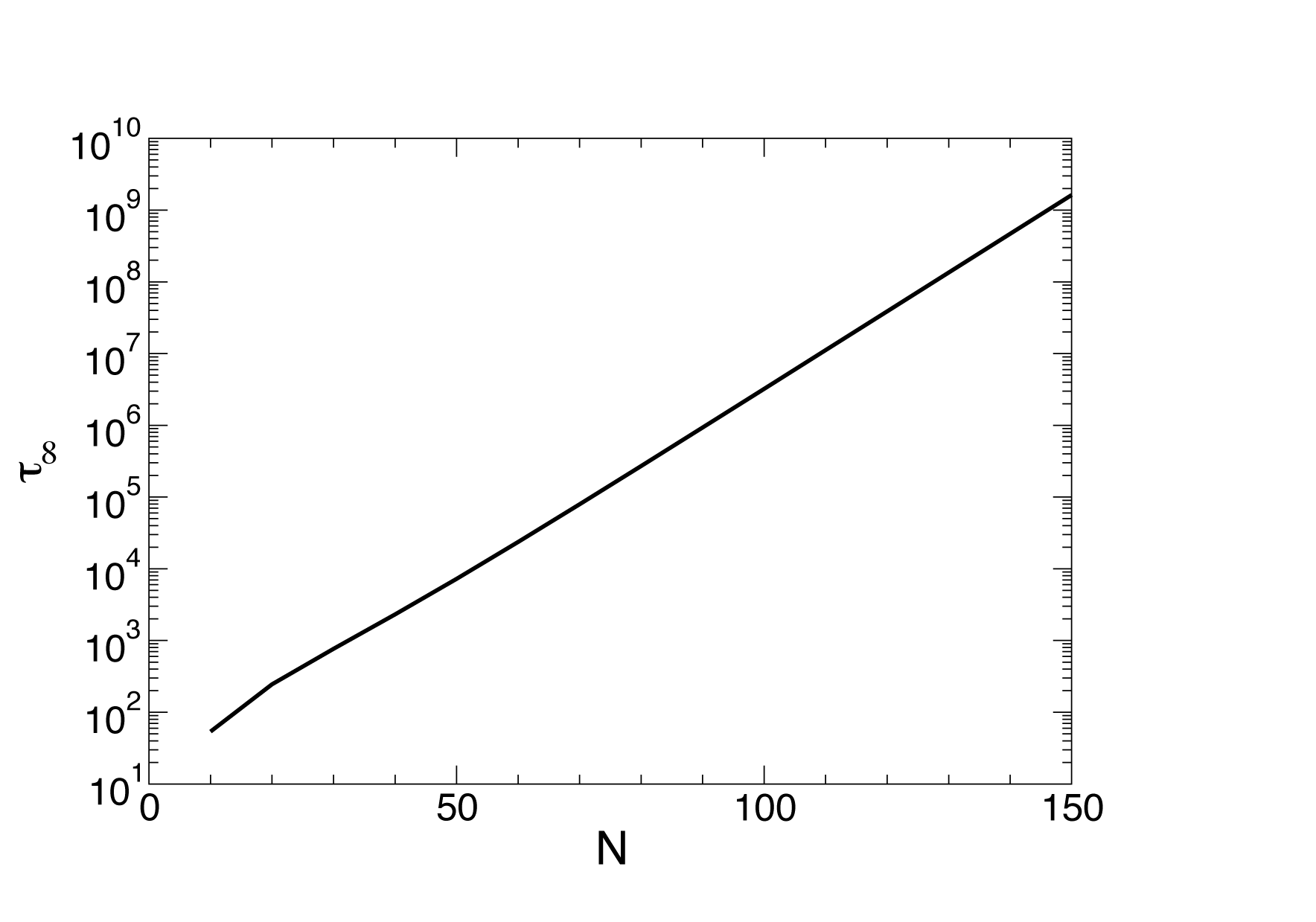}}
\end{center}
\caption[]{Absorption time starting from the state $n/N=0.5$ for a Snowdrift
game (payoffs
$W_{\text{CC}}=1$, $W_{\text{CD}}=0.2$, $W_{\text{DC}}=1.8$ and
$W_{\text{DD}}=0.01$), as a function of $s$ for population size $N=100$ (a) and
as a function of $N$ in the limit $s\to\infty$ (b). Note the logarithmic scale
for the absorption time.}
\label{fig:snowdrift}
\end{figure}

\begin{figure}[t]
\begin{center}
\includegraphics[width=85mm,clip=]{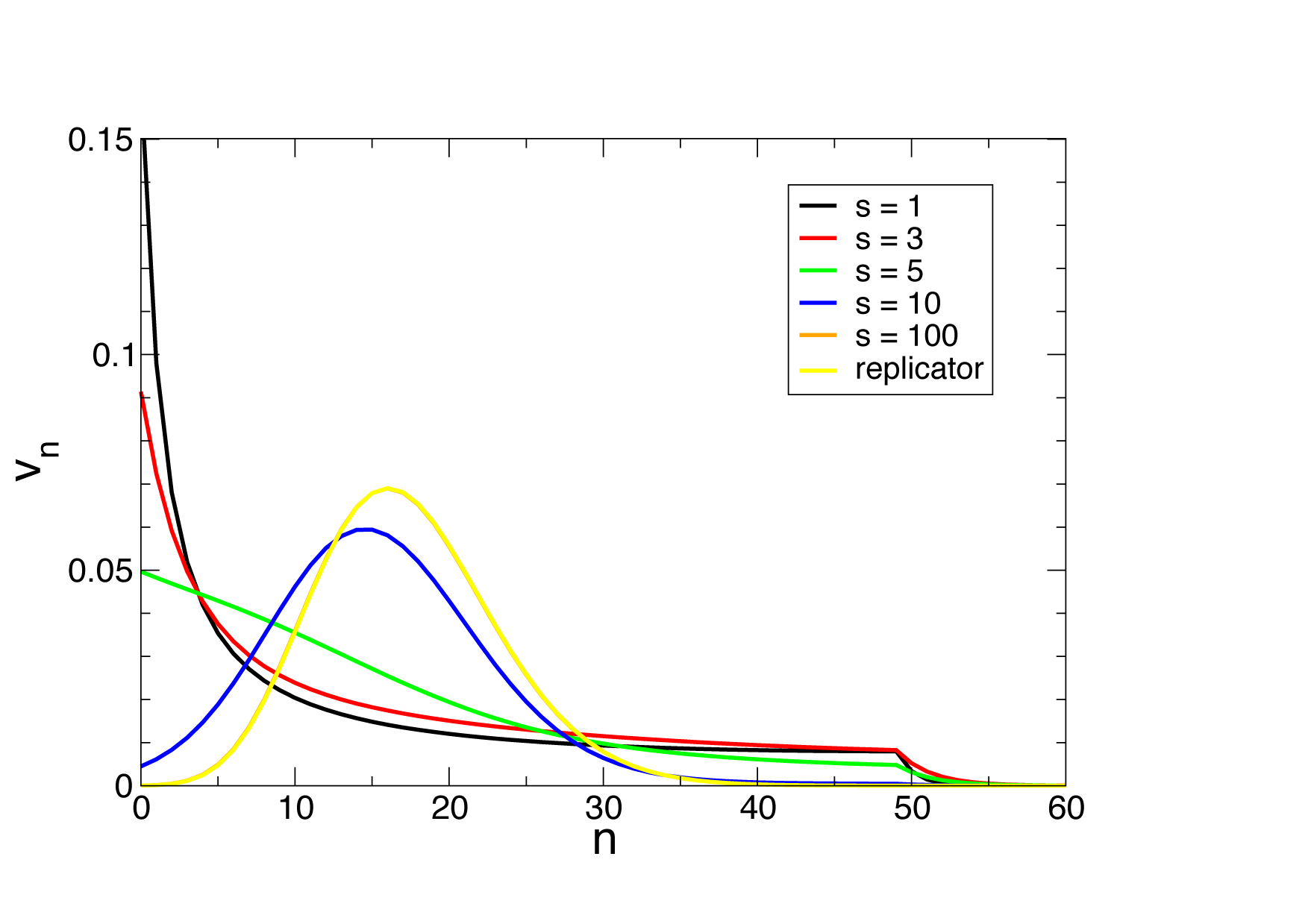}
\end{center}
\caption[]{Distribution of visits to state $n$ before absorption for population $N=100$,
initial number of cooperators $n=50$ and
several values of $s$. The game is the same Snowdrift game of Fig.~\ref{fig:snowdrift}.
The curve for $s=100$ is indistinguishable from the one for $s\to\infty$ (labeled `replicator').}
\label{fig:visits}
\end{figure}

Finally, a representative of category (iii) is the Snowdrift game, for which we
will choose the
payoffs $W_{\text{CC}}=1$, $W_{\text{CD}}=0.2$, $W_{\text{DC}}=1.8$ and $W_{\text{DD}}=0.01$.
For these values,
the replicator dynamics predicts that both strategies coexist with fractions of population given
by $x^*$ in (\ref{eq:xstar}), which for these parameters takes the value $x^*\approx 0.19$.
However, a birth-death process for finite $N$ always ends up in absorption into
one of the absorbing states. In fact, for any $s$ and $N$ and this choice of
payoffs, the population always ends up
absorbed into the $n=0$ state ---except when it starts very close to $n=N$. But
this case has
a peculiarity that makes it entirely different from the previous ones. Whereas
for the former cases the absorption time (\ref{eq:tau}) is $\tau=O(N)$
regardless of the value of $s$, for Snowdrift the absorption time is $O(N)$ for
$s=1$ but  grows very fast with $s$ towards an asymptotic value
$\tau_{\infty}$ (see Fig.~\ref{fig:snowdrift}(a)) and $\tau_{\infty}$ grows
exponentially with
$N$ (see Fig.~\ref{fig:snowdrift}(b)). This means that, while for $s=1$ the process behaves
as in previous cases, being absorbed into the $n=0$ state, as $s$ increases there is a crossover
to a regime in which the transient states become more relevant than the absorbing state
because the population spends an extremely long time in them. In fact, the process oscillates around the
mixed equilibrium predicted by the replicator dynamics. This is illustrated by the distribution
of visits to states $0<n<N$ before absorption (\ref{eq:visits}), shown in Fig.~\ref{fig:visits}. Thus the
effect of fast selection on Snowdrift games amounts to a qualitative change from
the mixed
equilibrium to the pure equilibrium at $n=0$.

Having illustrated the effect of fast selection in these three representative
games, we can now present the general picture. Similar results arise in the
remaining $2\times 2$ games, fast selection favoring in all cases the strategy
with the highest payoff against the opposite strategy. For
the remaining five games of category (i) this means favoring the dominant
strategy (Prisoner's Dilemma is a prominent example of it). The other two cases
of category (ii) also
experience a change in the basins of attraction of the two equilibria. Finally, the remaining
two games of category (iii) experience the suppression of the coexistence state in favor of
one of the two strategies. The conclusion of all this is that fast selection changes completely the outcome of replicator dynamics. In terms of cooperation, as the terms in the off-diagonal of social dilemmas verify $W_{\text{DC}} > W_{\text{CD}}$, this change in outcome has a negative influence on cooperation, as we have seen in all the games considered. Even for some payoff matrices of a non-dilemma game such as Harmony, it can make defectors invade the population.

\begin{figure}
\begin{center}
\subfigure[]{\includegraphics[width=75mm,clip=]{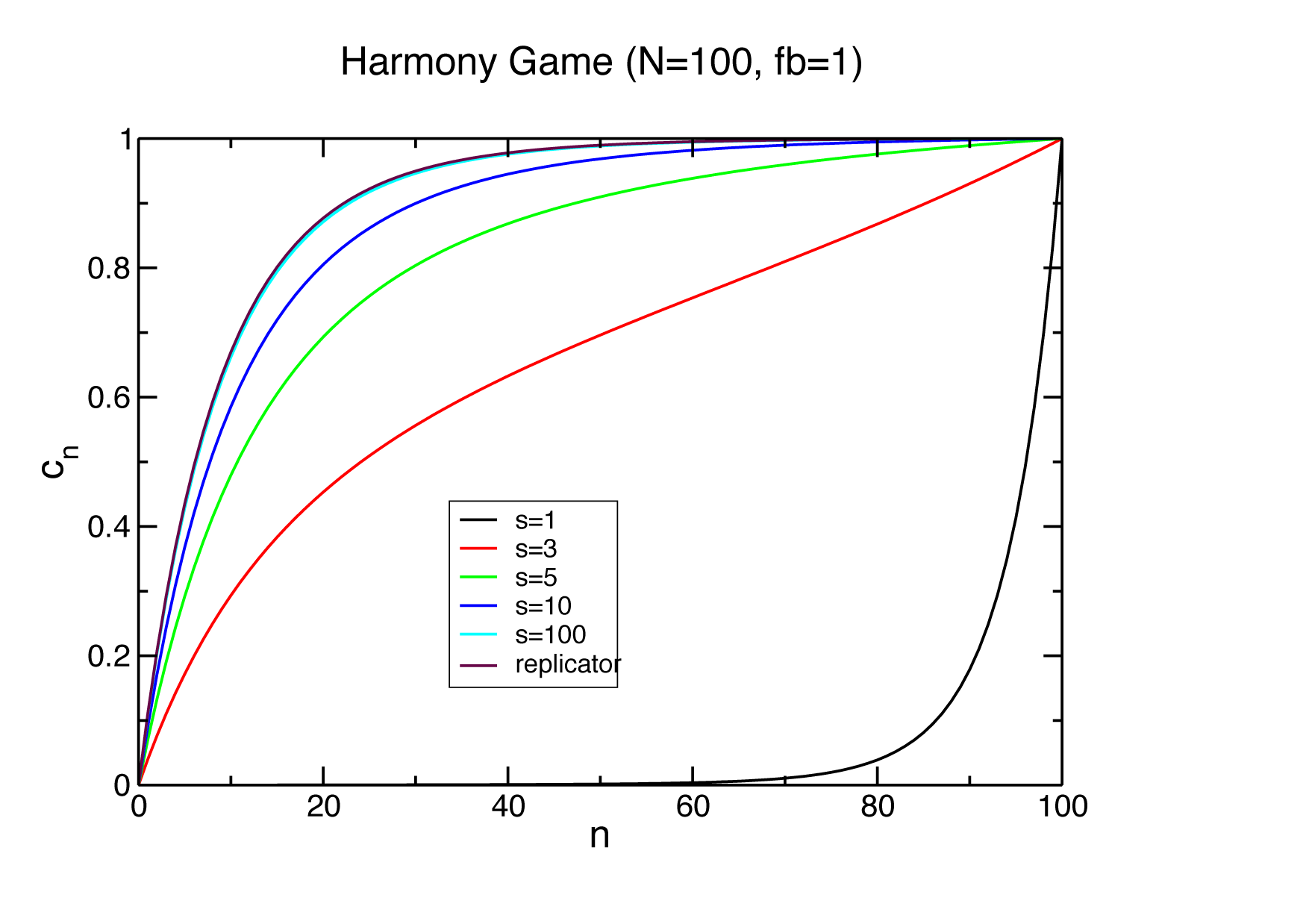}}\qquad
\subfigure[]{\includegraphics[width=75mm,clip=]{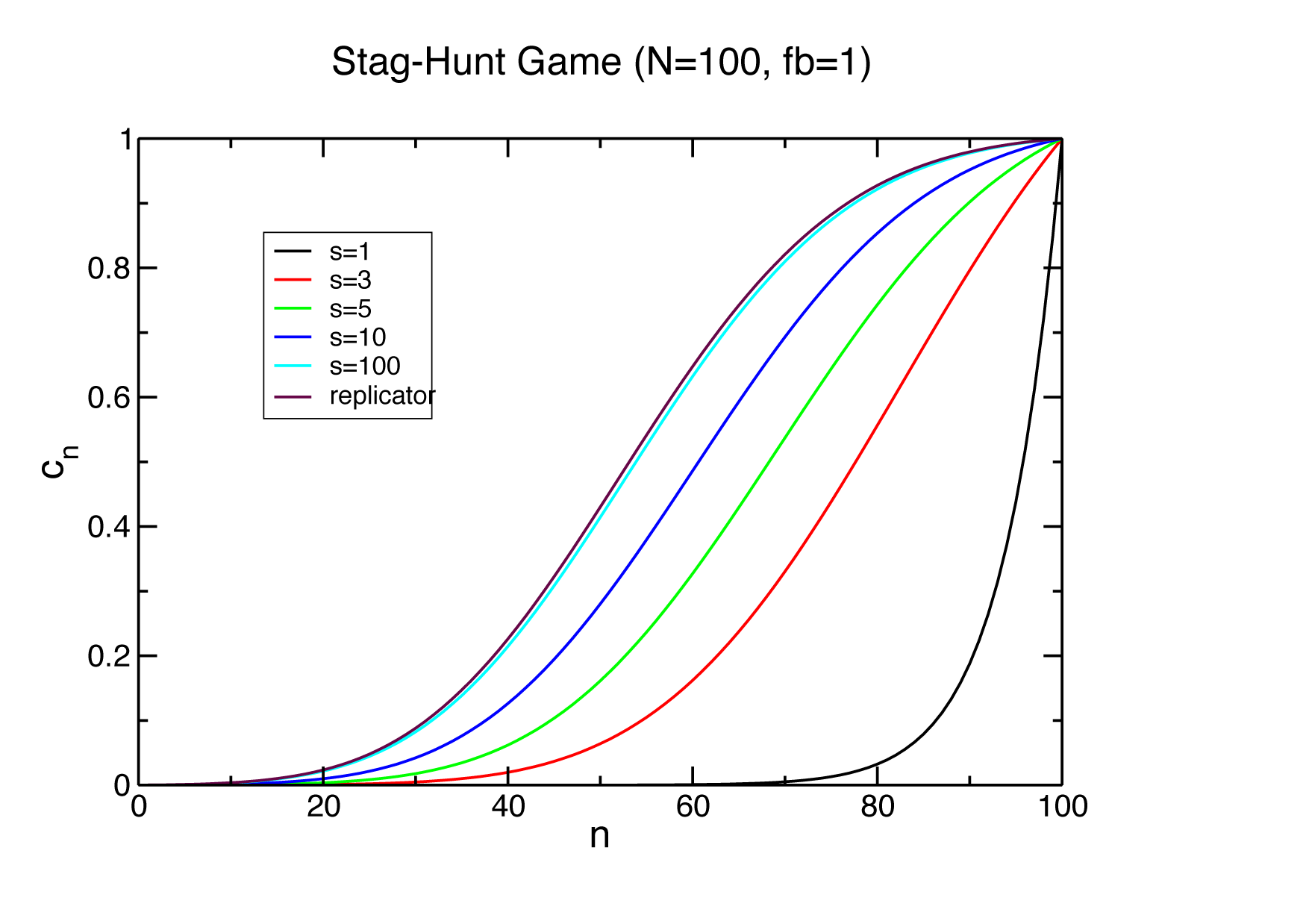}}
\end{center}
\caption[]{Absorption probability starting from state $n$ for the Harmony game
of Fig.~\ref{fig:harmony} (a) and the Stag Hunt game of Fig.~\ref{fig:stag-hunt}
(b) when $N=100$ and baseline fitness $f_b=1$.}
\label{fig:baseline}
\end{figure}

\begin{figure}
\begin{center}
\subfigure[]{\includegraphics[width=75mm,clip=]{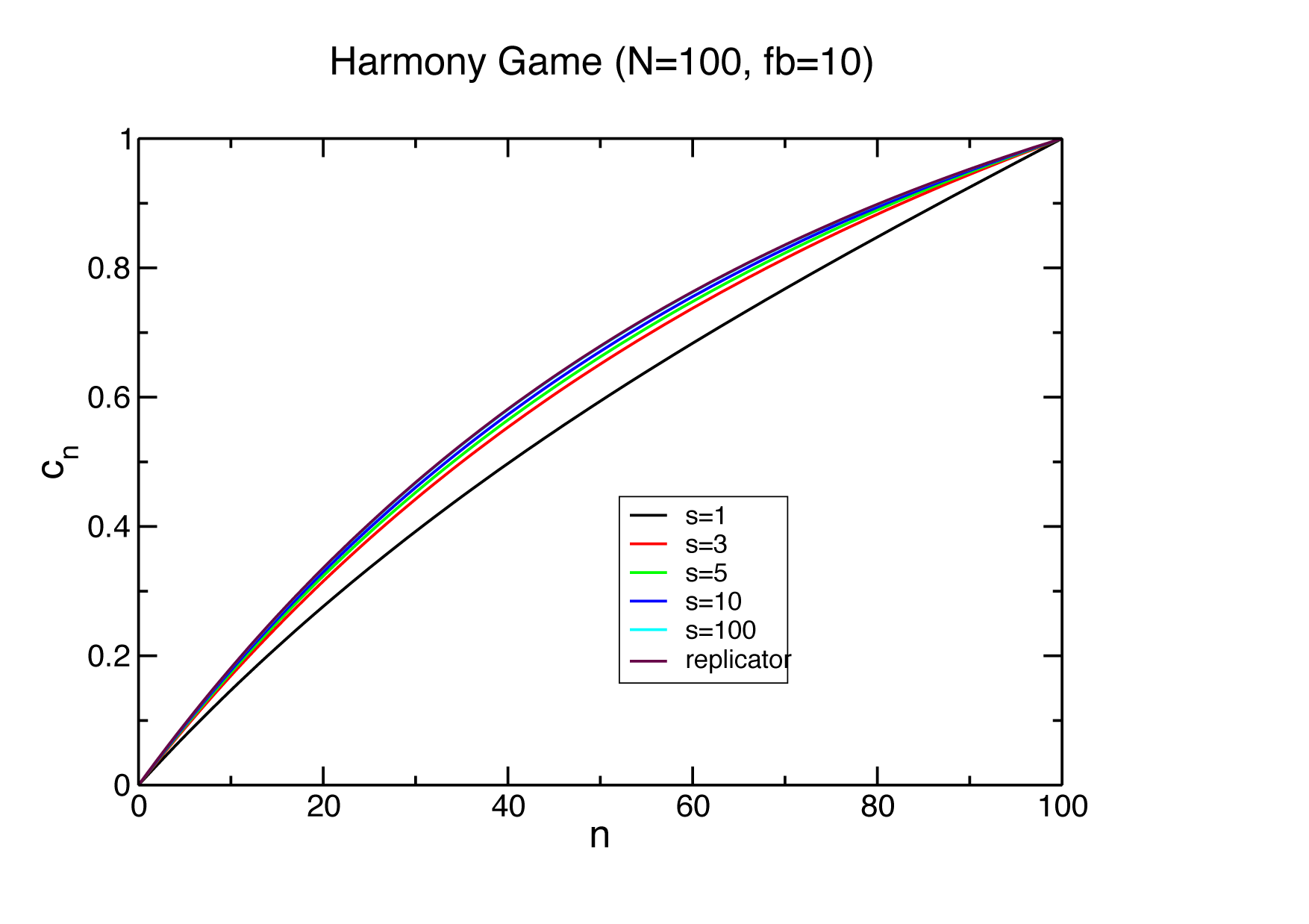}}\qquad
\subfigure[]{\includegraphics[width=75mm,clip=]{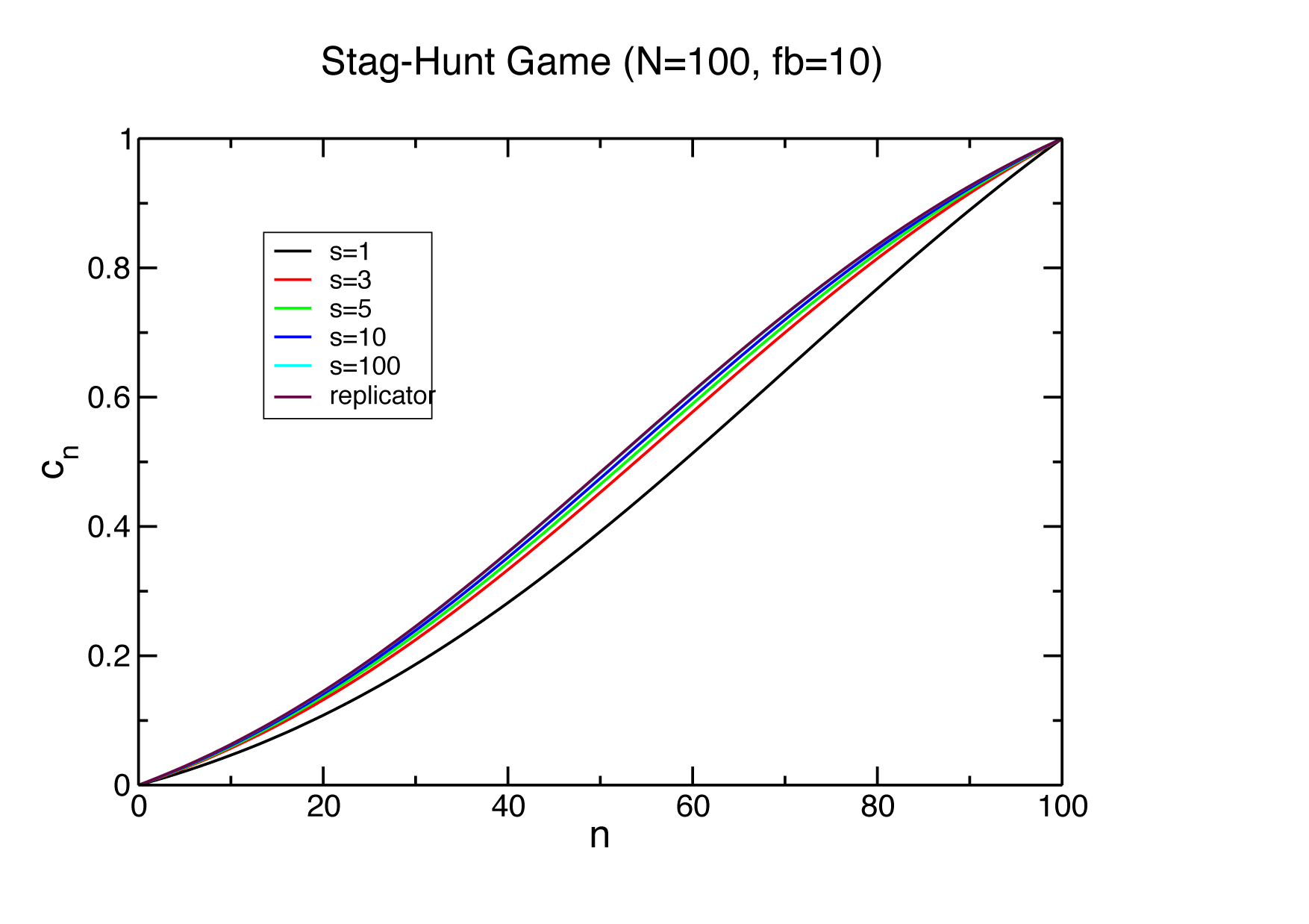}}
\end{center}
\caption[]{Same as Fig.~\ref{fig:baseline} for $f_b=10$.}
\label{fig:baseline10}
\end{figure}

Two final remarks are in order. First, these results do not change qualitatively with the population size. In fact, Eqs.~(\ref{eq:asympDp}) and~(\ref{eq:asympDm}) and Fig.~\ref{fig:asympt} very clearly illustrate this.
Second, there might be some concern about this analysis which the extreme $s=1$
case puts forward: All this might just be an effect of the fact that most
players do not play and therefore have no chance to be selected for
reproduction. In order to sort this out we have made a similar analysis but
introducing a baseline fitness for all players, so that even if
a player does not play she can still be selected for reproduction. The
probability will be, of course, smaller than the one of the players who do
play; however, we should bear in
mind that when $s$ is very small, the overwhelming majority of players are of this type
and this compensates for the smaller probability. Thus, let $f_b$ be the total
baseline fitness that all players share per round, so that $sf_b/N$ is
the baseline fitness every player has at the time reproduction/selection occurs.
This choice implies that if $f_b=1$ the overall baseline fitness and that
arising from the game are similar, regardless of $s$ and $N$. If $f_b$ is very
small ($f_b\lesssim 0.1$), the result is basically the same as that for $f_b=0$.
The effect for $f_b=1$ is illustrated in Fig.~\ref{fig:baseline} for Harmony and
Stag Hunt games. Note also that at very large baseline fitness ($f_b\gtrsim 10$)
the evolution is almost neutral, although the small deviations induced by the
game ---which are determinant for the ultimate fate of the population--- still
follow the same pattern (see Fig.~\ref{fig:baseline10}). Interestingly, Traulsen {\em et al.}
\cite{traulsen:2007} arrive at similar results by using a Fermi like rule (see Sec.\ 4.1 below)
to introduce noise (temperature) in the selection process, and a interaction probability
$q$ of interactions between individuals leading to heterogeneity in the payoffs, i.e.,
in the same words as above, to fluctuations, that in turn reduce the intensity of
selection as is the case when we introduce a very large baseline fitness.

\clearpage

\section{Structured populations}
\label{sec:4}

Having seen the highly non-trivial effects of considering temporal fluctuations
in evolutionary games, in this section we are going to consider the effect of
relaxing the well-mixed hypothesis by allowing the existence of spatial
correlations in the population. Recall from Section~2.2 that a well-mixed
population presupposes that every individual interacts with equal probability
with every other one in the population, or equivalently that each individual
interacts with the ``average'' individual. It is not clear,
however, that this hypothesis holds in many practical situations. Territorial or
physical constraints may limit the interactions between individuals, for
example. On the other hand, an all-to-all network of relationships does not seem
plausible in large societies; other key phenomena in social life, such as
segregation or group formation, challenge the idea of a mean player that
everyone interacts with.

It is adequate, therefore, to take into consideration the existence
of a certain network of relationships in the population, which determines
who interacts with whom. This network of relationships is what we will
call from now on the \emph{structure} of the population. Consistently,
a well-mixed population will be labeled as \emph{unstructured} and will
be represented by a complete graph. Games on many different types of networks have been investigated, examples of which include regular lattices \cite{nowak:1992,hauert:2002}, scale-free networks \cite{santos:2006a}, real social networks \cite{lozano:2008}, etc. This section is not intended to be an
exhaustive review of all this existent work and we refer the reader to \cite{szabo:2007} for such a
detailed account. We rather want to give a panoramic and a more personal and
idiosyncratic view of the field, based on the main available results and our own
research.

It is at least reasonable to expect that the existence of structure in a
population could give rise to the appearance of correlations and that they
would have an impact on the evolutionary outcome. For more than fifteen
years investigation into this phenomena has been a hot topic of research, as the
seminal result by Nowak and May \cite{nowak:1992}, which reported an impressive
fostering of cooperation in Prisoner's Dilemma on spatial lattices, triggered a
wealth of work focused on the extension of this effect to other games, networks
and strategy spaces. On the other hand, the impossibility in most cases of
analytical approaches and the complexity of the corresponding numerical
agent-based models have made any attempt of exhaustive approach very demanding.
Hence most studies have concentrated on concrete settings with a particular
kind of game, which in most cases has been the Prisoner's Dilemma
\cite{nowak:1992,nowak:1994,lindgren:1994,hutson:1995,grim:1996,nakamaru:1997,
szabo:1998,brauchli:1999,abramson:2001,cohen:2001,vainstein:2001,lim:2002,
schweitzer:2002,ifti:2004,tang:2006,perc:2008}. Other games has been much less
studied in what concerns the influence of population structure, as show the
comparatively much smaller number of works about Snowdrift or Hawk-Dove games
\cite{killingback:1996,hauert:2004,sysi-aho:2005,kun:2006,tomassini:2006,
zhong:2006}, or Stag Hunt games \cite{blume:1993,ellison:1993,kirchkamp:2000}. 
Moreover, comprehensive studies in the space of $2 \times 2$ games are very
scarce \cite{hauert:2002,santos:2006a}. As a result, many interesting features
of population structure and its influence on evolutionary games have been
reported in the literature, but the scope of these conclusions is rather
limited to particular models, so a general understanding of these issues, in the
broader context of $2 \times 2$ games and different update rules, is generally
missing.

However, the availability and performance of computational resources in
recent years have allowed us to undertake a systematic and exhaustive simulation
program \cite{roca:2009a,roca:2009b} on these evolutionary models. As a result
of
this study we have reached a number of conclusions that are obviously in
relation with previous research and that we will discuss in the following. In
some cases, these are generalizations of known results to wider sets of games
and update rules, as for example for the issue of the synchrony of the updating
of strategies
\cite{nowak:1992,nowak:1994,lindgren:1994,kun:2006,
tomassini:2006,kirchkamp:2000,huberman:1993} or the effect of small-world
networks vs regular
lattices \cite{abramson:2001,tomassini:2006,masuda:2003,tomochi:2004}. In
other cases, the more general view of our analysis has allowed us to
integrate apparently contradictory results in the literature, as the cooperation
on Prisoner's Dilemma vs.\ Snowdrift games
\cite{nowak:1992,killingback:1996,hauert:2004,sysi-aho:2005,tomassini:2006}, or
the importance of clustering in spatial lattices
\cite{cohen:2001,ifti:2004,tomassini:2006}. Other conclusions of ours, however,
refute what seems to be established opinions in the field, as the alleged
robustness of the positive influence of spatial structure on Prisoner's
Dilemma \cite{nowak:1992,hauert:2002,nowak:1994}. And finally, we have
reached novel conclusions that have not been highlighted by previous research,
as the robustness of the influence of spatial structure on coordination games,
or the asymmetry between the effects on games with mixed equilibria
(coordination and anti-coordination games) and how it varies with the intensity
of selection. 

It is important to make clear from the beginning that evolutionary games on
networks may be sensitive to another source of variation with respect
to replicator dynamics besides the introduction of spatial correlations. This
source is the update rule, i.e.\ the rule that defines the evolution dynamics of
individuals' strategies, whose influence seems to have been overlooked
\cite{hauert:2002}. Strictly speaking, only when the model implements the
so-called \emph{replicator rule} (see below) one is considering the effect of
the restriction of relationships that the population structure implies, in
comparison with standard replicator dynamics. When using a different update
rule, however, we are adding a second dimension of variability, which amounts to
relax another assumption of replicator dynamics, namely number 4, which posits a
population variation linear in the difference of payoffs (see
Section~\ref{sec:2}). We will show extensively that this issue may have a huge
impact on the evolutionary outcome.

In fact, we will see that there is not a general influence of population
structure on evolutionary games. Even for a particular type of network, its
influence on cooperation depends largely on the kind of game and the
specific update rule. All one can do is to identify relevant
\emph{topological characteristics} that have a consistent effect on a broad
range of games and update rules, and explain this influence in terms of the same
basic principles. To this end, we will be looking
at the asymptotic states for different values of the game parameters, and not at how
the system behaves when the parameters are varied, which would be an approach
of a more statistical mechanics character. In this respect, it is worth pointing out that
some studies did use this perspective: thus, it has been shown that the extinction
transitions  when the temptation parameter varies within the Prisoner's Dilemma game
and the evolutionary dynamics is stochastic fall in the directed percolation universality
class, in agreement with a well known conjecture \cite{hinrichsen:2000}. In particular,
some of the pioneering works in using a physics viewpoint on evolutionary games
\cite{szabo:1998,chiappin:1999}
have verified this result for specific models. The behavior changes under deterministic
rules such as {\em unconditional imitation} (see below), for which this extinction
transition is discontinuous.

Although our ultimate interest may be the effect on the evolution of
cooperation, measuring to which extent cooperation is enforced or inhibited
is not enough to clarify this effect. As in previous sections, our basic
observables will be the dynamical equilibria of the model, in comparison with
the equilibria of our reference model with standard replicator dynamics
--which, as we have explained in Section~\ref{sec:2}, are closely related to
those of the basic game--. The understanding of how the population structure
modifies qualitatively and quantitatively these equilibria will give us a much
clearer view on the behavior and properties of the model under study, and hence
on its influence on cooperation.

\subsection{Network models and update rules}

Many kinds of networks have been considered as models for population structure
(for recent reviews on networks, see \cite{newman:2003,boccaletti:2006}). A
first class includes networks that introduce a spatial arrangement of
relationships, which can represent territorial or physical constraints in the
interactions between individuals. Typical examples of this group are regular
lattices, with different degrees of neighborhood. Other important group is that
of synthetic networks that try to reproduce important properties that have been
found in real networks, such as the small-world or scale-free properties.
Prominent examples among these are Watts-Strogatz small-world networks
\cite{watts:1998} and Barab\'asi-Albert scale-free networks \cite{albert:2002}.
Finally, ``real'' social networks that come directly from experimental data have
also been studied, as for example in \cite{holme:2003,guimera:2003}.

As was stated before, one crucial component of the evolutionary models that we
are discussing in this section is the update rule, which determines how the
strategy of individuals evolves in time. There is a very large variety of update rules that have been used in the literature, each one arising from different backgrounds. The most important for our purposes is the \emph{replicator rule}, also known as the \emph{proportional imitation rule}, which is inspired on replicator dynamics and we describe in the following.\footnote{To our knowledge, Helbing was the first to show that a macroscopic population evolution following replicator dynamics could be induced by a microscopic imitative update rule \cite{helbing:1992a,helbing:1992b}. Schlag proved later the
optimality of such a rule under certain information constraints, and named it
\emph{proportional imitation} \cite{schlag:1998}.} Let $i = 1 \ldots N$ label
the individuals in the population. Let $s_i$ be the
strategy of player $i$, $W_i$ her payoff and $N_i$ her neighborhood, with $k_i$
neighbors. One neighbor $j$ of player $i$ is chosen at random, $j \in N_i$. The
probability of player $i$ adopting the strategy of player $j$ is given by
\begin{equation}
\label{eq:repldyn}
p^t_{ij} \equiv \Probability{ \{ s_j^t \to s_i^{t+1} \} } =
\left\{ \begin{array}{ll}
  ( W_j^t - W_i^t ) / \Phi \,, & W_j^t > W_i^t \,, \\
  0 \,, & W_j^t \leq W_i^t \,,
\end{array} \right.
\end{equation}
with $\Phi $ appropriately chosen as a function of the payoffs to ensure $\Probability{ \{\cdot\} } \in [0,1]$.

\begin{figure}[t]
\centering
\includegraphics[width=0.9\textwidth]{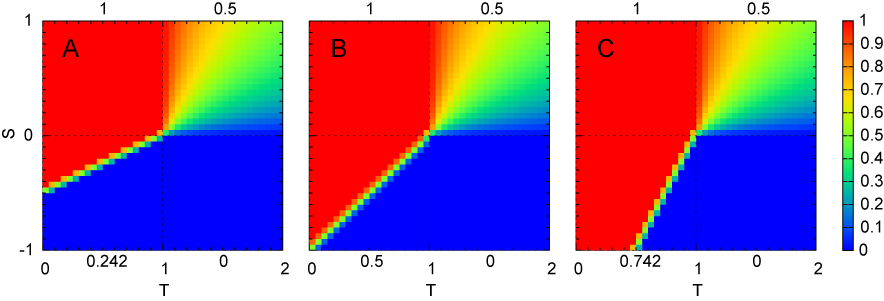}
\caption{Asymptotic density of cooperators $x^*$ with the replicator rule on a
complete network, when the initial density of cooperators is $x^0 =1/3$ (left,
A), $x^0 =1/2$ (middle, B) and $x^0 =2/3$ (right, C). This is the standard
outcome for a well-mixed population with replicator dynamics, and thus
constitutes the reference to assess the influence of a given population
structure (see main text for further details).}
\label{fig:compnet}
\end{figure}

The reason for the name of this rule is the fact that the equation of
evolution with this update rule, for large sizes of the population, is equal, up
to a time scale factor, to that of replicator dynamics
\cite{gintis:2000,hofbauer:1998}. Therefore, the complete network with
the replicator rule constitutes the finite-size, discrete-time version of replicator
dynamics on an infinite, well-mixed population in continuous time.
Fig.~\ref{fig:compnet} shows the evolutionary outcome of this model, in the same
type of plot as subsequent results in this section. Each panel of this figure
displays the asymptotic density of cooperators $x^*$ for a different initial
density $x^0$, in a grid of points in the $ST$-plane of games. The payoff
matrix of each game is given by
\begin{equation}
\label{eq:payoff-matrix}
\begin{array}{cc}
  & \begin{array} {cc} \!\! \mbox{C} & \mbox{D} \end{array} \\
  \begin{array}{c} \mbox{C} \\ \mbox{D} \end{array} &
  \left( \begin{array}{cc} 1 & S \\ T & 0 \end{array} \right).
\end{array}
\end{equation}
We will consider the generality of this choice of parameters at the end of this section, after introducing the other evolutionary rules. Note that, in the notation of Section~\ref{sec:3}, we have taken $W_{\text
{CC}}=1, W_{\text {CD}}=S, W_{\text {DC}}=T, W_{\text {DD}}=0$; note also that
for these payoffs,
the normalizing factor in the replicator rule can be chosen as
$\Phi = \max(k_i,k_j) ( \max(1,T) - \min(0,S) )$. In this manner,
we visualize the space of symmetric $2 \times 2$ games as a plane of
co-ordinates $S$ and $T$ --for \emph{Sucker's} and \emph{Temptation}--,
which are the respective payoffs of a cooperator and a defector when confronting
each other. The four quadrants represented correspond to the following
games: Harmony (upper left), Stag Hunt (lower left), Snowdrift or Hawk-Dove
(upper right) and Prisoner's Dilemma (lower right). As expected, these results
reflect the close relationship between the equilibria of replicator dynamics and
the equilibria of the basic game. Thus, all Harmony games end up in full
cooperation and all Prisoner's Dilemmas in full defection, regardless of the
initial condition. Snowdrift games reach a mixed strategy equilibrium, with
density of cooperators $x_e = S / (S + T - 1)$. Stag Hunt games are the only
ones whose outcome depends on the initial condition, because of their bistable
character with an unstable equilibrium also given by $x_e$. To allow a
quantitative comparison of the degree of cooperation in each game, we have
introduced a quantitative index, the average cooperation over the region
corresponding to each game, which appears beside each quadrant. The results in
Fig.~\ref{fig:compnet} constitute the reference against which the effect of
population structure will be assessed in the following.

One interesting variation of the replicator rule is the \emph{multiple replicator rule}, whose difference consists on checking simultaneously all the neighborhood and thus making more probable a strategy change. With this rule the probability that player $i$ maintains her strategy is
\begin{equation}
\label{eq:multirepldyn}
\Probability{ \{ s_i^t \to s_i^{t+1} \} } =
  \prod \limits_{j \in N_i} ( 1 - p_{ij}^t ) ,
\end{equation}
with $p^t_{ij}$ given by (\ref{eq:repldyn}). In case the strategy update takes place, the neighbor $j$ whose strategy is adopted by player $i$ is selected with probability proportional to $p^t_{ij}$.

A different option is the following \emph{Moran-like rule}, also called \emph{Death-Birth rule}, inspired on the Moran dynamics, described in Section~\ref{sec:3}. With this rule a player chooses the strategy of one of her neighbors, or herself's, with a probability proportional to the payoffs
\begin{equation}
\label{eq:moran}
\Probability{ \{ s_j^t \to s_i^{t+1} \} } =
  \frac {W_j^t - \Psi} {\sum \limits_{k \in N^*_i} ( W_k^t - \Psi )} ,
\end{equation}
with $N^*_i = N_i \cup \{i\}$.
Because payoffs may be negative in Prisoner's Dilemma and Stag Hunt games, the constant $\Psi = \max_{j \in N^*_i}(k_j) \min(0,S)$ is subtracted from them. Note that with this rule a player can adopt, with low probability, the strategy of a neighbor that has done worse than herself.

The three update rules presented so far are imitative rules. Another important example of this kind is the \emph{unconditional imitation rule}, also known as the \emph{``follow the best''} rule \cite{nowak:1992}. With this rule each player chooses the strategy of the individual with largest payoff in her neighborhood, provided this payoff is greater than her own. A crucial difference with the previous rules is that this one is a deterministic rule.

Another rule that has received a lot of attention in the literature, specially in economics, is the \emph{best response rule}. In this case, instead of some kind of imitation of neighbor's strategies based on payoff scoring, the player has enough cognitive abilities to realize whether she is playing an optimum strategy (i.e.\ a best response) given the current configuration of her neighbors. If it is not the case, she adopts with probability $p$ that optimum strategy. It is clear that this rule is innovative, as it is able to introduce strategies not present in the population, in contrast with the previous purely imitative rules.

Finally, an update rule that has been widely used in the literature, because of being analytically tractable, is the \emph{Fermi rule}, based on the Fermi
distribution function \cite{szabo:1998,blume:2003,traulsen:2006a}. With this
rule, a neighbor $j$ of player $i$ is selected at random (as with the replicator rule) and the probability of player $i$ acquiring the strategy of $j$ is given by
\begin{equation}
\label{eq:fermidyn}
\Probability{ \{ s_j^t \to s_i^{t+1} \} } =
\displaystyle \frac {1} {1 + \exp \left( - \beta \, ( W_j^t - W_i^t )
\right) } .
\end{equation}
The parameter $\beta$ controls the intensity of selection, and can be understood
as the inverse of temperature or noise in the update rule. Low $\beta$
represents high temperature or noise and, correspondingly, weak selection
pressure. Whereas this rule has been employed to study resonance-like behavior
in evolutionary games on lattices \cite{szabo:2005b}, we use it in this work to
deal with the issue of the intensity of selection (see
Subsection~\ref{subsec:weak-selection}).

Having introduced the evolutionary rules we will consider, it is important to
recall our choice for the payoff matrix (\ref{eq:payoff-matrix}), and discuss
its generality. Most of the rules (namely the replicator, the multiple
replicator, the unconditional imitation and the best response rules) are
invariant on homogeneous networks\footnote{The invariance under
translations of the payoff matrix does not hold if the network is
heterogenous. In this case, players with higher degrees receive
comparatively more (less) payoff under positive (negative) translations. Only
very recently has this issue been studied in the literature \cite{luthi:2009}.}
under translation and (positive) scaling of the payoff matrix. Among the
remaining rules, the dynamics changes upon translation for the Moran rule and
upon scaling
for the Fermi rule. The corresponding changes in these last two cases amount to
a modification of the intensity of selection, which we also treat in this
work. Therefore, we consider that the parameterization of
(\ref{eq:payoff-matrix}) is general enough for our purposes.

It is also important to realize that for a complete network, i.e.\ for a
well-mixed or unstructured population, the differences between update rules may
be not relevant, as far as they do not change in general the evolutionary
outcome \cite{roca:2009c}. These differences, however, become crucial when the
population has some structure, as we will point out in the following.

The results displayed in Fig.~\ref{fig:compnet} have been obtained analytically,
but the remaining results of this section come from the simulation of
agent-based models. In all cases, the population size is $N=10^4$ and the
allowed
time for convergence is $10^4$ time steps, which we have checked it is enough
to
reach a stationary state. One time step represents one update event for every individual in the population, exactly in the case of synchronous update and on average in the asynchronous case, so it could be considered as one generation. The asymptotic density of cooperators is obtained
averaging over the last $10^3$ time steps, and the values presented in the plots
are the result of averaging over 100 realizations. Cooperators and defectors are
randomly located at the beginning of evolution and, when applicable, networks
have been built with periodic boundary conditions. See \cite{roca:2009a} for
further details.

\subsection{Spatial structure and homogeneous networks}

In 1992 Nowak and May published a very influential paper \cite{nowak:1992}, where they showed the dramatic effect that the spatial distribution of a population could have on the evolution of cooperation. This has become the prototypical example of the promotion of cooperation favored by the structure of a population, also known as network reciprocity \cite{nowak:2006b}. They considered the following Prisoner's Dilemma:
\begin{equation}
\label{eq:nowak-pd}
\begin{array}{cc}
  & \begin{array} {cc} \! \mbox{C} & \! \mbox{D} \end{array} \\
  \begin{array}{c} \mbox{C} \\ \mbox{D} \end{array} &
  \left( \begin{array}{cc} 1 & 0 \\ T & \epsilon \end{array} \right),
\end{array}
\end{equation}
with $1 \le T \le 2$ and $\epsilon \lesssim 0$. Note that this one-dimensional parameterization corresponds in the $ST$-plane to a line very near the boundary with Snowdrift games.

\begin{figure}[t]
\centering
\includegraphics[width=0.45\textwidth]{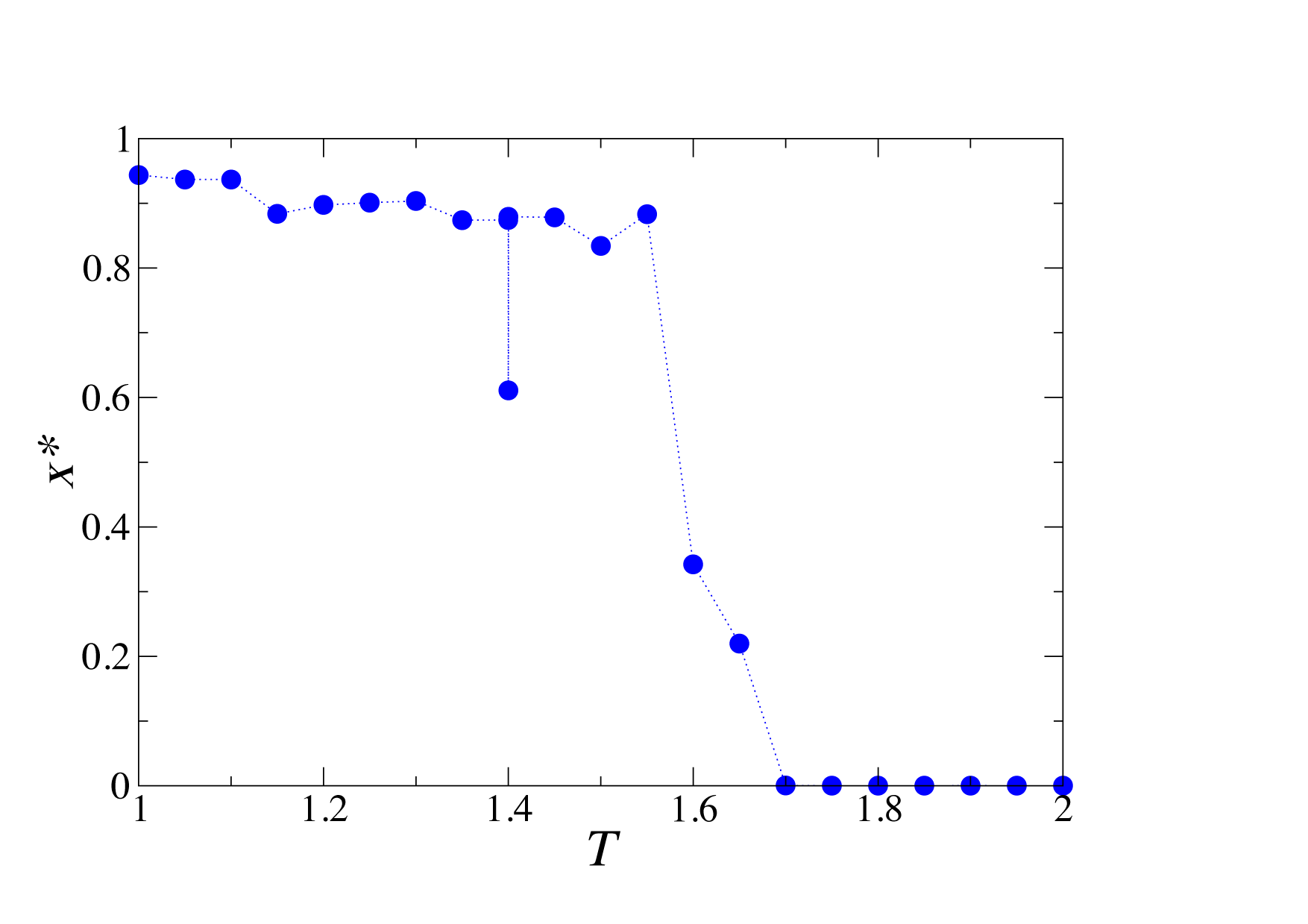}
\caption{Asymptotic density of cooperators $x^*$ in a square lattice with degree
$k=8$ and initial density of cooperators $x^0=0.5$, when the game is
the Prisoner's Dilemma as given by (\ref{eq:nowak-pd}), proposed by Nowak and
May \cite{nowak:1992}. Note that the outcome with replicator dynamics on a
well-mixed population is $x^*=0$ for all the displayed range of the temptation
parameter $T$. Notice also the singularity at $T=1.4$ with unconditional
imitation. The surrounding points are located at $T=1.3999$ and $T=1.4001$.}
\label{fig:nowak}
\end{figure}

Fig.~\ref{fig:nowak} shows the great fostering of cooperation reported by \cite{nowak:1992}. The authors explained this influence in terms of the formation of clusters of cooperators, which give cooperators enough payoff to survive even when surrounded by some defectors. This model has a crucial detail, whose importance we will stress later: The update rule used is unconditional imitation.

Since the publication of this work many studies have investigated related models
with different games and networks, reporting qualitatively consistent results
\cite{szabo:2007}. However, Hauert and Doebeli published in 2004 another
important result \cite{hauert:2004}, which casted a shadow of doubt on the
generality of the positive influence of spatial structure on cooperation. They
studied the following parameterization of Snowdrift games:
\begin{equation}
\label{eq:hauert-sd}
\begin{array}{cc}
  & \begin{array} {ccc} \!\!\!\!\! \mbox{C} & \mbox{          } &
\!\!\! \mbox{D  } \end{array} \\
  \begin{array}{c} \mbox{C} \\ \mbox{D} \end{array} &
  \left( \begin{array}{cc} 1 & 2-T \\ T & 0 \end{array} \right),
\end{array}
\end{equation}
with $1 \le T \le 2$ again.

\begin{figure}[t]
\centering
\includegraphics[width=0.45\textwidth]{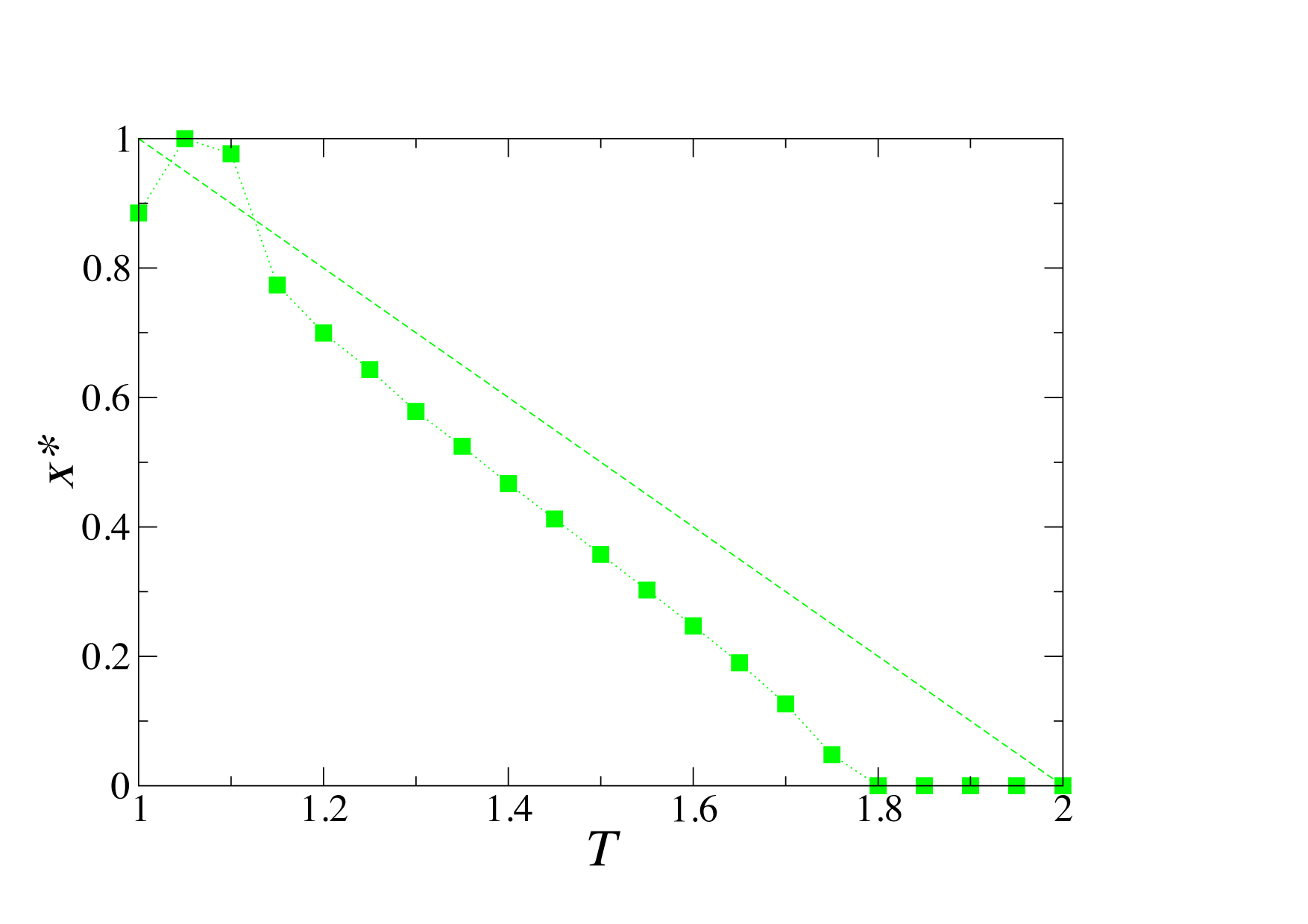}
\caption{Asymptotic density of cooperators $x^*$ in a square lattice with degree
$k=8$ and initial density of cooperators $x^0=0.5$, when the game is
the Snowdrift game as given by (\ref{eq:hauert-sd}), proposed by Hauert and
Doebeli \cite{hauert:2004}. The result for a well-mixed population is displayed
as a reference as a dashed line.}
\label{fig:hauert}
\end{figure}

The unexpected result obtained by the authors is displayed in Fig.~\ref{fig:hauert}. Only for low $T$ there is some improvement of cooperation,
whereas for medium and high $T$ cooperation is inhibited. This is a surprising
result, because the basic game, the Snowdrift, is in principle more favorable to
cooperation. As we have seen above, its only stable equilibrium is a mixed
strategy population with some density of cooperators, whereas the unique
equilibrium in Prisoner's Dilemma is full defection (see
Fig.~\ref{fig:compnet}). In fact, a previous paper by Killingback and Doebeli
\cite{killingback:1996} on the Hawk-Dove game, a game equivalent to
the Snowdrift game, had reported an effect of spatial structure equivalent to a
promotion of
cooperation.

Hauert and Doebeli explained their result in terms of the hindrance to cluster
formation and growth, at the microscopic level, caused by the payoff structure
of the Snowdrift game. Notwithstanding the different cluster dynamics in both
games, as observed by the authors, a hidden contradiction looms in their
argument, because it implies some kind of \emph{discontinuity} in the
microscopic dynamics in the crossover between Prisoner's Dilemma and Snowdrift
games ($S=0, 1 \le T \le 2$). However, the equilibrium structure of both basic
games, which drives this microscopic dynamics, is not discontinuous at this
boundary, because for both games the only stable equilibrium is full defection.
So, where does this change in the cluster dynamics come from?

\begin{figure}[t]
\centering
\subfigure[]{\includegraphics[width=0.45\textwidth,clip=]{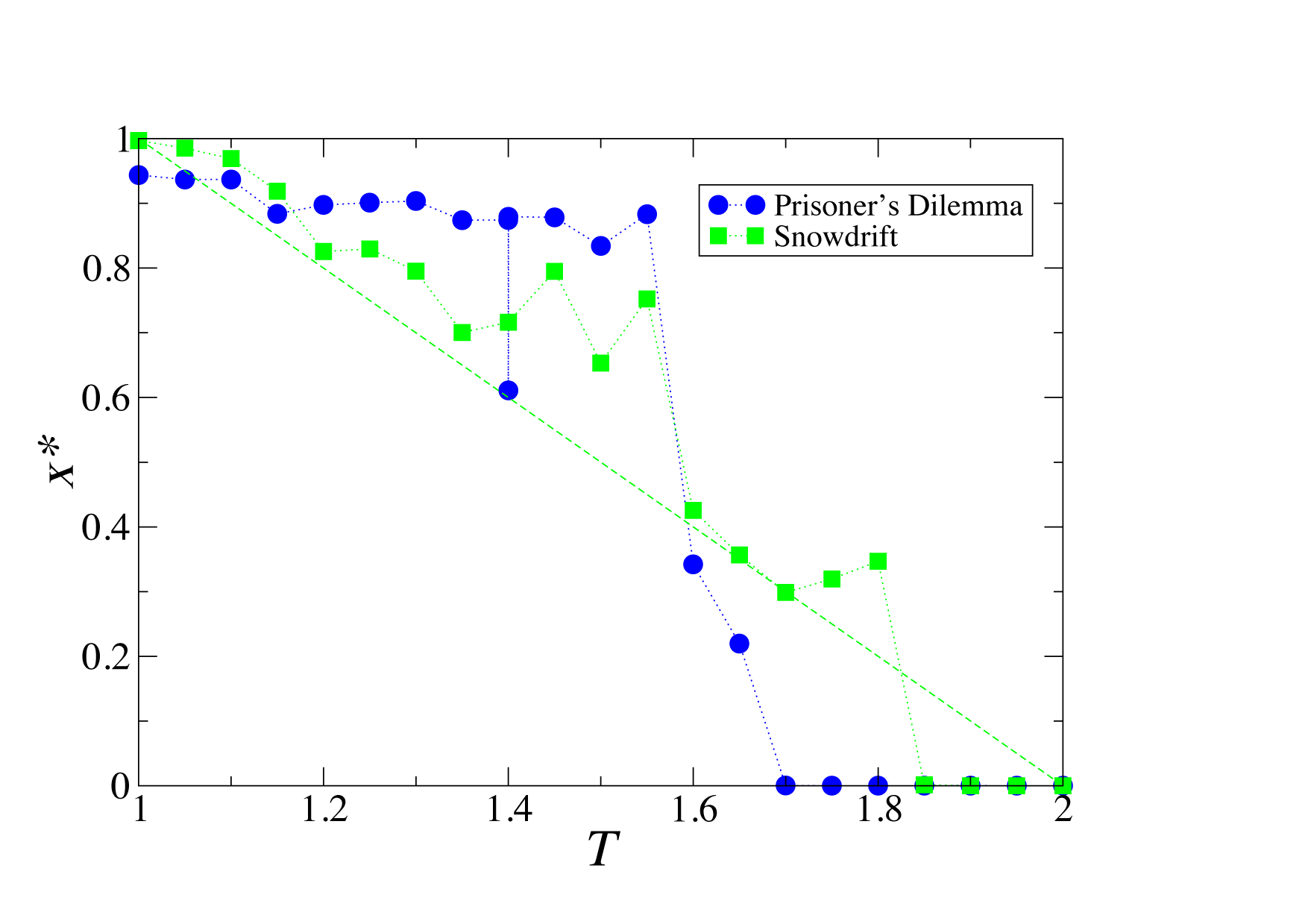}}\qquad
\subfigure[]{\includegraphics[width=0.45\textwidth,clip=]{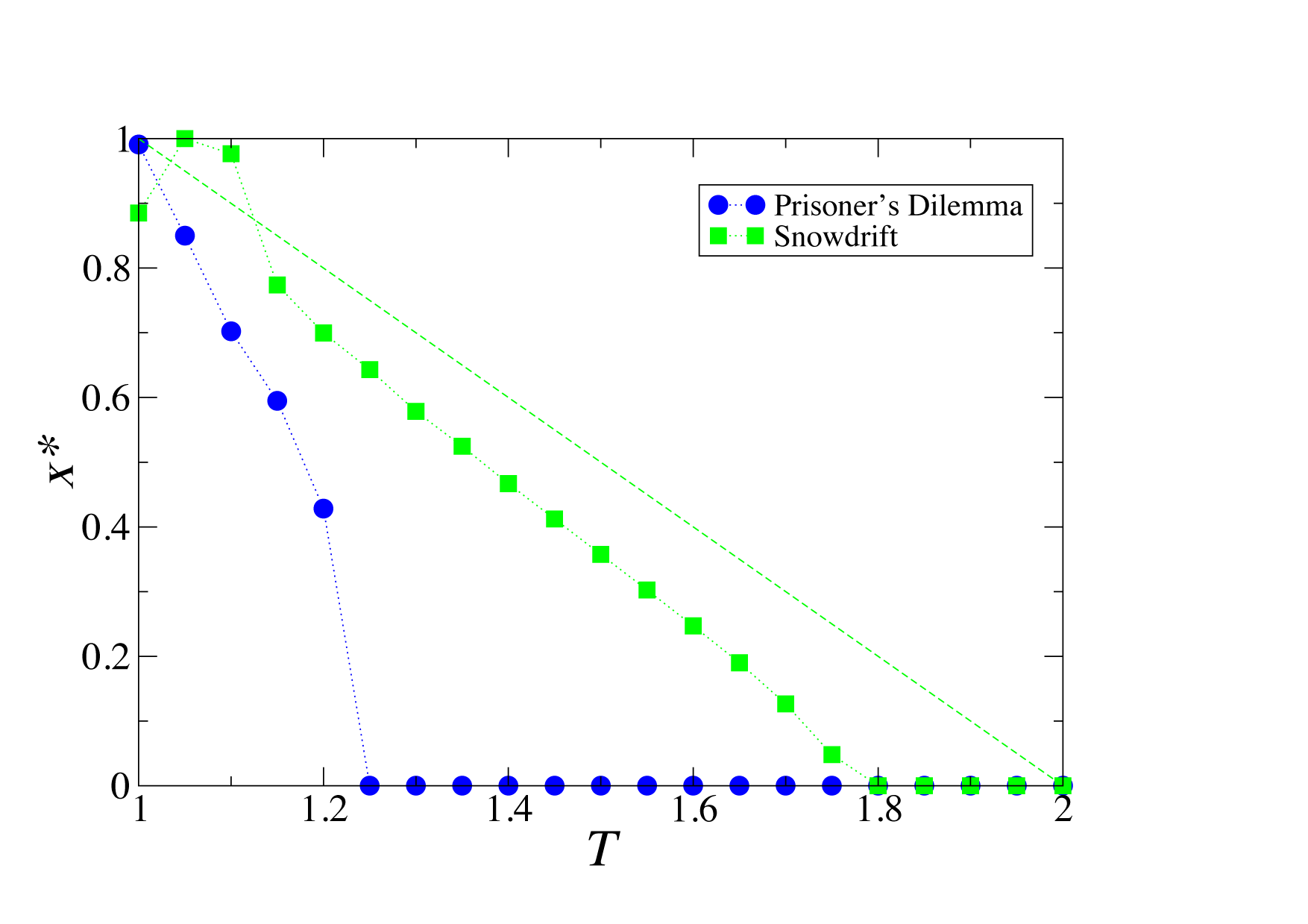}}
\caption{Asymptotic density of cooperators $x^*$ in square lattices with degree
$k=8$ and initial density of cooperators $x^0=0.5$, for both Prisoner's Dilemma
(\ref{eq:nowak-pd}) and Snowdrift games (\ref{eq:hauert-sd}), displayed
separately according to the update rule: (a) unconditional imitation (Nowak and
May's model \cite{nowak:1992}), (b) replicator rule (Hauert and Doebeli's model
\cite{hauert:2004}). The result for Snowdrift in a well-mixed population is
displayed as a reference as a dashed line. It is clear the similar influence of
regular lattices on both games, when the key role of the update rule is taken
into account (see main text for details).}
\label{fig:nowak-hauert}
\end{figure}

The fact is that there is not such a difference in the cluster dynamics between
Prisoner's Dilemma and Snowdrift games, but different update rules in the
models. Nowak and May \cite{nowak:1992}, and Killingback and Doebeli 
\cite{killingback:1996}, used the unconditional imitation rule, whereas Hauert
and Doebeli \cite{hauert:2004} employed the replicator rule. The crucial role
of the update rule becomes clear in Fig.~\ref{fig:nowak-hauert}, where results
in Prisoner's Dilemma and Snowdrift are depicted separately for each update
rule. It shows that, if the update rule used in the model is the same, the
influence on both games, in terms of promotion or inhibition of cooperation, has
a similar dependence on $T$. For both update rules, cooperation is fostered in
Prisoner's Dilemma and Snowdrift at low values of $T$, and cooperation is
inhibited at high $T$. Note that with unconditional imitation the crossover
between both behaviors takes place at $T \approx 1.7$, whereas with the
replicator rule it occurs at a much lower value of $T \approx 1.15$. The logic
behind this influence is better explained in the context of the full $ST$-plane,
as we will show later. 

The fact that this apparent contradiction has been resolved considering the role
of the update rule is a good example of its importance. This conclusion is in
agreement with those of \cite{tomassini:2006}, which performed an exhaustive
study on Snowdrift games with different network models and update rules, but
refutes those of \cite{hauert:2002}, which defended that the effect of
spatial lattices was almost independent of the update rule. In consequence, the
influence of the network models that we consider in the following is presented
separately for each kind of update rule, highlighting the differences in results
when appropriate. Apart from this, to assess and explain the influence of
spatial structure, we need to consider it along with games that have different
equilibrium structures, not only a particular game, in order to draw
sufficiently general conclusions. One way to do it is to
study their effect on the space of $2 \times 2$ games described by the
parameters $S$ and $T$ (\ref{eq:payoff-matrix}). A first attempt was done by
Hauert
\cite{hauert:2002}, but some problems in this study make it inconclusive (see
\cite{roca:2009a} for details on this issue).

Apart from lattices of different degrees (4, 6 and 8), we have also considered
homogeneous random networks, i.e.\ random networks where each node has exactly
the same number of neighbors. The aim of comparing with this kind of networks is
to isolate the effect of the spatial distribution of individuals from that of
the mere limitation of the number of neighbors and the \emph{context
preservation} \cite{cohen:2001} of a degree-homogeneous random network. The
well-mixed population hypothesis implies that every player plays with the
``average''
player in the population. From the point of view of the replicator rule this
means that every player samples successively the population in each evolution
step. It is not unreasonable to think that if the number of neighbors is
sufficiently restricted the result of this random sampling will differ from the
population average, thus introducing changes in the evolutionary outcome.

\begin{figure}[t]
\centering
\includegraphics[width=0.9\textwidth]{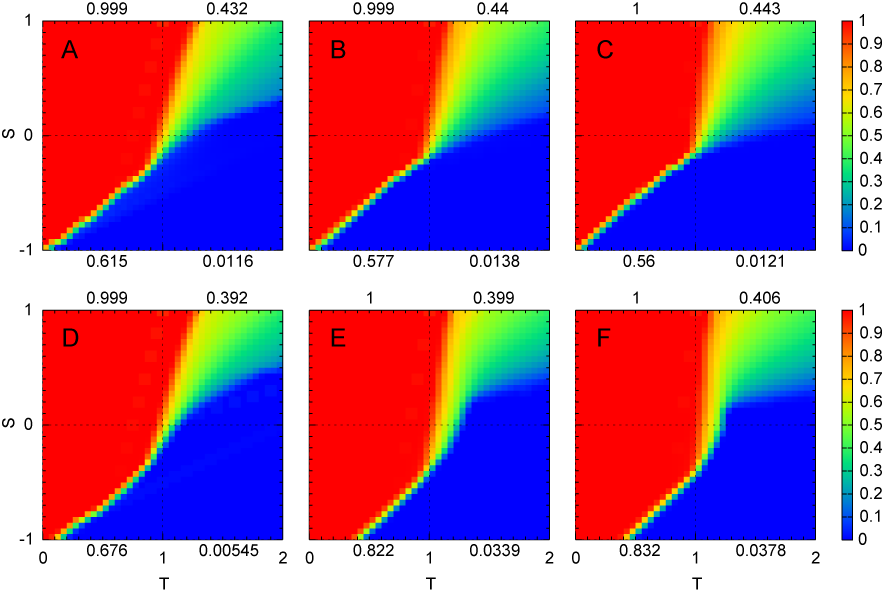}
\caption{Asymptotic density of cooperators $x^*$ in homogeneous random networks
(upper row, A to C) compared to regular lattices (lower row, D to F), with
degrees $k=$ 4 (A, D), 6 (B, E) and 8 (C, F). The update rule is the replicator
rule and the initial density of cooperators is $x^0 =0.5$. The plots show that
the main influence occurs in Stag Hunt and Snowdrift games, specially for
regular lattices with large clustering coefficient, $k=$ 6 and 8 (see main
text).}
\label{fig:spatial-replicator}
\end{figure}

Fig.~\ref{fig:spatial-replicator} shows the results for the replicator rule with
random and spatial networks of different degrees. First, it is clear that the
influence of these networks is negligible on Harmony games and minimal
on Prisoner's Dilemmas, given the reduced range of parameters where it is
noticeable. There is, however, a clear influence on Stag Hunt and Snowdrift
games, which is always of opposite sign: An enhancement of cooperation in Stag
Hunt and an inhibition in Snowdrift. Second, it is illuminating to consider the
effect of increasing the degree. For the random network, it means that its weak
influence vanishes. The spatial lattice, however, whose result is very similar
to that of the random one for the lowest degree ($k=4$), displays remarkable
differences for the greater degrees ($k=$ 6 and 8). These differences are a
clear promotion of cooperation in Stag Hunt games and a lesser, but measurable,
inhibition in Snowdrift games, specially for low $S$.

The relevant topological feature that underlies this effect is the existence of
clustering in the network, understood as the presence of triangles or,
equivalently, common neighbors
\cite{newman:2003, boccaletti:2006}. In regular lattices, for $k=4$ there is no
clustering, but there is for $k=6$ and $8$. This point explains the difference
between the conclusions of Cohen et al.\ \cite{cohen:2001} and those of Ifti et
al.\ \cite{ifti:2004} and Tomassini et al.\ \cite{tomassini:2006}, regarding
the role of network clustering in the effect of spatial populations. In
\cite{cohen:2001}, rectangular lattices of degree $k=4$ were
considered, which have strictly zero clustering because there are not closed
triangles in the network, hence finding no differences in outcome between the
spatial and the random topology. In the latter case, on the contrary, both
studies employed rectangular lattices of degree $k=8$, which do have clustering,
and thus they identified it as a key feature of the network, for particular
parameterizations of the games they were studying, namely Prisoner's Dilemma
\cite{ifti:2004} and Snowdrift \cite{tomassini:2006}.

An additional evidence for this conclusion is the fact that small-world
networks, which include random links to reduce the average path between nodes
while maintaining the clustering, produce almost indistinguishable results from
those of Fig.~\ref{fig:spatial-replicator}~D-F. This conclusion is in agreement
with existent theoretical work about small-world networks, on Prisoner's
Dilemma \cite{abramson:2001,masuda:2003,tomochi:2004} and its extensions
\cite{wu:2005,wu:2006}, on Snowdrift games \cite{tomassini:2006}, and also with
experimental studies on coordination games \cite{cassar:2007}. The difference
between the effect of regular lattices and small-world networks consists, in
general, in a \emph{greater efficiency} of the latter in reaching the
stationary state (see \cite{roca:2009a} for a further discussion on this
comparison).

\begin{figure}
\centering
\includegraphics[height=0.8\textheight]{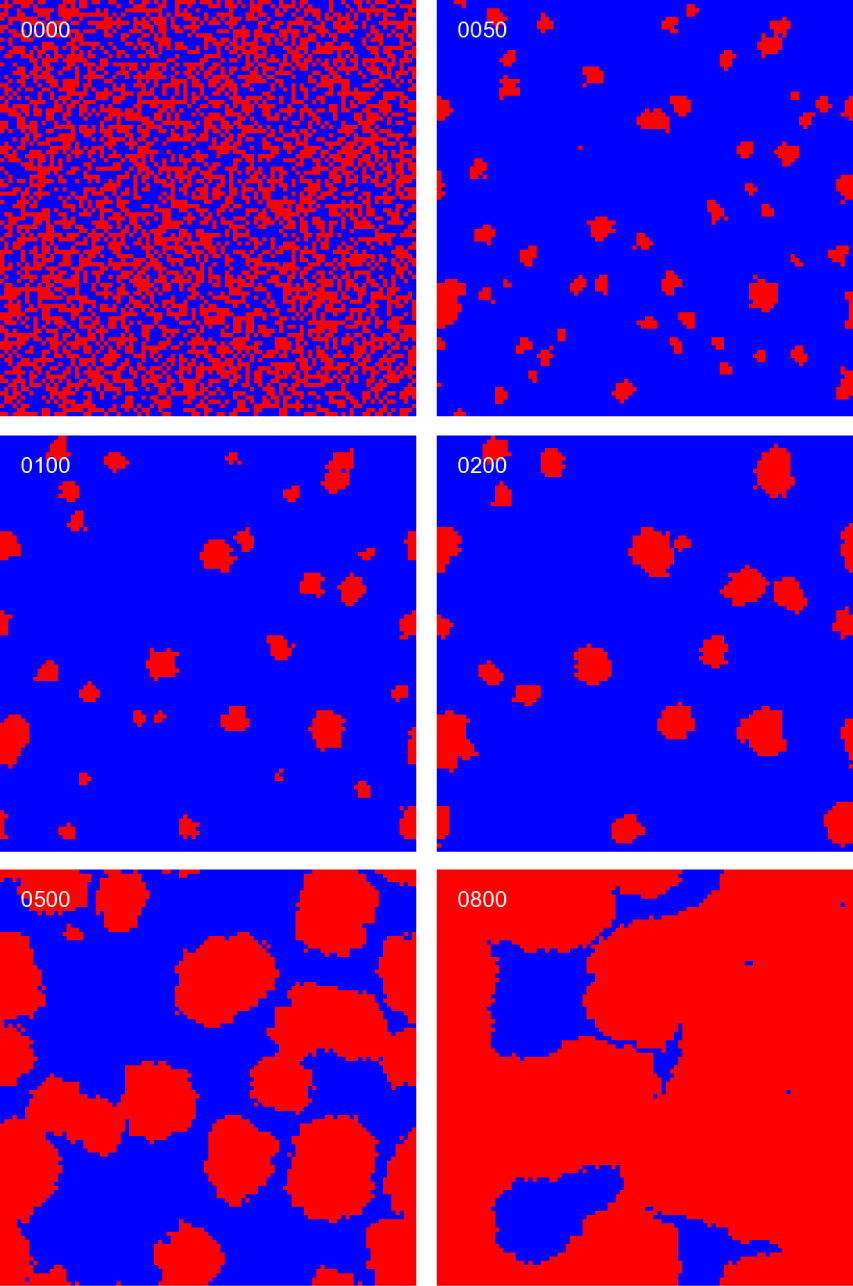}
\caption{Snapshots of the evolution of a population on a regular lattice of
degree $k=8$, playing a Stag Hunt game ($S=-0.65$ and $T=0.65$). Cooperators are
displayed in red and defectors in blue. The update rule is the replicator rule
and the initial density of cooperators is $x^0 =0.5$. The upper left label shows
the time step $t$. During the initial steps, cooperators with low local density
of cooperators in their neighborhood disappear, whereas those with high local
density grow into the clusters that eventually take up the complete population.}
\label{fig:spatial-replicator-snapshots}
\end{figure}

The mechanism that explains this effect is the formation and growth of clusters
of cooperators, as Fig.~\ref{fig:spatial-replicator-snapshots} displays for a
particular realization. The outcome of the population is then totally determined
by the stability and growth of these clusters, which in turn depend on the
dynamics of clusters interfaces. This means that the result is no longer
determined by the global population densities but by the local densities that
the players at the cluster interfaces see in their
neighborhood. In fact, the primary effect that the network clustering causes is
to favor, i.e.\ to maintain or to increase, the high local densities that were
present in the population from the random beginning. This favoring produces
opposite effects in Stag Hunt and Snowdrift games. As an illustrating example,
consider that the global density is precisely that of the mixed equilibrium of
the game. In Stag Hunt games, as this equilibrium is unstable, a higher local
density induces the conversion of nearby defectors to cooperators, thus making
the cluster grow. In Snowdrift games, on the contrary, as the equilibrium is
stable, it causes the conversion of cooperators to defectors. See
\cite{roca:2009a} for a full discussion on this mechanism.

In view of this, recalling that these are the results for the replicator rule,
and that therefore they correspond to the correct update rule to study the
influence of population structure on replicator dynamics, we can state that the
presence of clustering (triangles, common neighbors) in a network is a relevant
topological feature for the evolution of cooperation. Its main effects are, on
the one hand, a promotion of cooperation in Stag Hunt games, and, on the other
hand, an inhibition (of lower magnitude) in Snowdrift games.
We note, however, that clustering may not be the only relevant factor governing
the game
asymptotics: one can devise peculiar graphs, not representing proper spatial
structure, where
other influences prove relevant. This is the case of networks consisting of a
complete subgraphs
connected to each other by a few connections \cite{szabo:2005b}, a system
whose behavior, in spite of the high clustering coefficient, is similar to those
observed on the traditional square lattice where the clustering coefficient is
zero. This was subsequently related \cite{vukov:2006} to the existence of
overlapping triangles that support the spreading of cooperation. We thus see
that our claim about the outcome of evolutionary games on networks with
clustering is anything but general and depends on the translational invariance
of the network.

\begin{figure}[t]
\centering
\includegraphics[width=0.9\textwidth]{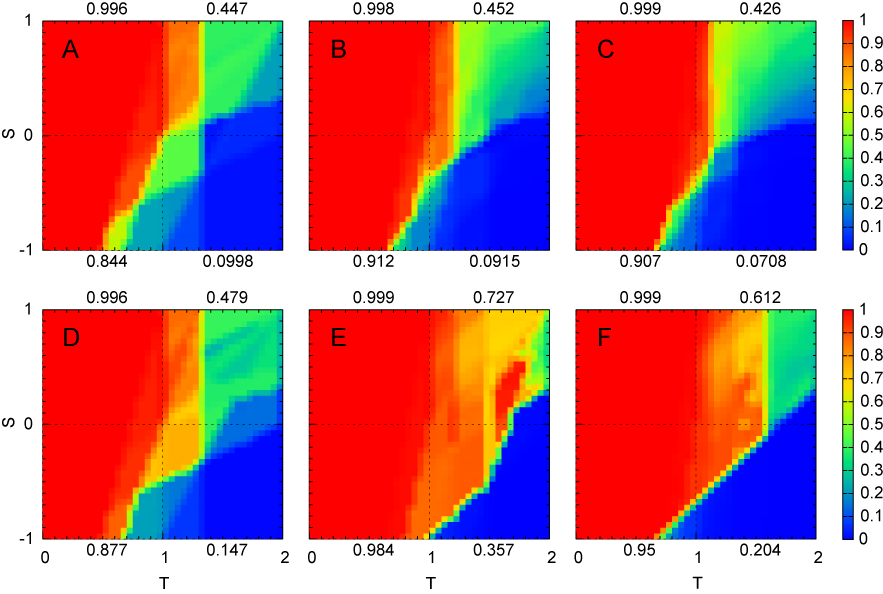}
\caption{Asymptotic density of cooperators $x^*$ in homogeneous random networks
(upper row, A to C) compared to regular lattices (lower row, D to F), with
degrees $k=$ 4 (A, D), 6 (B, E) and 8 (C, F). The update rule is unconditional
imitation and the initial density of cooperators is $x^0 =0.5$. Again as in
Fig.~\ref{fig:spatial-replicator}, spatial lattices have greater influence than
random networks when the clustering coefficient is high ($k=$ 6 and 8). In this
case, however, the beneficial effect for cooperation goes well into Snowdrift
and Prisoner's Dilemma quadrants.}
\label{fig:spatial-imitation}
\end{figure}

Other stochastic non-innovative rules, such as the multiple replicator and Moran
rules, yield similar results, without qualitative differences \cite{roca:2009a}.
Unconditional imitation, on the contrary, has a very different influence,
as can be seen in Fig.~\ref{fig:spatial-imitation}.

In the first place, homogenous random networks themselves have a marked
influence, that \emph{increases} with network degree for Stag Hunt and Snowdrift
games, but decreases for Prisoner's Dilemmas. Secondly, there are again no
important differences between random and spatial networks if there is no
clustering in the network (note how the transitions between the different
regions in the results are the same). There are, however, stark differences when
there is clustering in the network. Interestingly, these are the cases with an
important promotion of cooperation in Snowdrift and Prisoner's Dilemma games.

\begin{figure}
\centering
\includegraphics[height=0.8\textheight]{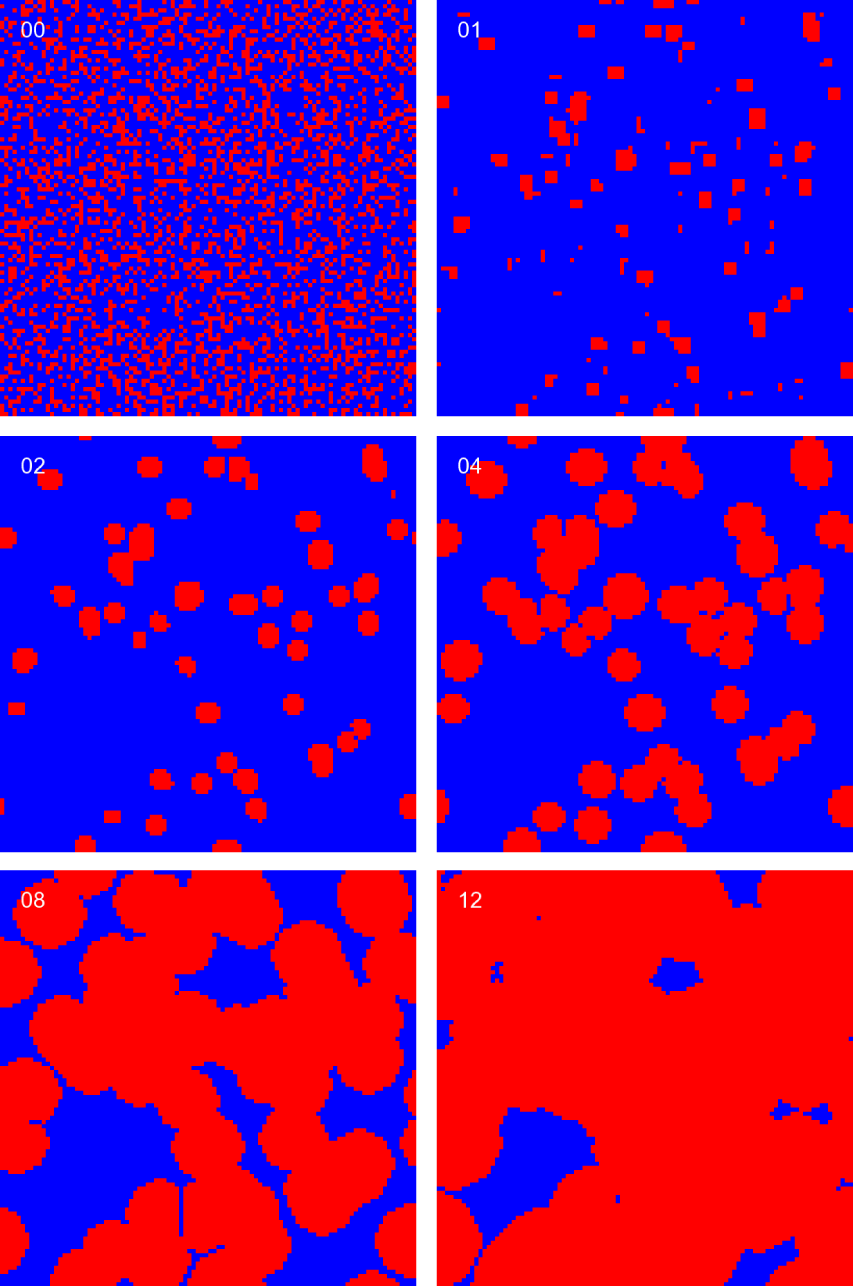}
\caption{Snapshots of the evolution of a population on a regular lattice of
degree $k=8$, playing a Stag Hunt game ($S=-0.65$ and $T=0.65$). Cooperators are
displayed in red and defectors in blue. The update rule is unconditional
imitation and the initial density of cooperators is $x^0 =1/3$ (this lower value
than that of Fig.~\ref{fig:spatial-replicator-snapshots} has been used to make
the evolution longer and thus more easily observable). The upper left label
shows the time step $t$. As with the replicator rule (see
Fig.~\ref{fig:spatial-replicator-snapshots}), during the initial time steps
clusters emerge from cooperators with high local density of cooperators in their
neighborhood. In this case, the interfaces advance deterministically at each
time step, thus giving a special significance to flat interfaces and producing a
much faster evolution than with the replicator rule (compare time labels with
those of Fig.~\ref{fig:spatial-replicator-snapshots})}
\label{fig:spatial-imitation-snapshots}
\end{figure}

In this case, the dynamical mechanism is the formation and growth of clusters of
cooperators as well, and the fate of the population is again determined by the
dynamics of cluster interfaces. With unconditional imitation, however, given its
deterministic nature, interfaces advance one link every time step. This makes
very easy the calculation of the conditions for their advancement, because these
conditions come down to those of a flat interface between cooperators and
defectors
\cite{roca:2009a}. See Fig.~\ref{fig:spatial-imitation-snapshots} for a typical
example of evolution.

\begin{figure}[t]
\centering
\includegraphics[width=0.9\textwidth]{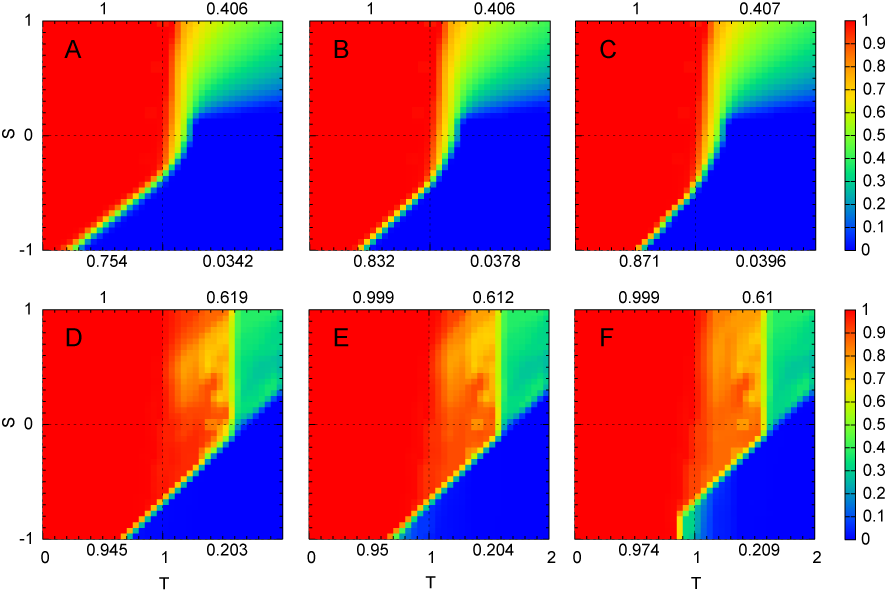}
\caption{Asymptotic density of cooperators $x^*$ in regular lattices of degree
$k=8$, for different initial densities of cooperators $x^0=$ 1/3 (A, D), 1/2 (B,
E) and 2/3 (C, F). The update rules are the replicator rule (upper row, A to C)
and unconditional imitation (lower row, D to F). With the replicator rule, the
evolutionary outcome in Stag Hunt games depends on the initial condition, as is
revealed by the displacement of the transition line between full cooperation and
full defection. However, with unconditional imitation this transition line
remains in the same position, thus showing the insensitivity to the initial
condition. In this case, the outcome is determined by the presence of small
clusters of cooperators in the initial random population, which takes place for
a large range of values of the initial density of cooperators $x^0$.}
\label{fig:spatial-initial-condition}
\end{figure}

An interesting consequence of the predominant role of flat interfaces with
unconditional imitation is that, as long as there is in the initial population a
flat interface (i.e.\ a cluster with it, as for example a $3 \times 2$ cluster
in a 8-neighbor lattice), the cluster will grow and eventually extend to the
entire population. This feature corresponds to the $3 \times 3$ cluster rule
proposed by Hauert \cite{hauert:2002}, which relates the outcome of the entire
population to that of a cluster of this size. This property makes the
evolutionary outcome quite independent of the initial density of cooperators,
because even for a low initial density the probability that a suitable small
cluster exists will be high for sufficiently large populations; see
Fig.~\ref{fig:spatial-initial-condition}~D-F about the differences in initial
conditions. Nevertheless, it is important to realize that this rule is based on
the dynamics of flat interfaces and, therefore, it is only valid for
unconditional imitation. Other update rules that also give rise to clusters, as
replicator rule for example, develop interfaces with different shapes, rendering
the particular case of flat interfaces irrelevant. As a consequence, the
evolution outcome becomes dependent on the initial condition, as
Fig.~\ref{fig:spatial-initial-condition}~A-C displays.

In summary, the relevant topological feature of these homogeneous networks, for
the games and update rules considered so far, is the clustering of the network.
Its effect depends largely on the update rule, and the most that can be said in
general is that, besides not affecting Harmony games, it consistently promotes
cooperation in Stag Hunt games.

\subsection{Synchronous vs asynchronous update}

Huberman and Glance \cite{huberman:1993} questioned the generality of the
results reported by Nowak and May \cite{nowak:1992}, in terms of the
synchronicity of the update of strategies. Nowak and May used synchronous
update, which means that every player is updated at the same time, so the
population evolves in successive generations. Huberman and Glance, in contrast, 
employed asynchronous update (also called random sequential update), in which
individuals are updated independently one by one, hence the neighborhood of each
player always remains the same while her strategy is being updated. They showed
that, for a particular game, the asymptotic cooperation obtained with
synchronous update disappeared. This has become since then one of the most
well-known and cited examples of the importance of synchronicity in the update
of strategies in evolutionary models. Subsequent works have, in turn,
critizised the importance of this issue, showing that the conclusions of
\cite{nowak:1992} are robust \cite{nowak:1994,szabo:2005}, or restricting the
effect reported by \cite{huberman:1993} to particular instances of Prisoner's
Dilemma \cite{lindgren:1994} or to the short memory of players
\cite{kirchkamp:2000}. Other works, however, in the different context of
Snowdrift games \cite{kun:2006,tomassini:2006} have found that the influence on
cooperation can be positive or negative, in the asynchronous case compared with
the synchronous one.

\begin{figure}[t!]
\centering
\includegraphics[width=0.6\textwidth]{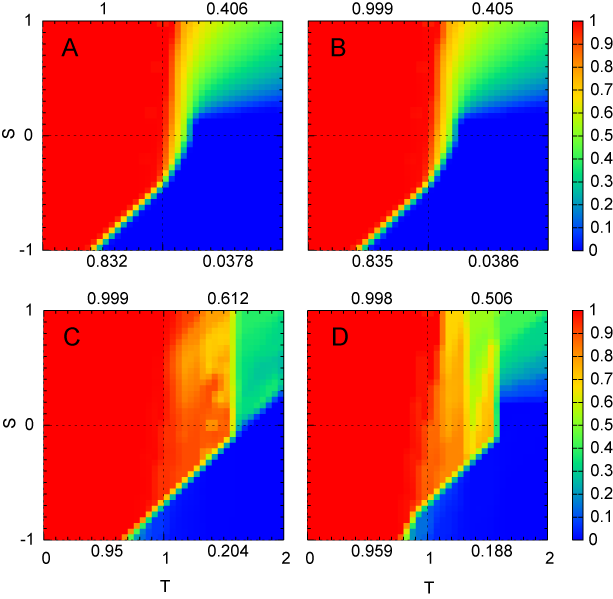}
\caption{Asymptotic density of cooperators $x^*$ in regular lattices of degree
$k=8$, with synchronous update (left, A and C) compared to asynchronous (right,
B and D). The update rules are the replicator rule (upper row) and unconditional
imitation (lower row). The initial density of cooperators is $x^0 =0.5$. For the
replicator rule, the results are virtually identical, showing the lack of
influence of the synchronicity of update on the evolutionary outcome. In the
case of unconditional imitation the results are very similar, but there are
differences for some points, specially Snowdrift games with $S \lesssim 0.3$ and
$T>5/3 \approx 1.67$. The particular game studied by Huberman and Glance
\cite{huberman:1993}, which reported a suppression  of cooperation due to
asynchronous update, belongs to this region.}
\label{fig:huberman-st}
\end{figure}

\begin{figure}[t!]
\centering
\includegraphics[width=0.99\textwidth]{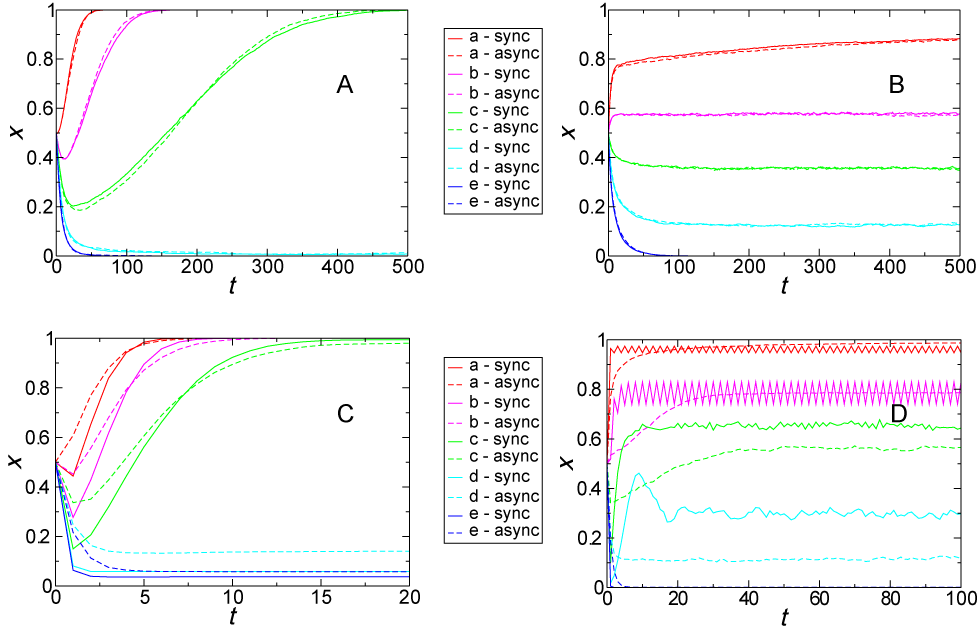}
\caption{Time evolution of the density of cooperators $x$ in regular lattices of
degree $k=8$, for typical realizations of Stag Hunt (left, A and C) and
Snowdrift games (right, B and D), with synchronous (continuous lines) or
asynchronous (dashed lines) update. The update rules are the replicator rule
(upper row) and unconditional imitation (lower row). The Stag Hunt games for the
replicator rule (A) are: a, $S=-0.4$, $T=0.4$; b, $S=-0.5$, $T=0.5$; c,
$S=-0.6$, $T=0.6$; d, $S=-0.7$, $T=0.7$; e, $S=-0.8$, $T=0.8$. For unconditional
imitation the Stag Hunt games (C) are: a, $S=-0.6$, $T=0.6$; b, $S=-0.7$,
$T=0.7$; c, $S=-0.8$, $T=0.8$; d, $S=-0.9$, $T=0.9$; e, $S=-1.0$, $T=1.0$. The
Snowdrift games are, for both update rules (B, D): a, $S=0.9$, $T=1.1$; b,
$S=0.7$, $T=1.3$; c, $S=0.5$, $T=1.5$; d, $S=0.3$, $T=1.7$; e, $S=0.1$, $T=1.9$.
The initial density of cooperators is $x^0 =0.5$. The time scale of the
asynchronous realizations has been re-scaled by the size of the population, so t
 hat for both kinds of update a time step represents the same number of update
events in the population. Figures A and B show that, in the case of the replicator
rule, not only the outcome but also the time evolution is independent of the
update synchronicity. With unconditional imitation the results are also very
similar for Stag Hunt (C), but somehow different in Snowdrift (D) for large $T$,
displaying the influence of synchronicity in this subregion. Note that in all
cases unconditional imitation yields a much faster evolution than the replicator
rule.}
\label{fig:huberman-time}
\end{figure}

We have thoroughly investigated this issue, finding that the effect of
synchronicity in the update of strategies is the exception rather than the
rule. With the replicator rule, for example, the evolutionary outcome in both
cases is virtually identical, as Fig.~\ref{fig:huberman-st}~A-B shows. Moreover,
in this case, the time evolution is also very similar
(see Fig.\ref{fig:huberman-time}~A-B). With unconditional imitation there are
important differences only in one particular subregion of the space of
parameters, corresponding mostly to Snowdrift games, to which the specific
game studied by Huberman and Glance belongs (see Fig.~\ref{fig:huberman-st}~C-D
and \ref{fig:huberman-time}~C-D).

\clearpage

\subsection{Heterogeneous networks}

\begin{figure}[t]
\centering
\includegraphics[width=0.9\textwidth]{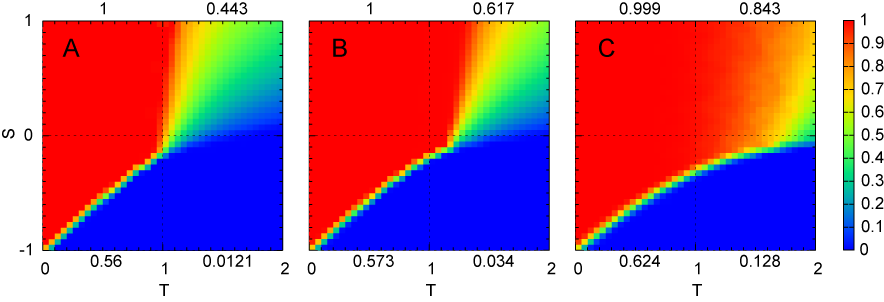}
\caption{Asymptotic density of cooperators $x^*$ with the replicator update
rule, for model networks with different degree heterogeneity: homogeneous random
networks (left, A), Erd\H{o}s-R\'enyi random networks (middle, B) and
Barab\'asi-Albert scale-free networks (right, C). In all cases the average
degree is $\bar{k}=8$ and the initial density of cooperators is $x^0=0.5$.
As degree heterogeneity grows, from left to right, cooperation in Snowdrift
games is clearly enhanced.}
\label{fig:santos-replicator}
\end{figure}

The other important topological feature for evolutionary games was introduced by
Santos and co-workers \cite{santos:2006a,santos:2005,santos:2006c}, who studied
the effect of degree heterogeneity, in particular scale-free networks. Their
main result is shown in Fig.~\ref{fig:santos-replicator}, which displays the
variation in the evolutionary outcome induced by increasing the variance of the
degree distribution in the population, from zero (homogeneous random networks)
to a finite value (Erd\H{o}s-R\'enyi random networks), and then to infinity
(scale-free networks). The enhancement of cooperation as degree heterogeneity
increases is very clear, specially in the region of Snowdrift games. The effect
is not so strong, however, in Stag Hunt or Prisoner's Dilemma games. Similar
conclusions are obtained with other scale-free topologies, as for example with
Klemm-Egu\'iluz scale-free networks \cite{klemm:2002}. Very recently, it has
been shown \cite{assenza:2009} that much as we discussed above for the case of
spatial structures, clustering is also a factor improving the cooperative
behavior in scale-free networks.

\begin{figure}[t]
\centering
\includegraphics[width=0.9\textwidth]{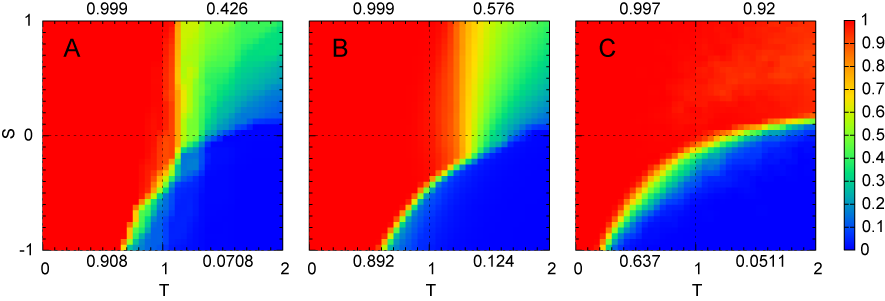}
\caption{Asymptotic density of cooperators $x^*$ with unconditional imitation as
update rule, for model networks with different degree heterogeneity: homogeneous
random networks (left, A), Erd\H{o}s-R\'enyi random networks (middle, B) and
Barab\'asi-Albert scale-free networks (right, C). In all cases the average
degree is $\bar{k}=8$ and the initial density of cooperators is $x^0=0.5$.
As degree heterogeneity grows, from left to right, cooperation in Snowdrift
games is enhanced again. In this case, however, cooperation is inhibited in Stag
Hunt games and reaches a maximum in Prisoner's Dilemmas for Erd\H{o}s-R\'enyi
random networks.}
\label{fig:santos-imitation}
\end{figure}

The positive influence on Snowdrift games is quite robust against changes in
network degree and the use of other update rules. On the other hand, the
influence on Stag Hunt and Prisoner's Dilemma games is quite restricted and very
dependent on the update rule, as Fig.~\ref{fig:santos-imitation} reveals. In
fact, with unconditional imitation cooperation is inhibited in Stag Hunt games
as the network becomes more heterogeneous, whereas in Prisoner's Dilemmas it
seems to have a maximum at networks with finite variance in the degree
distribution.

A very interesting insight from the comparison between the effects of network
clustering and degree heterogeneity is that they mostly affect games with one
equilibrium in mixed strategies, and that in addition the effects on these games
are different. This highlights the fact that they are different fundamental 
topological properties, which induce mechanisms of different nature. In the case
of network clustering we have seen the formation and growth of clusters of
cooperators. For network heterogeneity the phenomena is the bias and
stabilization of the strategy oscillations in Snowdrift games towards the
cooperative strategy
\cite{gomezgardenes:2007,poncela:2007}, as we explain in the following. The
asymptotic state of Snowdrift games in homogeneous networks consists of a mixed
strategy population, where every individual oscillates permanently between
cooperation and defection. Network heterogeneity tends to prevent this
oscillation, making players in more connected sites more prone to
be cooperators. At first, having more neighbors makes any individual receive
more payoff, despite her strategy, and hence she has an evolutionary advantage.
For a defector, this is a short-lived advantage, because it triggers the change
of her neighbors to defectors, thus loosing payoff. A high payoff cooperator, on
the contrary, will cause the conversion of her neighbors to cooperators,
increasing even more her own payoff. These highly connected cooperators
constitute the hubs that drive the population, fully or partially, to
cooperation. It is clear that this mechanism takes place when cooperators
collect more payoff from a greater neighborhood, independently of their
neighbors' strategies. This only happens when $S>0$, which is the reason why the
positive effect on cooperation of degree heterogeneity is mainly restricted to
Snowdrift games.

\clearpage

\subsection{Best response update rule}

So far, we have dealt with imitative update rules, which are non-innovative.
Here we present the results for an innovative rule, namely best response. With
this rule each player chooses, with certain probability $p$, the strategy that
is the best response for her current neighborhood. This rule is also referred to
as myopic best response, because the player only takes into account the last
evolution step to decide the optimum strategy for the next one. Compared to the
rules presented previously, this one assumes more powerful cognitive abilities
on the individual, as she is able to discern the payoffs she can obtain
depending on her strategy and those of her neighbors, in order to chose the best
response. From this point of view, it constitutes a next step in the
sophistication of update rules.

\begin{figure}[t]
\centering
\includegraphics[width=0.5\textwidth]{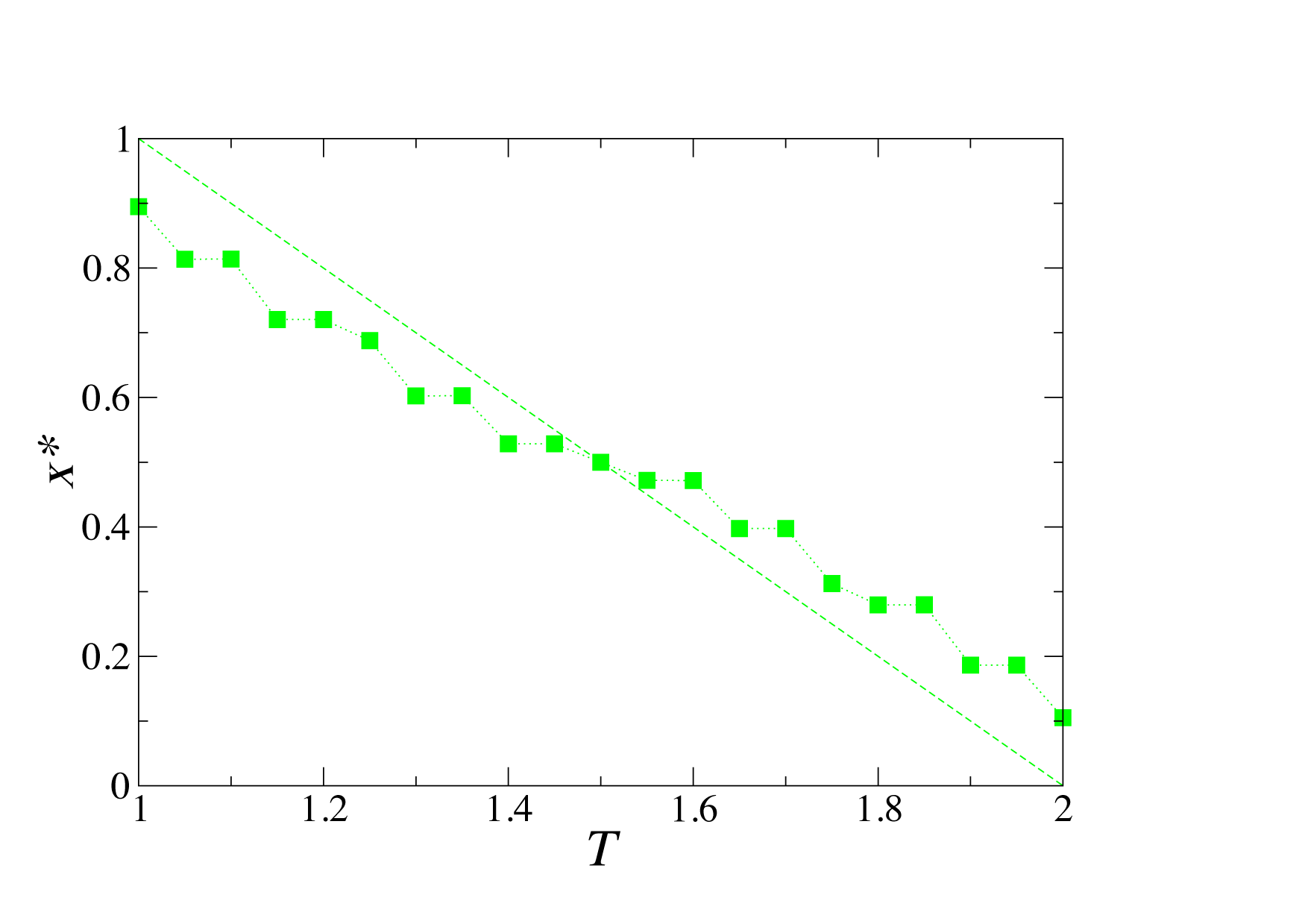}
\caption{Asymptotic density of cooperators $x^*$ in a square lattice with degree
$k=8$ and best response as update rule, in the model with Snowdrift
(\ref{eq:hauert-sd}) studied by Sysi-Aho and co-workers \cite{sysi-aho:2005}.
The result for a well-mixed population is displayed as a reference. Note how the
promotion or inhibition of cooperation does not follow the same variation as a
function of $T$ than in the case with the replicator rule studied by Hauert and
Doebeli \cite{hauert:2004} (Fig.~\ref{fig:hauert}).}
\label{fig:sysi-aho}
\end{figure}

An important result of the influence of this rule for evolutionary games was
published in 2005 by Sysi-Aho and co-workers \cite{sysi-aho:2005}. They studied
the combined influence of this rule with regular lattices, in the same
one-dimensional parameterization of Snowdrift games (\ref{eq:hauert-sd}) that
was employed by Hauert and Doebeli \cite{hauert:2004}. They reported a
modification in the cooperator density at equilibrium, with an increase for some
subrange of the parameter $T$ and a decrease for the other, as
Fig.~\ref{fig:sysi-aho} shows.

\begin{figure}
\centering
\includegraphics[width=0.9\textwidth]{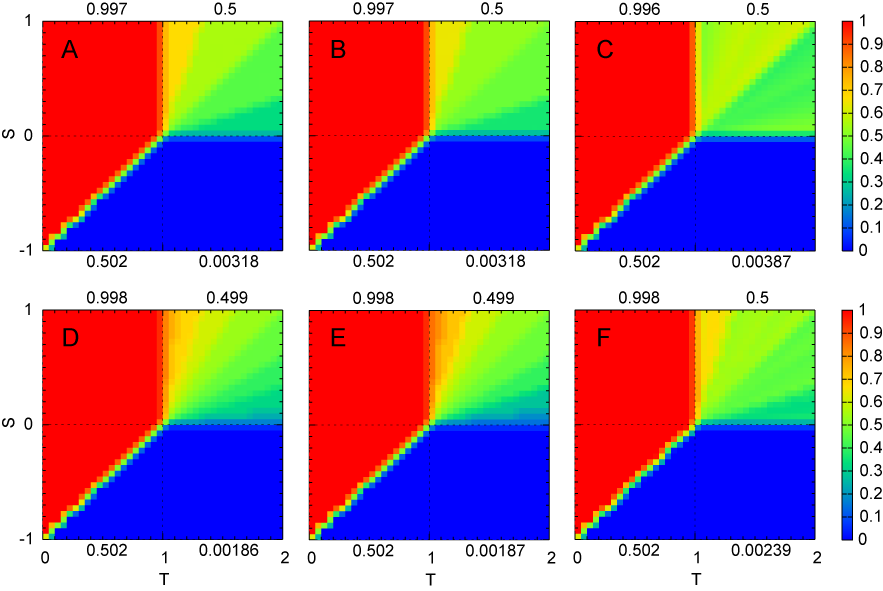}
\caption{Asymptotic density of cooperators $x^*$ in random (left, A and D),
regular (middle, B and E), and scale-free networks (right, C and F) with
average degrees $\bar{k}=4$ (upper row, A to C) and 8 (lower row, D to F). The
update rule is best response with $p=0.1$ and the initial density of cooperators
is $x^0 =0.5$. Differences are negligible in all cases; note, however, that the
steps appearing in the Snowdrift quadrant are slightly different.}
\label{fig:best-response}
\end{figure}

At the moment, it was intriguing that regular lattices had opposite effects
(promotion or inhibition of cooperation) in some ranges of the parameter $T$,
depending on the update rule used in the model. Very recently we have carried
out a thorough investigation of the influence of this update rule on a wide
range of networks \cite{roca:2009b}, focusing on the key topological properties
of network clustering and degree heterogeneity. The main conclusion of this
study is that, with only one relevant exception, the best response rule
suppresses the effect of population structure on evolutionary games.
Fig.~\ref{fig:best-response} shows a summary of these results. In all cases the
outcome is very similar to that of replicator dynamics on well-mixed populations
(Fig.~\ref{fig:compnet}), despite the fact that the networks studied explore
different options of network clustering and degree heterogeneity. The steps in
the equilibrium density of Snowdrift games, as those reported in
\cite{sysi-aho:2005}, show up in all cases, with slight variations which depend
mostly on the mean degree of the network.

\begin{figure}
\centering
\includegraphics[width=0.9\textwidth]{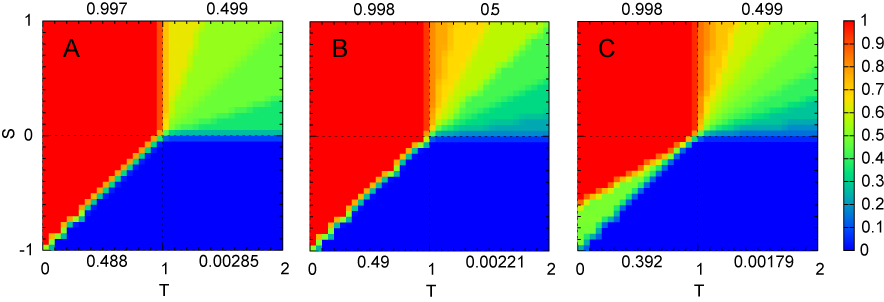}
\caption{Asymptotic density of cooperators $x^*$ in regular lattices with
initial density of cooperators $x^0 =1/3$. The degrees are $k=4$ (left, A),
$k=6$ (middle, B) and $k=8$ (right, C). The update rule is best response with
$p=0.1$. Comparing with Fig.~\ref{fig:compnet}~A, there is a clear displacement
of the boundary between full defection and full cooperation in Stag Hung games,
which amounts to a promotion of cooperation. The widening of the border in panel
C is a finite size effect, which disappears for larger populations. See main
text for further details.}
\label{fig:best-response-lattices}
\end{figure}

The exception to the absence of network influence is the case of regular
lattices, and consists of a modification of
the unstable equilibrium in Stag Hunt games, in the sense that it produces a
promotion of cooperation for initial densities lower than 0.5 and a
corresponding symmetric inhibition for greater densities. An example of this
effect is given in Fig.~\ref{fig:best-response-lattices}, where the outcome
should be compared to that of well-mixed populations in
Fig.~\ref{fig:compnet}~A. The reason for this effect is that the lattice creates
special conditions for the advancement (or receding) of the interfaces of
clusters of cooperators. We refer the interested reader to \cite{roca:2009b} for
a detailed description of this phenomena. Very remarkably, in this case network
clustering is not relevant, because the effect also takes place for degree
$k=4$, at which there is no clustering in the network.

\clearpage

\subsection{Weak selection}
\label{subsec:weak-selection}

This far, we have considered the influence of population structure in the case
of \emph{strong selection pressure}, which means that the fitness of individuals
is totally determined by the payoffs resulting from the game. In general this
may not be the case, and then to relax this restriction the fitness can be
expressed as $f = 1 - w + w \pi$ \cite{nowak:2004a}. The parameter $w$
represents the intensity of selection and can vary between $w=1$ (strong
selection limit) and $w \gtrsim 0$ (weak selection limit). With a different
parameterization, this implements the same idea as the baseline fitness
discussed in Section~\ref{sec:3}. We note that another interpretation has been
recently proposed \cite{wild:2007}
for this limit, namely  $\delta$-weak selection, which assumes that the game
means
much to the determination of reproductive success, but
that selection is weak because mutant and wild-type strategies are very similar.
This second
interpretation leads to different results \cite{wild:2007} and we do not deal
with it here, but
rather we stick with the first one, which is by far the most generally used.

The weak selection limit has the nice property of been tractable analytically.
For instance, Ohtsuki and Nowak have studied evolutionary games on homogeneous
random networks using this approach \cite{ohtsuki:2006}, finding an interesting
relation with replicator dynamics on well-mixed populations. Using our
normalization of the game (\ref{eq:payoff-matrix}), their main result can be
written as the following payoff matrix
\begin{equation}
\left(
  \begin{array}{cc} 1 & S + \Delta \\ T - \Delta & 0 \end{array}
\right).
\end{equation}
This means that the evolution in a population structured according to a random
homogeneous network, in the weak selection limit, is the same as that of a
well-mixed population with a game defined by this modified payoff matrix. The
effect of the network thus reduces to the term $\Delta$, which depends on the
game, the update rule and the degree $k$ of the network. With respect to the
influence on cooperation it admits a very straightforward interpretation: If
both the original and the modified payoff matrices correspond to a Harmony or
Prisoner's Dilemma game, then there is logically no influence, because the
population ends up equally in full cooperation or full defection; otherwise,
cooperation is enhanced if $\Delta > 0$, and inhibited if $\Delta < 0$.

The actual values of $\Delta$, for the update
rules Pairwise Comparison (PC), Imitation (IM) and
Death-Birth (DB) (see \cite{ohtsuki:2006} for full details), are
\begin{eqnarray}
\Delta_{PC} &=& \frac {S - (T-1)} {k-2} \\
\Delta_{IM} &=& \frac {k + S - (T-1)} {(k+1)(k-2)} \\
\Delta_{DB} &=& \frac {k + 3(S - (T-1))} {(k+3)(k-2)} ,
\end{eqnarray}
$k$ being the degree of the network. A very remarkable feature of these
expressions is that for every pair of games with parameters $(S_1,T_1)$ and
$(S_2,T_2)$, if $S_1-T_1 = S_2-T_2$ then $\Delta_1 = \Delta_2$. Hence the
influence on cooperation for such a pair of games, even if one is a Stag Hunt
and the other is a Snowdrift, will be the same. This stands in stark contrast to
all the reported results with strong selection, which generally exhibit
different, and in many cases opposite, effects on both games. Besides this, as
the term $S - (T-1)$ is negative in all Prisoner's Dilemmas and half the
cases of Stag Hunt and Snowdrift games, the beneficial influence on cooperation
is quite reduced for degrees $k$ as those considered above \cite{roca:2009a}.

\begin{figure}
\centering
\includegraphics[width=\textwidth]{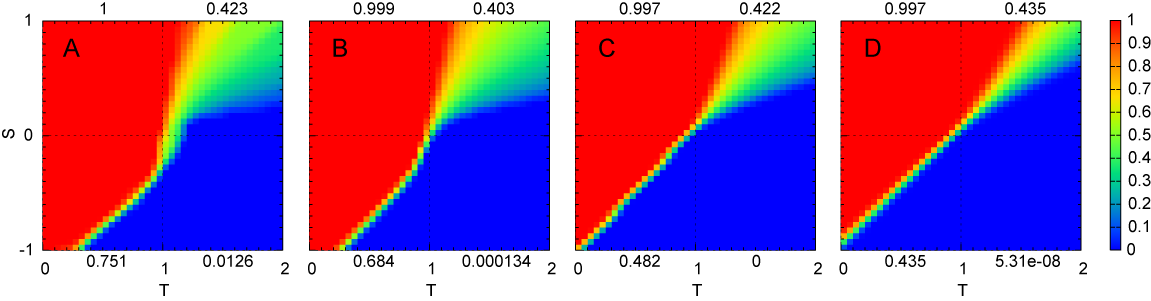}
\caption{Asymptotic density of cooperators $x^*$ in regular lattices of degree $k=8$, for the Fermi update rule with $\beta$ equal to 10 (A), 1 (B), 0.1 (C) and 0.01 (D). The initial density of cooperators is $x^0=0.5$. For high $\beta$ the result is quite similar to that obtained with the replicator rule (Fig.~\ref{fig:spatial-replicator}~F). As $\beta$ decreases, or equivalently for weaker intensities of selection, the influence becomes smaller and more symmetrical between Stag Hunt and Snowdrift games.}
\label{fig:fermi-lattice}
\end{figure}

\begin{figure}
\centering
\includegraphics[width=\textwidth]{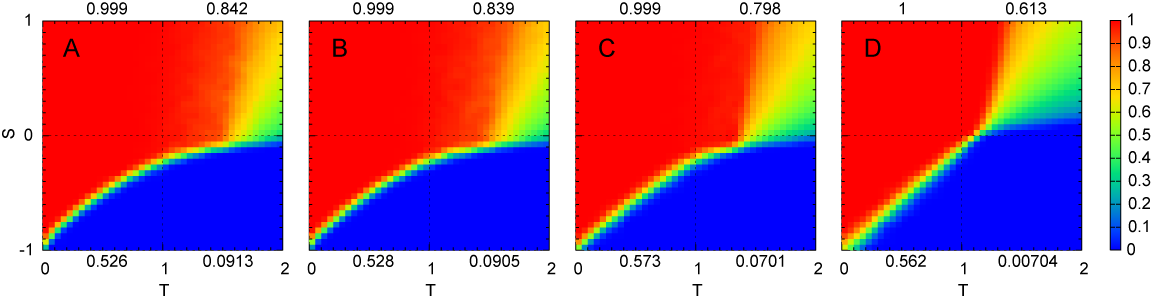}
\caption{Asymptotic density of cooperators $x^*$ in Barab\'asi-Albert scale-free
networks of average degree $\bar{k}=8$, for the Fermi update rule with
$\beta$ equal to 10 (A), 1 (B), 0.1 (C) and 0.01 (D). The initial density of
cooperators is $x^0=0.5$. As in Fig.~\ref{fig:fermi-lattice}, for high $\beta$
the result is quite similar to that obtained with the replicator rule
(Fig.~\ref{fig:santos-replicator}~C), and analogously, as $\beta$ decreases the
influence of the network becomes smaller and more symmetrical between Stag Hunt
and Snowdrift games.}
\label{fig:fermi-scalefree}
\end{figure}

Another way to investigate the influence of the intensity of selection if to employ the Fermi update rule, presented above, which allows to study numerically the effect of varying the intensity of selection on any network model. Figs.~\ref{fig:fermi-lattice} and~\ref{fig:fermi-scalefree} display the results obtained, for different intensities of selection, on networks that
are prototypical examples of strong influence on evolutionary games, namely regular lattices with high clustering and scale-free networks, with large degree heterogeneity. In both cases, as the intensity of selection is reduced, the effect of the network becomes weaker and more symmetrical between Stag Hunt and Snowdrift games. Therefore, these results show that the strong and weak selection limits are not comparable from the viewpoint of the evolutionary outcome, and that weak selection largely inhibits the influence of population structure.

\clearpage

\section{Conclusion and future prospects}
\label{sec:5}

In this review, we have discussed non-mean-field effects on evolutionary game dynamics. Our
reference framework for comparison has been the replicator equation, a pillar of modern evolutionary
game theory that has produced many interesting and fruitful insights on different fields. Our purpose
here has been to show that, in spite of its many successes, the replicator equation is only a part of
the story, much in the same manner as mean-field theories have been very important in physics
but they cannot (nor are they intended to) describe all possible phenomena. The main issues we
have discussed are the influence of fluctuations, by considering the existence of more than one
time scale, and of spatial correlations, through the constraints on interaction arising from an
underlying network structure.  In doing so, we have shown a wealth of evidence supporting our
first general conclusion: Deviations with respect to the hypothesis of a well-mixed population
(including nonlinear dependencies of the fitness on the payoff or not) have a large influence on
the outcome of the evolutionary process and in a majority of cases do change the equilibria
structure, stability and/or basins of attraction.

The specific question of the existence of different time scales was discussed in
Section~\ref{sec:3}. This is a problem that has received some attention in economics but
otherwise it has been largely ignored in biological contexts. In spite of this, we have
shown that considering fast evolution in the case of Ultimatum game may lead to
a non-trivial, unexpected conclusion: That individual selection may be enough to explain
the experimental evidence that people do not behave rationally. This is an important point
in so far as, to date, simple individual selection was believed not to provide an understanding
of the phenomena of altruistic punishment reported in many experiments \cite{camerer:2003}.
We thus see that the effect of different time scales might be determinant and therefore must be
considered among the relevant factors with an influence on evolutionary phenomena.

This conclusion is reinforced by our general study of symmetric $2 \times 2$
symmetric games, that
shows that the equilibria of about half of the possible games change when considering
fast evolution. Changes are particularly surprising in the case of the Harmony game, in
which it turns out that when evolution is fast, the selected strategy is the ``wrong'' one,
meaning that it is the less profitable for the individual and for the population. Such a
result implies that one has to be careful when speaking of adaptation through natural
selection, because in this example we have a situation in which selection leads to a
bad outcome through the influence of fluctuations. It is clear that similar instances may
arise in many other problems. On the other hand, as for the particular question of the
emergence of cooperation,
our results imply that in the framework of the classical $2 \times 2$ social
dilemmas, fast
evolution is generally bad for the appearance of cooperative behavior.

The results reported here concerning the effect of time scales on evolution are only the
first ones in this direction and, clearly, much remains to be done. In this respect, we
believe that it would be important to work out the case of asymmetric $2 \times
2$ games,
trying to reveal possible general conclusions that apply to families of them. The work
on the Ultimatum game \cite{sanchez:2005} is just a first example, but no systematic
analysis of asymmetric games has been  carried out. A subsequent extension to
games with more strategies would also be desirable; indeed, the richness of the
structures arising in those games (such as, e.g., the rock-scissors-papers game
\cite{hofbauer:1998}) suggests that considering fast evolution may lead to quite
unexpected results. This has been very recently considered in the framework of the
evolutionary minority game \cite{challet:1997} (where many strategies are possible,
not just two or three) once again from an economics perspective
\cite{zhong:2009}; the conclusion of this paper, namely that there is a phase transition as
a function of the time scale parameter that can be observed in the predictability of market behavior is a further hint of the interest of this problem.

In Section~\ref{sec:4} we have presented a global view of the influence of
population structure on evolutionary games. We have seen a rich variety of
results, of unquestionable interest, but that on the downside reflect the
non-generality of this kind of evolutionary models. Almost every detail in the
model matters on the outcome, and some of them dramatically.

We have provided evidence that population structure
does not necessarily promote cooperation in evolutionary game theory, showing instances in which
population structure enhances or inhibits it. Nonetheless, we have identified
two topological properties, network clustering and degree heterogeneity, as
those that allow a more unified approach to the characterization and
understanding of the influence of population structure on evolutionary games.
For certain subset of update rules, and for some subregion in the space of
games, they induce consistent modifications in the outcome. In summary, network
clustering has a positive impact on cooperation in Stag Hunt games and degree
heterogeneity in Snowdrift games. Therefore, it would be reasonable to expect
similar effects in other networks which share these key topological properties.
In fact, there is another topological feature of networks that conditions
evolutionary games, albeit of a different type: The community structure
\cite{newman:2003,boccaletti:2006}. Communities are subgraphs of densely
interconnected nodes, and they represent some kind of mesoscopic organization. A
recent study \cite{lozano:2008} has pointed out that communities may have their
own effects on the game asymptotics in otherwise similar graphs, but more work
is needed to assess this influence.

On the other hand, the importance of the update rules cannot be overstated. We
have seen that for the best response and Fermi rules even these ``robust''
effects of population structure are greatly reduced. It is very remarkable from
a application point of view that the influence of population structure is
inhibited so greatly when update rules more sophisticated than merely imitative
ones are considered, or when the selection pressure is reduced. It is evident
that a sound justification of several aspects of the models is mandatory for
applications. Crucial details, as the payoff structure of the game, the
characteristics of the update rule or the main topological features of the
network are critical for obtaining significant results. For the same reasons,
unchecked generalizations of the conclusions obtained from a particular model,
which go beyond the kind of game, the basic topology of the network or the
characteristics of the updated rule, are very risky in this field of research.
Very easily the evolutionary outcome of the model could change dramatically,
making such generalizations invalid.

This conclusion has led a number of researchers to address the issue from a further
evolutionary viewpoint: Perhaps, among the plethora of possible networks one can
think of, only some of them (or some values of their magnitudes) are really important,
because the rest are not found in actual situations. This means that networks themselves
may be subject to natural selection, i.e., they may co-evolve along with the game under
consideration. This promising idea has already been proposed
\cite{zimmermann:2004,marsili:2004,eguiluz:2005,santos:2006b,pacheco:2006,ohtsuki:2007,fosco:2007,pacheco:2008}
and a number of interesting results, which would deserve a separate review on their own
right\footnote{For a first attempt, see Sec.~5 of \cite{gross:2008}.}, have been
obtained regarding the emergence of cooperation. In this respect, it has been
observed that co-evolution seems to favor the
stabilization of cooperative behavior, more so if the network is not rewired
from a
preexisting one but rather grows upon arrival of new players \cite{poncela:2008}.
A related approach, in which the dynamics of the interaction network results from the mobility of
players over a spatial substrate, has been the focus of recent works \cite{sicardi:2009,helbing:2009}.
Albeit these lines of research are appealing and natural when one thinks of possible
applications, we believe the same caveat applies: It is still too early to draw general
conclusions and it might be that details would be again important. Nevertheless,
work along these lines is needed to assess the potential applicability of these
types of models. Interestingly, the same approach is also being introduced to
understand which strategy update rules should be used, once again as a manner
to discriminate among the very many possibilities. This was pioneered by
Harley \cite{harley:1981} (see also the book by Maynard Smith
\cite{maynard-smith:1982},
where the paper by Harley is presented as a chapter) and a few works have appeared
in the last few years
\cite{kirchkamp:1999,moyano:2009,szabo:2008,szolnoki:2008,szolnoki:2009};
although the available results are too specific to allow for a glimpse of any general
feature, they suggest that continuing this research may render fruitful results.

We thus reach our main conclusion: The outcome of
evolutionary game theory depends to a large extent on the details,
a result that has very important implications
for the use of evolutionary game theory to model actual biological, sociological or
economical systems. Indeed, in view of this lack of generality, one has to look carefully
at the main factors involved in the situation to be modeled because they need to
be included as close as necessary to reality to produce conclusions relevant for
the case of interest. Note that this does not mean that it is not possible to
study evolutionary
games from a more general viewpoint; as we have seen above, general conclusions
can be drawn, e.g., about the beneficial effects of clustering for cooperation or the
key role of hubs in highly heterogeneous networks. However, what we do mean is
that one should not take such general conclusions for granted when thinking of a
specific problem or phenomenon, because it might well be that some of its specifics
render these abstract ideas unapplicable. On the other hand, it might be possible that
we are not looking at the problem in the right manner; there may be other magnitudes
we have not identified yet that allow for a classification of the different games and
settings into something similar to universality classes. Whichever the case, it seems
clear to us that much research is yet to be done along the lines presented here.
We hope that this review encourages others to walk off the beaten path in order
to make substantial contributions to the field.

\section*{Acknowledgements}

This work has been supported by projects MOSAICO, from the Spanish Ministerio de Educaci\'on y
Ciencia, and MOSSNOHO and SIMUMAT, from the Comunidad Aut\'onoma de Madrid.

\bibliography{evol-coop}

\appendix

\section{Characterization of birth-death processes}
\label{app:A}

One of the relevant quantities to determine in a birth-death process is the probability $c_n$ that,
starting from state $n$, the process ends eventually absorbed into the absorbing state $n=N$. There is a simple relationship between $c_n$ and the stochastic matrix $P$,
namely
\begin{equation}
c_n=P_{n,n-1}c_{n-1}+P_{n,n}c_n+P_{n,n+1}c_{n+1}, \qquad 0<n<N,
\label{eq:absorption}
\end{equation}
with the obvious boundary conditions $c_0=0$ and $c_N=1$. The solution to this equation is
\cite{karlin:1975}
\begin{equation}
c_n=\frac{Q_n}{Q_N}, \qquad Q_n=\sum_{j=0}^{n-1}q_j, \qquad
q_0=1, \qquad q_j=\prod_{i=1}^j\frac{P_{i,i-1}}{P_{i,i+1}} \quad (j>0).
\label{eq:cQq}
\end{equation}
Another relevant quantity is $v_{k,n}$, the expected number of visits that, starting from state $k$, the
process pays to site $n$ before it enters one absorbing state. If $V=(v_{k,n})$, with $0<k,n<N$,
then
\begin{equation}
V=I+R+R^2+\cdots=(I-R)^{-1},
\label{eq:visits}
\end{equation}
where $I$ is the identity matrix and $R$ is the submatrix of $P$ corresponding to the
transient (non-absorbing) states. The series converges because $R$ is substochastic
\cite{grinstead:1997}. Thus $V$ fulfills the equation $V=VR+I$, which amounts to an equation
similar to (\ref{eq:absorption}) for every row of $V$, namely
\begin{equation}
v_{k,n}=v_{k,n-1}P_{n-1,n}+v_{k,n}P_{n,n}+v_{k,n+1}P_{n+1,n}+\delta_{k,n}, \qquad 0<k,n<N,
\end{equation}
where $\delta_{k,n}=1$ if $k=n$ and $0$ otherwise. Contrary to what happens with
Eq.~(\ref{eq:absorption}), this equation has no simple solution and it is better solved as
in (\ref{eq:visits}). Finally, $\tau_k$, the number of steps before absorption occurs into any absorbing state, when starting at state $k$, is obtained as
\begin{equation}
\label{eq:tau}
\tau_k=\sum_{n=1}^{N-1}v_{k,n}.
\end{equation}

\section{Absorption probability in the hypergeometric case}
\label{app:B}

For the special case in which
\begin{equation}
\frac{P_{n,n-1}}{P_{n,n+1}}=\frac{\alpha n+\beta}{\alpha(n+1) + \gamma}
\end{equation}
the absorption probability into state $n=N$, $c_n$, can be obtained in closed form. According to
(\ref{eq:cQq}) the sequence $q_j$ fulfills the hypergeometric relation
\begin{equation}
\frac{q_j}{q_{j-1}}=\frac{\alpha j+\beta}{\alpha (j+1)+\gamma},
\end{equation}
from which
\begin{equation}
(\alpha (j+1)+\gamma)q_j=(\alpha j+\beta)q_{j-1}.
\end{equation}
Adding this equation up for $j=1,\dots,n-1$ we get
\begin{equation}
\alpha\sum_{j=1}^{n-1}(j+1)q_j+\gamma(Q_n-1)=\alpha\sum_{j=0}^{n-2}(j+1)q_j+\beta(Q_n-q_{n-1}),
\end{equation}
and therefore
\begin{equation}
\label{eq:Qq}
(\gamma-\beta)Q_n=\gamma+\alpha-(\beta+\alpha n)q_{n-1}.
\end{equation}
Thus, provided $\gamma\ne\beta$, we obtain
\begin{equation}
Q_n=\frac{\gamma+\alpha}{\gamma-\beta}\left[1-\prod_{j=1}^n\frac{\alpha j+\beta}{\alpha j+\gamma}\right].
\end{equation}
If $\alpha=0$ this has the simple form
\begin{equation}
Q_n=\frac{\gamma}{\gamma-\beta}\big[1-(\beta/\gamma)^n\big].
\end{equation}
If $\alpha\ne 0$, then we can rewrite
\begin{equation}
Q_n=\frac{\gamma+\alpha}{\gamma-\beta}\left[1-\frac{\Gamma(\beta/\alpha+n+1)\Gamma(\gamma/\alpha+1)}
{\Gamma(\gamma/\alpha+n+1)\Gamma(\beta/\alpha+1)}\right].
\label{eq:Qnfinal}
\end{equation}
The case $\gamma=\beta$ can be obtained from (\ref{eq:Qq}) or as the limit of
expression (\ref{eq:Qnfinal}) when
$\gamma\to\beta$. Both ways yield
\begin{equation}
Q_n=\sum_{j=1}^n\frac{\alpha+\gamma}{\alpha j+\gamma}.
\end{equation}

\end{document}